\pdfoutput=1 
\documentclass[12pt,letterpaper]{JHEP3}

\usepackage{amsmath,amssymb,amsfonts,slashed,xspace,undertilde}

\usepackage{youngtab}

\usepackage{mathrsfs}
\usepackage{dsfont}
\usepackage{amssymb}
\usepackage{wasysym}
\usepackage{graphicx}

\title{Extremal black holes, nilpotent orbits \vspace{2mm} \\* and the true fake superpotential}

\author{Guillaume Bossard$^1$, Yann Michel, Boris Pioline$^{2}$,  \\

$^1$AEI, Max-Planck-Institut f\"{u}r Gravitationsphysik\\
Am M\"{u}hlenberg 1, D-14476 Potsdam, Germany\\
\\
$^2$ Laboratoire de Physique Th\'eorique et Hautes
Energies\footnote{Unit\'e mixte de recherche du CNRS UMR 7589},\\
Universit\'e Pierre et Marie Curie - Paris 6,
4 place Jussieu, F-75252 Paris cedex 05 \\
\\
{\tt bossard@aei.mpg.de, ymichelpro@gmail.com, pioline@lpthe.jussieu.fr}}

\preprint{AEI-2009-076\\arXiv:0908.1742v2}

\abstract{
Dimensional reduction along time offers a powerful way to study
stationary solutions of 4D symmetric supergravity models
via group-theoretical methods. We apply this approach 
systematically to extremal, BPS and non-BPS,  
spherically symmetric  black holes, and obtain  
their ``fake superpotential" $W$. The latter provides first order 
equations for the radial problem, governs the mass and entropy formula and 
gives the semi-classical approximation to the radial wave function. 
To achieve this goal, we note that  the Noether charge for the radial
evolution must lie in a certain Lagrangian submanifold of a nilpotent orbit of the 3D 
continuous duality group, and construct  a suitable parametrization of this  Lagrangian.
For general non-BPS extremal black holes in $\N=8$ supergravity, 
$W$ is obtained by solving a non-standard diagonalization problem, which reduces
to a sextic polynomial in $W^2$ whose coefficients are $SU(8)$ invariant functions
of the central charges. 
By consistent truncation we obtain $W$ for other supergravity models with a symmetric moduli space. In particular, for the one-modulus $S^3$ model, $W^2$ is given
explicitely as the root of a cubic polynomial.  The STU model is investigated in detail and the nilpotency of the Noether charge is checked on explicit solutions. }

\newcommand{\color}[6]{}
\allowdisplaybreaks[1]

\def\bea{\begin{eqnarray}}
\def\eea{\end{eqnarray}}
\def\be{\begin{equation}}
\def\ee{\end{equation}}
\def\ba{\begin{align}}
\def\ea{\end{align}}
\def\bse{\begin{subequations}}
\def\ese{\end{subequations}}

\renewcommand{\Im}{{\rm Im}}
\renewcommand{\Re}{{\rm Re}}
\newcommand{\Pfaff}{{\rm Pfaff}}
\newcommand{\bZ}{{\bar Z}}

\newcommand{\pa}{\partial}
\newcommand{\Tr}{{\rm Tr}}
\newcommand{\nn}{\nonumber}

\newcommand{\vareps}{\varepsilon}
\newcommand{\IR}{\mathbb{R}}
\newcommand{\IC}{\mathbb{C}}

\newcommand{\IZ}{\mathbb{Z}}

\newcommand{\cA}{\mathcal{A}}

\newcommand{\cN}{\mathcal{N}}
\newcommand{\cM}{\mathcal{M}}
\newcommand{\cO}{\mathcal{O}}
\newcommand{\cS}{\mathcal{S}}
\newcommand{\tzeta}{{\tilde\zeta}}

\newcommand{\fg}{\mathfrak{g}}

\newcommand{\I}{{i}}

\def\un{{\mathpzc{1}}}
\def\deux{{\mathpzc{2}}}
\def\trois{{\mathpzc{3}}}
\def\quatre{{\mathpzc{4}}}

\newcommand{\sfrac}[2]{{\scriptstyle \frac{#1}{#2}}}

\newcommand{\Scal}[1]{\Bigl ({#1} \Bigr )}
\newcommand{\scal}[1]{\bigl ({#1} \bigr )}
\newcommand{\CR}{\nonumber \\*}

\newcommand{\trace}{\hbox {Tr}~}

\def\C{\mathscr{C}}

\DeclareMathAlphabet{\mathpzc}{OT1}{pzc}{m}{it}

\DeclareMathOperator{\ad}{ad}
\DeclareMathOperator{\Ad}{Ad}

\DeclareMathOperator{\Esh}{Esh}

\newcommand{\gra}[2]{{\scriptscriptstyle (#1 , #2 )}}
\newcommand{\ord}[1]{{\scriptscriptstyle (#1)}}

\def\ie{{\it i.e.}\ }
\def\eg{{\it e.g.}\ }
\def\nn{\nonumber}

\def\N{\mathcal{N}}

\def\C{{\mathscr{C}}}
\def\V{{\mathcal{V}}}
\def\w{{\scriptstyle W}}

\def\P{{\mathfrak{P}}}
\def\M{{\mathcal{M}}}

\def\k{\mathfrak{k}}

\def\Nil{\mathfrak{N}}

\def\Spin{Spin^*_{\scriptscriptstyle \rm c}(16)}

\def\m{{\mathpzc{m}}}
\def\n{{\mathpzc{n}}}
\def\p{{\mathpzc{p}}}

\def\zero{{\mathpzc{0}}}
\def\un{{\mathpzc{1}}}
\def\deux{{\mathpzc{2}}}
\def\trois{{\mathpzc{3}}}
\def\quatre{{\mathpzc{4}}}

\def\invo{{\APLstar}}

\def\DJo{$\;$\kern-.4em \hbox{D\kern-.8em\raise.15ex\hbox{--}\kern.35em okovi\'c}}

\newcommand{\eprint}[1]{{\href{http://arxiv.org/abs/#1}{\texttt{[#1}]}}}
\newcommand{\eprintN}[1]{{\href{http://arxiv.org/abs/#1}{\texttt{#1 [hep-th]}}}}

\def\e{\boldsymbol{e}}

\def\asym{{\scriptscriptstyle 0}}

\def\w{{\scriptstyle W}}

\def\gl{\mathfrak{gl}}
\def\sl{\mathfrak{sl}}
\def\so{\mathfrak{so}}
\def\su{\mathfrak{su}}

\def\e{\mathfrak{e}}

\def\SU{SU_{\scriptscriptstyle \rm c}(8)}
\def\Sp{USp_{\scriptscriptstyle \rm c}(8)}

\def\nn{\nonumber}
\def\N{\mathcal{N}}

\def\DJo{$\;$\kern-.4em \hbox{D\kern-.8em\raise.15ex\hbox{--}\kern.35em okovi\'c}}

\def\DEVIII#1#2#3#4#5#6#7#8{{\tiny $ { \left[ \begin{array}{ccccccc}  & & \mathfrak{#2} \hspace{-1.2mm}&&&& \vspace{ -1.2mm} \\ \mathfrak{#1}\hspace{-1.2mm} &  \mathfrak{#3} \hspace{-1.2mm}& \mathfrak{#4} \hspace{-1.2mm} & \mathfrak{#5}\hspace{-1.2mm}&\mathfrak{#6}\hspace{-1.2mm}&\mathfrak{#7}\hspace{-1.2mm}&\mathfrak{#8} \end{array}\right] }$}}
\def\DSOXVI#1#2#3#4#5#6#7#8{{\tiny $ {  \vspace{-2mm} \left[ \begin{array}{ccccccccc}  && \mathfrak{#8} \hspace{-1.2mm}&&&&&& \vspace{ -1.3mm} \\ \cdot \hspace{-1.0mm}& \mathfrak{#7}\hspace{-1.2mm} &\mathfrak{#6}\hspace{-1.2mm} &  \mathfrak{#5} \hspace{-1.2mm}& \mathfrak{#4} \hspace{-1.2mm} & \mathfrak{#3}\hspace{-1.2mm}&\mathfrak{#2}\hspace{-1.2mm}&\mathfrak{#1}  \hspace{-1.2mm}\end{array}\right] }$}}

\def\DEVII#1#2#3#4#5#6#7{{\tiny $ { \left[ \begin{array}{cccccc}  & & \mathfrak{#2} \hspace{-1.2mm}&&& \vspace{ -1.0mm} \\ \mathfrak{#1}\hspace{-1.2mm} &  \mathfrak{#3} \hspace{-1.2mm}& \mathfrak{#4} \hspace{-1.2mm} & \mathfrak{#5}\hspace{-1.2mm}&\mathfrak{#6}\hspace{-1.2mm}&\mathfrak{#7} \end{array}\right] }$}}
\def\DSOXII#1#2#3#4#5#6#7{{\tiny $ {   \left[ \begin{array}{ccccccc}  && & \mathfrak{#6} \hspace{-1.2mm}&&& \vspace{ -1.3mm} \\  \mathfrak{#1}\hspace{-1.2mm} &\mathfrak{#2}\hspace{-1.2mm} &  \mathfrak{#3} \hspace{-1.2mm}& \mathfrak{#4} \hspace{-1.2mm} & \mathfrak{#5}\hspace{-1.2mm}&\cdot \hspace{-1.0mm}& \mathfrak{#7} \end{array}\right] }$}}

\def\DEVIISL#1#2#3#4#5#6#7#8{{\tiny $ { \left[ \begin{array}{cccccccc}  & & \mathfrak{#2} \hspace{-1.2mm}&&&&& \vspace{ -1.0mm} \\ \mathfrak{#1}\hspace{-1.2mm} &  \mathfrak{#3} \hspace{-1.2mm}& \mathfrak{#4} \hspace{-1.2mm} & \mathfrak{#5}\hspace{-1.2mm}&\mathfrak{#6}\hspace{-1.2mm}&\mathfrak{#7}\hspace{-1.2mm}&\cdot \hspace{-1.0mm}& \mathfrak{#8}  \end{array}\right] }$}}

\def\DSOVIII#1#2#3#4{{\tiny $ {   \left[ \begin{array}{ccc}  &&\mathfrak{#2}  \vspace{ -1.4mm} \\  \mathfrak{#1}\hspace{-1.2mm} &\mathfrak{#4} \hspace{-1.3mm}&\vspace{-1.4mm}\\ && \mathfrak{#3}  \end{array}\right] }$}}

\begin{document}

\section{Introduction}

In trying to extend our  string-theoretic understanding of black holes
away from the supersymmetric regime, extremality is often a key simplifying assumption. 
Firstly, it 
eliminates Hawking radiation and ensures that the solution is semi-classically
stable. Secondly, it implies the existence of 
an $AdS$ region which may admit a dual conformal
field theory description. Thirdly, it guarantees that the near-horizon 
solution is entirely determined by the conserved charges measurable at spatial infinity,
and therefore insensitive (away from lines of marginal stability) 
to variations to the moduli at infinity. This attractor behavior,
first discovered for supersymmetric (BPS) black holes \cite{Ferrara:1995ih,Ferrara:1996um},
holds for all extremal solutions \cite{Ferrara:1997tw,Sen:2005wa,
Goldstein:2005hq}, and is arguably responsible for the validity 
of certain weakly coupled description of non-BPS black hole micro-states \cite{Dabholkar:2006tb}. 

However, while the most general BPS solution is known explicitly \cite{Denef:2000nb,Bates:2003vx}, 
our ability to construct non-BPS extremal solutions is quite limited. Early
solutions were found in \cite{Khuri:1995xq, Ortin:1996bz} by an astute embedding
of the Reissner-Nordstr\"om solution in $\N=8$ supergravity, while some 
solutions were studied numerically in \cite{Tripathy:2005qp}. More recently, it was shown how
to deduce first-order equations for non-BPS extremal solutions from 
a ``fake superpotential" \cite{Ceresole:2007wx,Andrianopoli:2007gt,
Lopes Cardoso:2007ky,Perz:2008kh}, and some solutions were obtained. Unfortunately, this strategy has suffered
from the lack of a general method to construct the fake superpotential. 
A generic 5-parameter  seed solution was obtained and analyzed 
in \cite{Lopes Cardoso:2007ky,Hotta:2007wz,Gimon:2007mh,Bellucci:2008sv,Gimon:2009gk}.
Using dimensional reduction 
along the time direction, non-BPS
solutions of the one-modulus $\N=2$ supergravity model were obtained
in \cite{Gaiotto:2007ag} via the determination of their nilpotent Noether charge. 
The nilpotent orbits\footnote{The relation between extremal black holes and nilpotent orbits was first uncovered in the BPS case in \cite{Gunaydin:2005mx}, generalized
to 5D black holes in \cite{Berkooz:2008rj}, and has been further developed since 
in \cite{Michel:2008bx,Bossard:2009at}.} associated to extremal black holes have been constructed 
in \cite{Bossard:2009my,Bossard:2009mz}.

In this paper, we build upon the insights of \cite{Ceresole:2007wx,Gaiotto:2007ag,Bossard:2009my} 
and give a systematic method to construct the 
fake superpotential  for non-BPS extremal solutions in supergravity
models with a symmetric moduli space. In particular, we derive the 
fake superpotential for extremal non-BPS black holes in $\N= 8$ 
supergravity in full generality, and in magic $\cN=2$ supergravity
models by consistent truncation. Before giving the outline of the paper,
we start by briefly reviewing the 
dimensional reduction, fake superpotential and nilpotent orbit
techniques which underlie our approach.

\subsection{Radial evolution and geodesic motion\label{secgeo}}

Stationary solutions in $D=4$ supergravity are efficiently studied by reduction
along the time direction \cite{Breitenlohner:1987dg,Gunaydin:2005mx, Clement:1996nh,
Bergshoeff:2008be}
(see e.g. \cite{Pioline:2006ni} for a review). 
After dualizing  the one-forms in three dimensions and restricting to 
weakly extremal\footnote{We define weak extremality as 
the condition that the three-dimensional spacial slices be flat. In order to be extremal,
such a solution must also be smooth.}   solutions, one obtains a non-linear sigma model 
with pseudo-Riemannian target space 
\be
\cM_3^* \sim \IR^+ \times \cM_4 \times T \times S^1\ , \label{Mstructure}
\ee 
where  
\begin{itemize}
\item $\IR^+$ is parametrized by the scale function $U$ in the metric ansatz
\be
\label{4dmetric}
ds^2 = -e^{2U} (dt+\omega)^2+ e^{-2U} \scal{ dr^2 + r^2 (d\theta^2
+\sin^2\theta \, d\phi^2)}\ ,
\ee
\item 
$\cM_4$ is the moduli space of massless scalar fields in 4 dimensions, 
with coordinates $\phi^i$ and metric $g_{ij}$, 
\item $T$ is a $2n_V-$ dimensional symplectic torus parametrized by 
the Wilson lines $\zeta^\Lambda,\tzeta_\Lambda$ of $A^\Lambda$ and its magnetic dual $A_\Lambda$
around the time direction, and 
\item $S^1$ is the fibre of a circle bundle over $T$
parametrized by the NUT potential $\sigma$ dual to the off-diagonal metric one-form $\omega$.  
Its first Chern class is proportional to the canonical
symplectic form $d\zeta^\Lambda \wedge d\tzeta_\Lambda$.
\end{itemize}
We shall denote the coordinates $U,\phi^i,\zeta^\Lambda,\tzeta_\Lambda,\sigma$ on $\cM_3^*$ collectively
as $\phi^\mu$, and $U,\phi^i$ as $\phi^a$. As indicated by the $\sim$ sign, the metric on $T\times S^1$
varies over the base $\IR^+ \times \cM_4$,\footnote{Whereas this fibration over  $\IR^+ \times \cM_4$ is globally defined on the Riemannian target space $\cM_3$ which appears in the space-like reduction, it only holds on a dense open set of $\cM_3^*$  homeomorphic to $\cM_3$. Nevertheless, the complement of this open set has support at $U = - \infty$, and this subtlety appears to be irrelevant for the purposes of this paper.} 
being positive
definite along $S^1$ and negative definite along $T$,
while the metric on the base 
is $\frac12 dU^2 + g_{ij} d\phi^i d\phi^j\equiv g_{ab} d\phi^a d\phi^b$, positive
definite. The negative signature along $T$ can be traced to the negative signature
of the time direction along which   
the dimensional reduction is carried out.  
 In the context of $\N=2$ supergravity, $\cM_4$ is special K\"ahler, and $\cM_3^*$,
 the ``$c^*$-map" of $\cM_4$, is  related to 
the ``$c$-map" of \cite{Ferrara:1989ik} by analytic continuation $(\zeta,\tzeta,\sigma)
\mapsto (i \zeta,i \tzeta,- \sigma)$. Under this dimensional reduction, stationary solutions of 4D supergravity become harmonic maps from $\IR^3$ to $\cM_3^*$, with pointwise
vanishing Lagrangian density (this latter condition follows from the restriction to
weakly extremal solutions).

Assuming in addition spherical symmetry, the supergravity equations of motion become equivalent to light-like geodesic motion on $\cM_3^*$, with the affine parameter $\tau$ identified as the inverse
radial distance $\tau=1/r$. The conserved Noether charges along the twisted torus
$T\times S^1$ are identified as the electric, magnetic and NUT charges $q_\Lambda, p^\Lambda, k$, respectively. 
Static solutions have zero NUT charge $k=0$. In this case the Hamiltonian 
for light-like geodesic motion on $\cM_3^*$ becomes independent of 
$\zeta^\Lambda,\tzeta_\Lambda$, and reduces to the Hamiltonian
for the motion of a fiducial particle on $\IR^+ \times \cM_4$ subject to the 
potential $V$, 
\be
\label{ham0}
H = \frac12 p_a g^{ab} p_b + V(p,q; \phi^a)\equiv 0\ .
\ee
Here $p_a=g_{ab} \dot \phi^b$ is the momentum conjugate to $\phi^a$, 
the dot denotes the derivative
with respect to $\tau$, and $V\equiv -e^{2U} V_{\rm \scriptscriptstyle BH}$ 
encodes the (negative definite) kinetic energy along $T$.
The latter depends quadratically on the charges $p,q$, and is proportional to 
$V_{\rm \scriptscriptstyle BH}$, which  is sometimes called the ``black hole potential". 

\subsection{Extremal solutions and fake superpotential}

While the condition that the spatial slices be flat is necessary for extremality, it is
by no means sufficient. In order that the solution be smooth, one must
fine-tune the 
boundary conditions at spatial infinity so that  the particle 
reaches the top of the potential hill in infinite proper time and with zero velocity,
$p_a(\tau=\infty)=0$.
For fixed electromagnetic charges, this fine-tuning holds only on a
Lagrangian subspace of the phase space $(\phi^a,p_a)$ of initial conditions
at~$\tau=0$. 

There can be different ways of performing this fine-tuning. In supergravity 
models with extended supersymmetry, one may impose the existence of Killing
spinors to obtain first order equations which relate the momentum $p_a$ to 
the coordinate $\phi^a$, and guarantee that the second order equations of 
motion are obeyed. The resulting solution then preserves some fraction
of supersymmetry and, if smooth, will also be extremal. This is most familiar
in the framework of $D=4, \cN=2$ supergravity, where BPS black holes 
satisfy the``attractor flow'' equations
\be
\label{attflow}
p_a = - \pa_{\phi^a} \cS_{\rm \scriptscriptstyle BPS} \ ,
\ee
\ie follow the gradient flow of the potential
\be
\label{BPSS}
\cS_{\rm \scriptscriptstyle BPS} (\phi^a)= e^{U} W_{\rm \scriptscriptstyle BPS} \ ,\qquad W_{\rm \scriptscriptstyle BPS} = |Z_{p,q}(\phi^i)| \ .
\ee
The potential $\cS_{\rm \scriptscriptstyle BPS} $ and ``superpotential" $W_{\rm \scriptscriptstyle BPS}$ satisfy 
\be
\label{vbhsuperbps}
V  = - g^{ab}\, \pa_a \cS_{\rm \scriptscriptstyle BPS} \, \pa_b \cS_{\rm \scriptscriptstyle BPS} 
= - e^{2U} \left(  W_{\rm \scriptscriptstyle BPS}^2+ 2 g^{ij}\pa_i W_{\rm \scriptscriptstyle BPS} \pa_{j} W_{\rm \scriptscriptstyle BPS} \right)\ ,
\ee
such that on solutions of \eqref{attflow}, the positive kinetic energy compensates the 
potential term $V(\phi_a)<0$, ensuring that the spatial slices $ds_3^2$ are flat. 
Moreover, they guarantee that a maximum of $V$
(or minimum of $V_{\rm \scriptscriptstyle BH}$) 
is reached at zero momentum, provided this extremum occurs at
a regular attractor point $|Z_*|>0$ \cite{Moore:1998pn}. 
The restriction of  \eqref{attflow}  at $\tau=0$ define the ``BPS"
component of the  Lagrangian subspace of extremal solutions,
and corresponds to BPS black holes with ADM mass and 
Bekenstein-Hawking entropy given by
\be
\label{massent}
2 G M =  W_{\rm \scriptscriptstyle BPS}(\tau=0) \ ,\qquad
S_{\rm \scriptscriptstyle BH} = \pi W_{\rm \scriptscriptstyle BPS}^2(\tau=\infty)\ .
\ee

There may however exist other disconnected components of the Lagrangian subspace
of  extremal solutions corresponding to non-BPS black holes. As shown 
in \cite{Ceresole:2007wx,Andrianopoli:2007gt}, some non-BPS solutions can be
obtained in a similar way as the BPS ones, provided there 
exists another function $\cS(\phi^a)=  e^{U} W(\phi^i)$, where $W(\phi^i)$
is dubbed the ``fake superpotential", 
such that  the potential $V$ can be written as in \eqref{vbhsuperbps},
\be
\label{vbhsuper}
V=  - e^{2U} \left( W^2+2 g^{ij}\pa_i W\pa_{j}W \right) = - g^{ab}\pa_a \cS \, \pa_b \cS \ . 
\ee
The first order equations
\be
\label{attflow2}
p_a = - \pa_{\phi^a} \cS(\phi^a)\ ,
\ee
then imply, just as in the BPS case, that the 
second order equations of motion are satisfied, that the kinetic and potential
energy compensate each other, and that 
the solutions reach a critical point of the potential 
at zero velocity. The first order equations \eqref{attflow2} at  $\tau=0$ 
therefore provide another component of the  Lagrangian subspace of extremal solutions,
and correspond to extremal non-BPS black holes with
mass and Bekenstein-Hawking entropy given by \eqref{massent}
where $W_{\rm \scriptscriptstyle BPS}$ is replaced by $W$. In contrast to the BPS case, the 
critical point of $W$ is not guaranteed to be an isolated maximum, but could exhibit
flat directions or even saddle behavior;
in the presence of  flat directions, some of the 
scalars at the horizon are determined uniquely by the conserved charges, although
the entropy will be independent of the asymptotic value of the scalars \cite{Sen:2005wa}. 
Using this method, non-BPS extremal black holes for the $STU$ model
were obtained in  \cite{Ceresole:2007wx} in the axion-free case\footnote{A fake superpotential 
for non-zero axion was postulated in \cite{Bellucci:2007zi},
but it depends explicitely on the flat directions and its
status is unclear to us.}.
Unfortunately, there has been no systematic way of computing $W$
without solving for the full problem (although, in some cases, one may 
engineer different fake superpotentials $W$ for the same potential $V$
using discrete symmetries). The purpose of this paper is to give a method to determine 
$W$ {\it a priori} for symmetric supergravity models.

\subsection{Fake superpotential and radial wave function}

As a side remark, we note that the potential $ \cS(\phi^a)$ may be identified, 
by virtue of  \eqref{attflow2}, as the generating function of  the Lagrangian subspace 
of the ``small phase space" $T_*(\IR^+\times \cM_4)$ corresponding 
to smooth  extremal solutions with fixed values of the
electromagnetic charges (see  \cite{Andrianopoli:2009je}
for a related discussion). By construction, $\cS$
solves the Hamilton-Jacobi equation\footnote{Hamilton's principle function is usually a function of the position variables at time $t$, 
canonical momenta at initial time $0$, and time $t$ itself. The electric and magnetic charges $p^{\Lambda}, q_{\Lambda}$ in \eqref{defS}
can be regarded as the values of the canonical momenta at $t=0$, while the absence
of explicit time dependence is a consequence of weak extremality, $H=0$. } 
associated to the Hamiltonian \eqref{ham}
\be
H( \pa_{\phi^a} \cS , \phi^a) = \pa_{t} \cS = 0\ .
\ee
Equivalently,
both the first order equations  \eqref{attflow2} and the relations\footnote{The numerical factors in the
forthcoming relations are convention-dependent, and have been chosen consistently with the coordinates used for the $STU$ model in 
Section \ref{secstu}.}
\be
p^\Lambda  =  -\frac{\sqrt2}{2}  
( p_{\tzeta_\Lambda}-\zeta^\Lambda \, p_\sigma )\ ,\qquad
q_\Lambda  = -\frac{\sqrt2}{2} 
( p_{\zeta^\Lambda}+\tzeta_\Lambda \, p_\sigma ) \ ,\qquad
k=-2 p_\sigma
\ee
between the charges $q_\Lambda, p^\Lambda,k$ and the canonical momenta $p_{\zeta^\Lambda},
p_{\tzeta_\Lambda}, p_\sigma$ can be derived from the generating function 
on the ``large phase space" $T_*(\cM_3^*)$
\be
\label{defS}
\tilde \cS(\phi^\mu) =  -4  \cS(\phi^a) 
+\sqrt2 (q_{\Lambda}\zeta^{\Lambda}+p^{\Lambda}\tilde{\zeta}_{\Lambda})\ ,
\ee
via $p_{\phi^\mu}=\pa_{\phi^\mu}  \tilde \cS$.
Upon quantization of the radial evolution of the 
scalars by replacing $p_i=\frac{\I}{\hbar}\pa_{\phi^i}$  \cite{Gunaydin:2005mx,
Gunaydin:2007bg}, 
this generating function determines the semi-classical form of the radial wave function, i.e.
\be
\label{wavef}
\Psi(\phi^a) \sim \exp\left(\frac{\I}{\hbar} \tilde\cS(\phi^a)\right) \sim \exp\left(\frac{\I}{\hbar} 
e^U W(\phi^i) +i \sqrt2 (q_{\Lambda}\zeta^{\Lambda}+p^{\Lambda}\tilde{\zeta}_{\Lambda})
\right)\ .
\ee
For BPS black holes with superpotential \eqref{BPSS}, one recovers
the semi-classical BPS wave function found in  \cite{Gunaydin:2007bg,Neitzke:2007ke}. 
In addition to the usefulness of $W$ for determining the mass, entropy and fine-tuning
at infinity, this relation to the radial wave function
provides extra incentive to study  the fake superpotential for non-BPS
extremal black holes.

\subsection{Extremal black holes and nilpotent orbits}

In this note, we focus on the special case of $\cN\geq 2$ supergravity theories with 
a symmetric moduli space, where group theoretical methods can be used to bear
on this problem. In these cases, both $\cM_4$ and $\cM_3^*$ are symmetric spaces,
\be
\cM_4= K_4 \backslash G_4\ ,\qquad
\cM_3^*=K_3^*\backslash G_3\ .
\ee
Here $K_4$ is the maximal compact subgroup of the continuous 4D duality group $G_4$, while 
$K_3^*$ is a non-compact real form of the maximal compact subgroup of the 
continuous 3D duality group $G_3$. The latter acts isometrically on $\cM_3^*$ by 
right-multiplication, and yields a conserved Noether charge $Q$ valued in the Lie 
algebra\footnote{$Q$ is more naturally valued in the dual Lie algebra $\mathfrak{g}_3^*$, 
but we can identify the two using the Killing form.}
$\mathfrak{g}_3$.
Those include not only the conserved charges for translations along the twisted torus 
$T\times S_1$, \ie the electromagnetic and NUT charges, but also additional
charges corresponding to Ehlers and Harrison transformations, as well as 4D duality
rotations. The geodesic motion on $\cM_3^*$ is integrable, and in fact 
all geodesics on $\cM_3^*$ can be obtained by exponentiating
a generator $ - P_\asym \tau  \in \mathfrak{g}_3 \ominus \mathfrak{k}_3^*$, where
$P_\asym$ determines the momentum along the trajectory. $P_\asym$ is conjugate to the Noether
charge $Q$ via the coset representative $\V$ in $G_3$,
\be 
P \equiv -  \Scal{Ê\dot{\V} \, \V^{-1} }  \big|_{\fg\ominus \mathfrak{k}^*} = \V Q \V^{-1} \ .
\ee

Extremal solutions correspond to  special geodesics which reach the boundary $U=-\infty$ 
in infinite proper time \cite{Breitenlohner:1987dg}. As already mentioned, it
is necessary but not sufficient that the geodesic  be light-like.
For BPS black holes, it was observed in \cite{Gunaydin:2005mx} that the Noether
charge must satisfy $[\ad(Q)]^5=0$, \ie belong to a nilpotent orbit
of degree 5 (See Appendix A for a summary of useful facts about nilpotent orbits).
More precisely, it was shown that the Noether charge defines a 5-grading\footnote{To 
match standard conventions in the mathematics literature on nilpotent orbits, we rescale
the Cartan generator ${\bf h}$ by a factor of 2 compared to \cite{Gunaydin:2005mx}, such that the 5-grading
becomes an even 9-grading;  the original 5-grading with charges ranging from
-2 to 2 corresponds to the minimal orbit.}
of the Lie algebra $\fg_3=\fg^\ord{-4}\oplus \fg^\ord{-2}\oplus\fg^\ord{0}\oplus \fg^\ord{2}\oplus\fg^\ord{4}$
where the top spaces $\fg^\ord{4}$ is one-dimensional.
Upon quantization, the BPS phase space becomes the Hilbert space of the
quaternionic discrete series  of $G_3$ \cite{Gunaydin:2007bg}, closely related 
to the quasiconformal realization \cite{Gunaydin:2007qq}.

For what concerns extremal non-BPS black holes, it was later shown
in the special case of the one-modulus $S^3$ model 
that extremality requires the condition $[Q_{|\bf 7}]^3=0$,
where $\bf 7$ denotes the 7-dimensional representation 
of the 3D duality group $G_{2(2)}$ \cite{Gaiotto:2007ag}. This condition is equivalent to 
$[\ad(Q)]^5=0$ in this particular case. 
$G_{2(2)}$ admits two distinct nilpotent orbits
of degree 5 with the same dimension 10, corresponding to extremal
BPS and non-BPS black holes, respectively.

More recently, the supersymmetry and extremality conditions on the Noether charge 
for symmetric supergravity models were re-analyzed in \cite{Bossard:2009at}.
It was shown in all cases where $G_3$ is simple 
that extremality requires\footnote{An important assumption in \cite{Bossard:2009at}
is that all extremal solutions can be obtained as limits of non-extremal
black hole solutions. Irrespective of this, the condition $[Q_{|\bf R}]^3=0$
must be supplemented by a condition on $P$ as discussed at the end of Section A.1.}
to $[Q_{|\bf R}]^3=0$, where $\bf R$ denotes the ``fundamental representation" of $G_3$:
for example  the spinor representation
if $G_3$ is an orthogonal group $SO(2+m,2+n)$ or $SO^*(2m+4)$.
The only exception is for $G_3=E_{8(8)}$ or $E_{8(-24)}$, where
the condition becomes $[Q_{|{\bf 3875}}]^5=0$, with $\bf 3875$ being the $3875$-dimensional irreducible representation appearing in the symmetric tensor product of two adjoints. 

More precisely, any generic extremal spherically symmetric black hole (\ie with a non-zero horizon area) is characterized by a nilpotent Noether charge $Q$ which lies inside the grade-two component 
$ \mathfrak{l}_4^\ord{2}$ of $\fg_3$ with respect to the 5-grading (more appropriately, even 9-grading)
which arises in the reduction from 4 to 3 dimensions:
\be \fg_3 \cong {\bf 1}^\ord{-4}Ê\oplus \mathfrak{l}_4^\ord{-2}Ê\oplus \scal{Ê\gl_1 \oplus \fg_4}^\ord{0} \oplus  \mathfrak{l}_4^\ord{2} \oplus  {\bf 1}^\ord{4} \ .
\label{DimRedGrade} 
\ee
The nilpotent orbit $\cO_{G_3}$ of $Q\in \fg_3$ under $G_3$ is  
characterized by the isotropy subgroup of $Q$ 
in $G_4$.  For extremal black holes, this isotropy subgroup coincides with the isotropy subgroup of the electromagnetic charges in the four-dimensional duality group $G_4$ computed in 
\cite{Ferrara:1997uz,Bellucci:2006xz}. On the other hand, the momentum $P_\asym$ is valued in the coset 
$\fg_3 \ominus \k_3^*$, and therefore defines a $K_3^*$-orbit $\cO_{K_3^*}$ 
inside $\fg_3 \ominus \k_3^*$. As explained in \cite{Bossard:2009my}, $\cO_{K_3^*}$
 is a Lagrangian submanifold of $\cO_{G_3}$ equipped with its  
 canonical Kirillov--Kostant symplectic form. Parametrizing this Lagrangian will
 be a key step towards computing the fake superpotential.

\subsection{Strategy and main results}

Since the coset component of the Maurer--Cartan form 
is conjugate to the Noether charge via $P = \V Q \V^{-1}$, 
it defines a representative ${\bf e} \equiv P$ of the corresponding nilpotent orbit inside the coset component $\fg_3 \ominus \mathfrak{k}^*$, and therefore defines a $K^*$-orbit inside this coset. A general fact
about nilpotent elements is that one can always find another nilpotent element ${\bf f}$ and
a semi-simple generator ${\bf h}$ such that the triplet
$({\bf e},\, {\bf f},\, {\bf h})$ defines an $\mathfrak{sl}_2$ subalgebra
of $\fg_3$, \ie 
\be
\label{triple}
[{\bf e},{\bf f}]={\bf h}\ , \qquad \framebox{ $[{\bf h},{\bf e}]=2{\bf e} \ ,  $}\qquad [{\bf h},{\bf f}]=- 2{\bf f}\ .
\ee
The eigenspaces of ${\bf h}$ furnish a graded decomposition of $\fg$ which uniquely 
characterizes the complex nilpotent $G_\IC$ orbit \cite{Collingwood}. Extremal solutions are such that the $K^*$-orbit of $P$ is characterized by a graded decomposition of $\mathfrak{k}^*$ of the 
 same form as (\ref{DimRedGrade}) \cite{Bossard:2009my}, 
\be \mathfrak{k}^* \cong \mathfrak{k}^\ord{-4} \oplus \mathfrak{k}^\ord{-2} \oplus \gl_1 \oplus \mathfrak{k}^\ord{0} \oplus  \mathfrak{k}^\ord{2} \oplus  \mathfrak{k}^\ord{4} \ .
\ee
As we shall show explicitly in the framework of $\N=8$ supergravity, for static solutions 
(\ie with zero NUT charge) the semi-simple  element $\bf h$ associated to 
the nilpotent element $P$ can be computed in terms of the central charges $Z_{ij}$ alone, and more generally, in terms of the central and matter charges which we write collectively $Z_I$\footnote{Here $Z_I$ are the scalar field dependent linear combinations of the electromagnetic charges, 
transforming in a complex representation of $K_4$ and such that $V_{\rm BH}=Z_I Z^I$.}. 
Decomposing  $P\in \fg\ominus \mathfrak{k}^*$ with respect to  the Ehlers  $U(1)$ and the four-dimensional R-symmetry group $K_4$,
 \be 
P = - \dot{U}\,  {\bf H} \, + \, e^U Z_I\,  {\bf L}^I  \, - \,  {e_i}^j \dot{\phi}^i \, {\bf G}_j   
\quad \in \quad  \mathds{C} \oplus \mathfrak{l}_4 \oplus \scal{Ê\fg_4 \ominus \mathfrak{k}_4} \ ,
\ee
where ${e_i}^j$ is a vielbein for the metric $g_{ij}$, 
one may recast the middle equation in \eqref{triple}
into a system of first order differential 
equations of the form
\be \dot{U} =  - e^{U} W\, , \qquadÊ 
g_{ij}Ê\,  \dot{\phi}^j =  - e^{U} W_i\ ,
\label{FirstOrderSys} 
\ee
where $W$ and $W_i$ depend on the moduli $\phi^i$ and electromagnetic charges $Q_I$ through
the charges $Z_I$ only; moreover, we shall  prove that 
\be
 W_i = \pa_{\phi^i} W\ .
 \ee
Thus, extremal solutions attached to the given nilpotent orbit 
satisfy  a gradient flow under the fake superpotential $W$. 
In particular, it follows from the nilpotency of $P$ that 
\be \trace P^2 = 0 = e^{2U} \Scal{ÊW^2 - Z_I Z^I + 2 g^{ij} W_i W_j} = 0\ , 
\ee
and therefore that \eqref{vbhsuper} is obeyed. 

Applying this strategy to $\N=8$ supergravity with $G_3=E_{8(8)}$, we are able to determine the fake superpotentials for both BPS and non-BPS extremal black holes, and express them in terms of the $SU(8)$ invariant 
combinations of the central charges:
\begin{itemize}
\item  In the BPS case, we find that $W$ is the 
modulus of the largest skew eigenvalue of the central charge matrix (in particular, $W^2$ is
largest root of a quartic polynomial whose coefficients are  polynomials  in $Z_{ij}$).
This reproduces the result of \cite{Andrianopoli:2007gt}.   
\item In the non-BPS case, we find that $W=2\varrho$, where $\varrho$
is obtained from the non-standard diagonalization problem \eqref{Zdiag}. This problem is
solved in Appendix B, where $W^2$ is expressed as a 
particular root of an irreducible sextic polynomial\footnote{By irreducible, we mean as a polynomial with coefficients defined as rational functions of the $SU(8)$ invariant polynomials in $Z_{ij}$, or more formally, within the field extension of $\mathds{Q}$ generated by these invariants.}. 
This polynomial becomes reducible at particular values of the central charges, at which points $W$ can be computed in closed form. In particular, on the semi-line 
$\Pfaff(Z) \in \IR_-$ we recover the result of \cite{Andrianopoli:2007gt}.
\end{itemize}

Our expression for the  fake superpotential in fact extends straightforwardly to all theories with a symmetric scalar manifold whose isometry group acts faithfully on the electromagnetic charges \cite{Breitenlohner:1987dg}, as discussed in Section 3 below. In particular, we obtain the fake superpotential for all magic $\cN=2$ supergravity
models, and in fact for all supergravity theories with $\N \ge 2$ with a symmetric moduli space. 
In the one-modulus case, $W$ takes a completely explicit form given in \eqref{W1mod} below.



\subsection{Outline}

In Section \ref{secmax}, we apply the above strategy and find the 
complete fake superpotential for BPS and non-BPS extremal black
holes in $\N=8$ supergravity. In Section \ref{sectrunc}, we extend
these results to $\N=4$ and  symmetric $\N=2$ supergravity models.
In Section \ref{secstu}, we analyze the $STU$ model in more detail,
rephrase the BPS and non-BPS, $Z_*=0$ solutions in terms of the
para-quaternionic geometry of $\cM_3^*$, and check the nilpotency
of the Noether charge on explicit solutions. Appendix A contains a
detailed discussion of the real nilpotent orbits of $E_{8(8)}$ and 
$SO(4,4)$, relevant for maximal supergravity and the STU model.
In Appendix B we discuss how to evaluate the fake superpotential 
for non-BPS, $Z\neq 0$ black holes at various loci in 
the space of central charges corresponding to consistent
truncations. Appendix C  records some
extremal solutions  of the STU model, which provide a useful
testing bench for our analysis.

\vskip 5mm

\noindent 
{\it Note added:} the fake superpotential \eqref{W1mod} for the $S^3$ model was derived independently
in \cite{Ceresole:2009iy}, which appeared on arXiv after the present work had been completed.


\section{Extremal black holes in $\N=8$ supergravity}
\label{secmax}

In this section, we parametrize the $Spin^*(16)$ orbits of generic extremal spherically 
symmetric black holes of $\N=8$ supergravity, as Lagrangian submanifolds of certain 
nilpotent orbits of $E_{8(8)}$ in $\e_{8(8)} \ominus \so^*(16)$. 
For static solutions, this  parametrization determines $P$ in terms of the central charges and 
allows us to identify the fake superpotential. We reproduce the known result 
for the fake superpotential for $1/8$-BPS black holes \eqref{WBPS}, and obtain 
a new expression \eqref{WNB} for the fake superpotential  for non-BPS extremal black holes,
valid everywhere in moduli space.

\subsection{Generalities on $\N=8$ supergravity}

We first set up our notations for $\N=8$ supergravity in 4 dimensions. The massless scalar 
fields take values in  the symmetric space \cite{CremmerJulia}
\be 
\label{M4N8}
\cM_4 \cong \SU \backslash E_{7(7)} \ ,
\ee
where $\SU$ is the quotient of $SU(8)$ by the $\mathds{Z}_2$ centre leaving invariant the representations of even rank. According to the conventions of \cite{de Wit:1982ig} (up to normalization factors), we write the coset representative $v$ as
\be v \, \hat{=} \,  \left(\begin{array}{cc} {u_{ij}}^{IJ} & v_{ijKL} \\ v^{klIJ} & {u^{kl}}_{KL} \end{array}\right) 
\ ,
\ee
where little Latin letters are associated to the $SU(8)$ gauge symmetry, whereas capital Latin letters refer to the global $\SU \subset E_{7(7)}$. They both run from $1$ to $8$, and raising or lowering indices corresponds to complex conjugation (\eg $\Phi^{IJ} = ( \Phi_{IJ} )^*$ and $Z^{ij} = ( Z_{ij} )^*$). 
The invariant metric on $\cM_4$ can be written as 
\be 
ds^2_{\cM_4} =  \frac{1}{24} V_{ijkl} V^{ijkl} \ ,
\ee
where 
\be Ê
V_{ijkl} =Ê{u_{ij}}{}^{IJ} d v_{klIJ} - v_{ijIJ} d {u_{kl}}{}^{IJ}  
 \label{Evielbeins} 
\ee
is  the $\SU \backslash E_{7(7)}$ vielbein, which is automatically a complex self-dual antisymmetric tensor by property of the $\e_{7(7)}$ Lie algebra. In the symmetric gauge, $v$ can be written in terms of a complex self-dual tensor
$\phi_{ijkl}= \frac{1}{24} \varepsilon_{ijklmnpq} \phi^{mnpq}$,
\be 
v \, \hat{=} \,  \exp   \left(\begin{array}{cc} \hspace{3mm} 0 \hspace{3mm} & \phi_{ijkl} \\ \phi^{ijkl}  & \hspace{3mm} 0 \hspace{3mm}   \end{array}\right)  =    \left(\begin{array}{cc}  \cosh(\phi)_{ij}{}^{kl}  \, & \, \sinh(\phi)_{ijkl}  \vspace{3mm}  \\ \overline{\sinh} (\phi)^{ijkl} & \overline{ \cosh}(\phi)^{ij}{}_{kl}  \end{array}\right)  \label{SymEsept} 
\ee
The hyperbolic functions in this expression are Taylor series in $\phi_{ijkl}$ 
with $SU(8)$ covariant contractions \cite{de Wit:1982ig}, 
\be
\begin{split}  
\cosh(\phi)_{ij}{}^{kl} &= \delta_{ij}^{kl} + \frac{1}{2}Ê\phi_{ijpq} \phi^{pqkl} + \mathcal{O}(\phi^4) \ ,\\
\sinh(\phi)_{ijkl} &= \phi_{ijkl} + \frac{1}{6} \phi_{mn[ij} \phi^{mnpq} \phi_{kl]pq} + \mathcal{O}(\phi^5)\ . 
\end{split}  
\ee 
The $\SU \backslash E_{7(7)}$ vielbein is then given by 
\be 
V_{ijkl}  = \Esh_\phi (d\phi)_{ijkl} \ ,
\ee
where $\Esh_\phi$ is the linear operator
\be 
\Esh_\phi (X)_{ijkl} \equiv \frac{ \sinh\sqrt{{\rm A}_\phi}}{\sqrt{ {\rm A}_\phi}} (X)_{ijkl} 
= X_{ijkl} + \frac{1}{6}  {\rm A}_\phi (X)_{ijkl} + \frac{1}{120} { {\rm A}_\phi }^2(X)_{ijkl} 
+ \mathcal{O}( Ê\phi^6) \ .
\ee
Here 
\be
 {\rm A}_\phi (X)_{ijkl} \equiv 2 \phi_{mn[ij} \phi^{mnpq} X_{kl]pq} - 2 \phi_{mn[ij} \phi_{kl]pq} X^{mnpq}\ . \ee
is defined such that
\be \left[Ê \ad  \left(\begin{array}{cc} \hspace{3mm} 0 \hspace{3mm} & \phi_{mnpq} \\ \phi^{mnpq}  & \hspace{3mm} 0 \hspace{3mm}   \end{array}\right) \right]^2  \,  \left(\begin{array}{cc} \hspace{3mm} 0 \hspace{3mm} & X_{ijkl} \\ X^{ijkl}  & \hspace{3mm} 0 \hspace{3mm}   \end{array}\right)  =  \left(\begin{array}{cc} \hspace{3mm} 0 \hspace{3mm} &  {\rm A}_\phi (X)_{ijkl} \\  {\rm A}_\phi (X)^{ijkl}  & \hspace{3mm} 0 \hspace{3mm}   \end{array}\right) \ ,
\ee

The central charge of $\N=8$ supergravity is a complex antisymmetric tensor $Z_{ij}$. It can 
always be rotated via a suitable $SU(8)$ transformation into the form
\be 
R^k{}_i R^l{}_j Z_{kl} \, \, \hat{=}  \, \, \frac{1}{2}Êe^{i\varphi}Ê\, \left( \begin{array}{cc}0 & \, \, 1  \\- 1 & \, \, 0 \end{array}\right) \otimes  \left(
\begin{array}{cccc} 
\, \, \rho_\zero \, \, Ê& 0 & 0 & 0 \\
0& \, \, \rho_\un \, \,  & 0 & 0 \\ 
0 & 0 &\, \,  \rho_\deux \, \, & 0 \\ 
0 & 0 & 0 & \rho_\trois 
\end{array} \right)  \label{Zdecompose} \ee
such that $\rho_\zero \ge \rho_\un \ge \rho_\deux \ge \rho_\trois$ are positive real numbers. 
$\varphi$ is defined  (modulo $\frac{\pi}{2}$) as the $SU(8)$ invariant function
\be \varphi = \frac{1}{4} \arg\bigl[ \Pfaff(Z) \bigr] 
\ ,\qquad
\Pfaff(Z) \equiv \frac{1}{16 \cdot 4!} \varepsilon^{ijklmnpq}  Z_{ij} Z_{kl} Z_{mn} Z_{pq}\ ,
\ee
and  the four $SU(8)$ invariant functions ${\rho_\zero}^2,\,  {\rho_\un}^2, \, { \rho_\deux}^2 ,\,  {\rho_\trois}^2  $ are the four roots of the polynomial 
\begin{multline}  
\lambda^4 - 2 Z_{ij} Z^{ij} \lambda^3 + \Scal{ 2 \scal{ÊZ_{ij} Z^{ij} }^2 - 4 Z_{ij} Z^{jk} Z_{kl} Z^{li}Ê} \lambda^2 \\* 
- \left( \frac{4}{3}  \scal{ÊZ_{ij} Z^{ij} }^3 - 8 Z_{ij} Z^{ij} \, Z_{kl} Z^{lp} Z_{pq} Z^{qk} - \frac{32}{3} Z_{ij} Z^{jk} Z_{kl} Z^{lp} Z_{pq} Z^{qi} \right) \lambda 
+ 16 \bigl| \Pfaff(Z) \bigr|^2 \\*
= \scal{Ê\lambda - {\rho_\zero}^2 } \scal{Ê\lambda - {\rho_\un}^2 } \scal{Ê\lambda - {\rho_\deux}^2 } \scal{Ê\lambda - {\rho_\trois}^2 } \label{SUpoly}
\end{multline}
The  $E_{7(7)}$ quartic invariant \cite{CremmerJulia}
\be
\lozenge(Z) = 16 \left( Z_{ij} Z^{jk} Z_{kl} Z^{li} - \frac14 \scal{ÊZ_{ij} Z^{ij} }^2 + 4 
\Scal{Ê\Pfaff(Z) + \overline{\Pfaff( Z)} } \right)
\label{defloz}
\ee
can be expressed in terms of the $SU(8)$ invariants as
\be
 \label{Lozrho}
\lozenge(Z) =  \sum_{\n = \zero}^\trois {  \rho_\n}^4 - 2 \sum_{\m > \n} {\rho_\m}^2 {\rho_\n}^2 + 8 \, \rho_\zero  \rho_\un \rho_\deux \rho_\trois \cos(4\varphi)\ .
%
\ee
Being $E_{7(7)}$ invariant, it is a function of the electromagnetic charges alone and
independent of the moduli. 
In \eqref{Zdiag} below, we shall define a different parametrization of the central charge 
$Z_{ij}$, which plays the same role for non-BPS black holes as \eqref{Zdecompose}
for BPS ones.

\subsection{Spherically symmetric, weakly extremal solutions}

As explained in \cite{Breitenlohner:1987dg}, the dimensional 
reduction of $\N=8$ supergravity along the time direction leads
to a non-linear sigma model on

 \be 
 \cM_3^* \cong \Spin \backslash E_{8(8)} \ ,
 \ee
where $\Spin$ is the quotient of $Spin^*(16)$ by the $\mathds{Z}_2$ subgroup that acts trivially in the chiral Weyl representation. To parametrize this space in a way suited to the dimensional reduction, 
recall that 
%
%
the Lie algebra $\e_{8(8)}$ admits the real five-graded decomposition 
\be 
\e_{8(8)}Ê\cong {\bf 1}^{\ord{-2}} 
\oplus {\bf 56}^{\ord{-1}} \oplus \scal{Ê\gl_1 \oplus 
\e_{7(7)}}^\ord{0} \oplus {\bf 56}^\ord{1} 
\oplus {\bf 1}^\ord{2} \ ,
\label{five}
\ee
such that $\e_{7(7)}$ is the Lie algebra of the four-dimensional duality group, and $\sl_2 \cong  {\bf 1}^{\ord{-2}} \oplus {\gl_1}^\ord{0} \oplus {\bf 1}^\ord{2}$ the Lie algebra of the Ehlers duality group for stationary solutions. We write the generators of $\e_{7(7)} \cong \su(8) \oplus {\bf 70}$ as ${{\bf G}_I}^J,\, {\bf G}_{IJKL}$ and the ones of $\sl_2 \cong  {\bf 1}^{\ord{-2}} \oplus {\gl_1}^\ord{0} \oplus {\bf 1}^\ord{2}$ as ${\bf F},\, {\bf H} ,\, {\bf E}$, respectively. The generators of grade $1$ and $-1$ will be written as ${\bf E}_{IJ},\, {\bf E}^{IJ}$ and ${\bf F}_{IJ},\, {\bf F}^{IJ}$, such that they only appear in $\e_{8(8)}$ through the combinations 
\be X_{IJ} {\bf E}^{IJ} - X^{IJ} {\bf E}_{IJ} \, , \hspace{10mm}ÊY_{IJ} {\bf F}^{IJ} - Y^{IJ} {\bf F}_{IJ} \ee

The negative weight part of the the five-graded decomposition (\ref{five})
 \be 
 \mathfrak{p} \cong  {\bf 1}^\ord{-2} \oplus  {\bf 56}^\ord{-1}  \oplus  \scal{Ê\gl_1 \oplus \e_{7(7)}}^\ord{0}   
\label{parabolic}
\ee
defines the Lie algebra of a maximal parabolic subgroup $\P\subset E_{8(8)}$, 
also known as the Heisenberg parabolic. $\SU \backslash \P$ 
is isomorphic to the Riemannian symmetric space $\cM_3 \cong Spin_{\scriptscriptstyle \rm c}(16)\backslash
E_{8(8)}$ by the Iwasawa decomposition, and to a dense subset of 
the pseudo-Riemannian symmetric space $\cM_3^*$. A generic 
element of $\SU \backslash\P$ may be parametrized as
\be 
\V = \Ad(v)\, \exp\left(  U \,   {\bf H}\right) \,  \exp\Scal{Ê\sigma\, {\bf F} + 
\Phi_{IJ} {\bf F}^{IJ} - \Phi^{IJ} {\bf F}_{IJ}  } \ ,
 \label{Pgauge} 
\ee
where $U$ is identified as the scale factor in the metric ansatz \eqref{4dmetric},
$v$ is the coset representative in \eqref{M4N8}, $\sigma$ the NUT scalar, and $\Phi^{IJ}$
are linear combinations of the Wilson lines $\zeta,\tzeta$ transforming 
as an antisymmetric complex tensor of $\SU \subset E_{7(7)}$. 
The associated Maurer--Cartan form decomposes into its coset 
and $\so^*(16)$ components according to
\be
- \dot{\V} \, \V^{-1}  = B+P \quad, \qquad B \in \so^*(16) 
  \;\; , \;\;\; 
 P \in \e_{8(8)} \ominus\so^*(16) 
\ee
A straightforward computation gives
\begin{multline}  
\label{P}
-P =  \dot{U}  \, {\bf H}  + \frac{1}{2} e^{-2U}Ê \Scal{ \dot{\sigma} + \frac{i}{2} \scal{Ê \Phi^{IJ}  \dot{ \Phi}_{IJ} - \Phi_{IJ}  \dot{ \Phi}^{IJ}   } } ( {\bf F} + {\bf E})  \\* + \frac{1}{2}Êe^{-U}Ê  Ê\,\Bigl(  \scal{Ê{u_{ij}}^{IJ} \dot{\Phi}_{IJ} - v_{ijIJ} \dot{\Phi}^{IJ} } \scal{Ê{\bf F}^{ij} - {\bf E}^{ij} } -  \scal{Ê{u^{ij}}_{IJ} \dot{\Phi}^{IJ} - v^{ijIJ} \dot{\Phi}_{IJ} } \scal{Ê{\bf F}_{ij} - {\bf E}_{ij} } \Bigr) \\* + \frac{1}{24}Ê\Scal{Ê{u_{ij}}{}^{IJ} \dot{v}_{klIJ} - v_{ijIJ} {\dot{u}_{kl}}{}^{IJ}}   {\bf G}^{ijkl}   \end{multline}
where the $\e_{8(8)}$ generators with lowercase indices $i,\, j,\, \cdots$ satisfy
the same commutations rules as the ones with capital indices, and are related to the latter 
via a fixed vielbein $v$, chosen to be the vielbein $v_\asym$ 
at spatial infinity.

The equations of motion for  weakly extremal spherically symmetric solutions
(including those with non-zero NUT charge) then take the manifestly $E_{8(8)}$ invariant form
\be
\trace P^2 = 0 \ ,\qquad \frac{\partial\, }{\partial \tau} \,\Scal{Ê \V^{-1}ÊP \V } = 0 
\label{EinsteinE}
\ee
Defining the  $\e_{8(8)}$-valued Noether charge as 
\be 
Q  \equiv  \V^{-1}ÊP \V\ ,
\ee
we can characterize spherically symmetric weakly extremal black holes  by the
constant value of $Q$, subject to $\trace Q^2 = 0$, and the asymptotic value of 
the scalars fields $v_\asym \in E_{7(7)}$ 
at spatial infinity (\ie at $\tau = 0$). The condition of smoothness of the metric
puts additional restrictions on the Noether charge discussed in the next subsection. 
For our purposes it will be more convenient to characterize the solutions 
instead in terms of the value $P_\asym$ of $P$ at $\tau=0$. 

$P_\asym$ transforms  as a Majorana--Weyl spinor under $\Spin$. It can be  conveniently 
parametrized using a fermionic oscillator basis \cite{Bossard:2009at},
\begin{multline} 
\hspace{-4mm} 
|P_\asym \rangle = \Scal{Ê\w + Z_{ij} a^i a^j + \Sigma_{ijkl} a^i a^j a^k a^l + \frac{1}{6!} \varepsilon_{ijklmnpq} Z^{pq} \, a^i \cdots  a^n +  \frac{1}{8!} \varepsilon_{ijklmnpq} \bar \w  \, a^i \cdots  a^q } | 0 \rangle  \\*
= ( 1 + \invo ) \Scal{ÊÊ\w + Z_{ij} a^i a^j + 
\frac{1}{2}Ê\Sigma_{ijkl} a^i a^j a^k a^l } | 0 \rangle 
\end{multline} 
where $\invo$ is the anti-involution defining the chiral Majorana--Weyl representation 
of $Spin^*(16)$,  $\w = M + i k$ where $M$ is the mass and $k$ the NUT charge, 
$Z_{ij}$ are the supersymmetric central charges  and $\Sigma_{ijkl}$ are
the  ``scalar charges". 

We will focus on the case of static solutions, such that $k=0$ and the equations of the electromagnetic fields reduce to\footnote{Where we used the identities $u^{ij}{}_{IJ} u_{ij}{}^{KL} - v_{ijIJ}Êv^{ijKL} = \delta^{KL}_{IJ}$ and $u^{ij}{}_{IJ}Êv_{ijKL}  = v_{ijIJ} u^{ij}{}_{KL} $ \cite{de Wit:1982ig}.}
\be \frac{\partial\, }{\partial \tau} \, e^{-2U} \Scal{Ê\dot{\Phi}_{IJ} - 2 v_{ijIJ}  \scal{u^{ij}{}_{KL}  \dot{\Phi}^{KL}-  v^{ijKL}  \dot{\Phi}_{KL}   } } = 0 \ee
These integrals of motion define the complex electromagnetic charges $Q_{IJ} \in {\bf 28}_{\mathds{C}}$. This allows to replace the `electromagnetic component' of $P$ by the central charges $Z(v)_{ij}$,  as follows
\be - e^{-U} \scal{Ê{u_{ij}}^{IJ} \dot{\Phi}_{IJ} - v_{ijIJ} \dot{\Phi}^{IJ} }  =  e^{U} \scal{ {u_{ij}}^{IJ} Q_{IJ} + v_{ijIJ} Q^{IJ} }  = e^{U} Z(v)_{ij} \ee
Within the fermionic oscillator basis, $P$ then reduces to the following Majorana--Weyl spinor 
\be
\framebox{$ | P \rangle = ( 1 + \invo ) \Scal{Ê- \dot{U} + e^{U} Z(v)_{ij} \, a^i a^j - \frac{1}{24}Ê Ê\scal{ÊÊ{u_{ij}}{}^{IJ} \dot{v}_{klIJ} - v_{ijIJ} {\dot{u}_{kl}}{}^{IJ} } a^i a^j a^k a^l } | 0 \rangle$}
 \label{Pspinor} \ee

Smoothness of the metric requires $Q$, and therefore $P = \V Q \V^{-1}$, to be nilpotent of degree five 
when evaluated in the ${\bf 3875}$ representation of $\e_{8(8)}$ \cite{Bossard:2009at},
\be 
[P_{\, |\bf 3875}]^{\, 5} = 0  \label{Nilfive} \ .
\ee
This condition is invariant under the adjoint action $P \to g^{-1}ÊP g$ where $g \in E_{8(8)}$, 
and therefore defines an adjoint orbit of $E_{8(8)}$. Moreover, the $\Spin\subset E_{8(8)}$ orbit of 
$P$ defines a  Lagrangian subspace of the adjoint orbit, for the Kirillov-Kostant symplectic form 
induced by the Killing norm \cite{Bossard:2009at}. The adjoint orbits of elements of $\e_{8(8)}$ satisfying the nilpotency conditions (\ref{Nilfive}) are in one-to-one correspondence with the $ \Spin$ orbits of spherically symmetric black holes satisfying $v_\asym = \mathds{1}$. Other solutions with
general value of $v_\asym$ can be obtained by acting further with  $E_{7(7)}$.
The stratification of the moduli space of extremal black holes solutions with $v_\asym = \mathds{1}$
is identical to that
of the corresponding $E_{8(8)}$ nilpotent orbits \cite{Bossard:2009at,E8strat}, and is displayed
 in Figure \ref{E8Hasse}.


\FIGURE{\includegraphics[height=7cm]{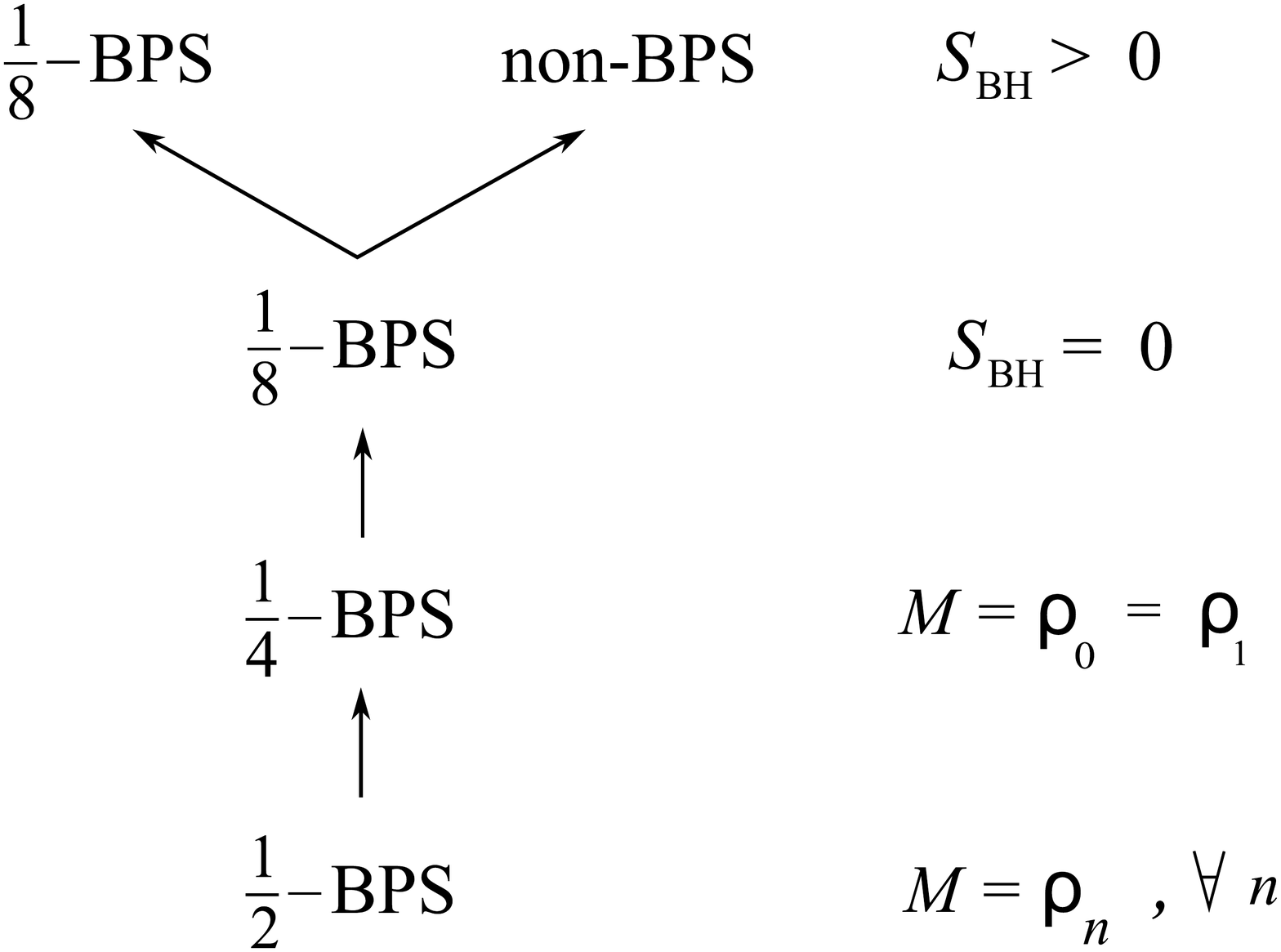}
\caption{Stratification of the moduli space 
of extremal solutions in $\N=8$ supergravity.       \label{E8Hasse}}}


There are two $E_{8(8)}$ orbits associated to the nilpotency condition \eqref{Nilfive}, whose union is dense in the space of solutions of this equation. They both lie in a single $E_8(\mathds{C})$ orbit,  labelled by the weighted Dynkin diagram \DEVIII00000002,\footnote{
In general, a complex nilpotent orbit of $G_\IC$ is uniquely labelled by a  weighted Dynkin diagram of $G$, which records the coordinates of the Cartan generator of the $SL(2,\IC)$ triplet defining
the orbit. Real orbits are generally uniquely labelled by a pair of weighted Dynkin diagrams for $G$
and its maximal compact subgroup. See e.g. \cite{Collingwood} for a thorough introduction
to nilpotent orbits.} associated to the same five graded decomposition as in (\ref{five}), 
up to a rescaling of the grading generator by two,
\be 
\e_{8(8)} \cong   {\bf 1}^{\ord{-4}} 
\oplus {\bf 56}^{\ord{-2}} \oplus \scal{Ê\gl_1 \oplus 
\e_{7(7)}}^\ord{0} \oplus {\bf 56}^\ord{2} 
\oplus {\bf 1}^\ord{4} \label{fiveTwo}
\ee
A representative ${\bf E}$ of the nilpotent orbit \DEVIII00000002 is a generic element of the grade two component ${\bf 56}^\ord{2}$. There are two classes of such elements which are distinguished by their isotropy subgroup inside $E_{7(7)}$, respectively $E_{6(2)}$ and $E_{6(6)}$ \cite{Levi,Ferrara:1997uz}\footnote{Note however that in the case of physical interest, the graded decomposition (\ref{fiveTwo}) is defined with respect to  a $\gl_1$ subalgebra of $\so^*(16)$ such that the corresponding component $\e_{7(7)}^\ord{0}$ is not the four-dimensional duality group. Nevertheless, one checks that the isotropy subgroup of the corresponding electromagnetic charges $Z_{ij}$ defining $P$ are also left invariant by the same subgroups of $E_{7(7)}$.}.
Each of these two  $E_{8(8)}$ orbits does not contain a unique  $\Spin$ orbit. Nonetheless, there is one single $\Spin$ orbit of regular spherically symmetric black holes in each $E_{8(8)}$ orbit \cite{Bossard:2009at}. As for the $E_{8(8)}$ orbits, we can label them by the associated $\so^*(16)$ weighted Dynkin diagrams. An $\so^*(16)$ weighted Dynkin diagram defines the coordinates of a $\gl_1$ generator ${\bf h}$ of a chosen Cartan subalgebra of $\so^*(16)$  which defines a corresponding graded decomposition of $\so^*(16)$ and its Majorana--Weyl representation ${\bf 128}_+$ such that a representative of the orbit lies in the component of grade two of ${\bf 128}_+$, in particular
\be 
\framebox{${\bf h} \, | P_\asym \rangle = 2 | P_\asym \rangle \ .$}
\label{GradeTwo}
\ee
The two orbits of $\Spin$ associated to generic extremal black holes (\ie black holes with a non-vanishing horizon area) are labelled by \DSOXVI02000000 and \DSOXVI00000020, respectively.
The aim of this section is to solve explicitly Eq. (\ref{GradeTwo}) at all values of $\tau$,
re-express it as a system of first order differential equations for $U$ and $v$,
and read off the fake superpotential for both BPS and non-BPS extremal black holes.

\subsection{$1/8-$BPS black holes}
For a generic $1/8-$BPS spherically symmetric black hole, $P_\asym$ lies in the $Spin^*(16)$ orbit \cite{Bossard:2009at}
\be  \frac{Spin^*(16)}{\scal{ÊSU(2) \times SU(6)} \ltimes \scal{ ( { \bf 2} \otimes {\bf 6})^\ord{2}Ê\oplus \mathds{R}^\ord{4}}}  \subset \frac{E_{8(8)}}{E_{6(2)}\ltimes \scal{Ê( \mathds{R} \oplus {\bf 27} )^\ord{2} \oplus \mathds{R}^\ord{4} }}  \label{BPSorbit}\ee
labelled by the weighted Dynkin diagram \DSOXVI02000000, associated to the five graded decomposition of $\so^*(16)$
\be \so^*(16) \cong {\bf 1}^\ord{-4} \oplus \scal{ {\bf 2} \otimes {\bf 12} }^\ord{-2}_\mathds{R} \oplus  \scal{\gl_1 \oplus Ê\su(2) \oplus \so^*(12) }^\ord{0} \oplus  \scal{ {\bf 2} \otimes {\bf 12} }^\ord{2}_\mathds{R} \oplus  {\bf 1}^\ord{4}  \label{susygrade}
\ee
In the symmetric gauge (in which $ \V  = \exp( - P_\asym \tau )$), $P = P_\asym$. Since
$P$ transforms as a Majorana--Weyl spinor with respect to $Spin^*(16)$, it follows that $P$ lies in the same $Spin^*(16)$ orbit for any value of $\tau$ in the parabolic gauge. 

The generator  ${\bf h}_\frac{1}{8}$ defining the grading (\ref{susygrade}) through $\gl_1 \equiv \mathds{R} {\bf h}_\frac{1}{8}$ associated to $1/8$-BPS solutions can be identified as follows. Let 
$\omega_{ij}$  be an antisymmetric tensor of rank two of $SU(8)$ such that $I_i^j\equiv \omega_{ik} \omega^{ jk}$ is a projector onto the two-dimensional subspace $\mathds{C}^2 \subset \mathds{C}^8$ 
of the preserved Killing  spinors at infinity (\ie at $\tau \rightarrow 0$),
\be 
\epsilon_\alpha^i + \varepsilon_{\alpha\beta} \omega^{ ij} \, \epsilon^\beta_j = 0 \hspace{10mm}Ê\Rightarrow \hspace{10mm} \epsilon_\alpha^i = I^i_j \, \epsilon_\alpha^j \ .
\ee
Here we have written the Killing spinor  as a complex $SU(2)$ spinor valued in the fundamental of $SU(8)$. It transforms as a real $({\bf 2},{\bf 16})$ under 
$SU(2) \times SO^*(16)$, where the real structure is given by the product of the pseudo-real
structures on both factors. It  can be then checked that the generator ${\bf h}_\frac{1}{8}$,
 \be {\bf h}_{\frac{1}{8}} \equiv \omega_{ij} a^i a^j - \omega^{ ij} a_i a_j  \label{HBPS} 
 \ee
 generates the 5-grading (\ref{susygrade}). 
 
Solutions to (\ref{GradeTwo}) are $Spin^*(12)$ Majorana--Weyl spinors
of the form
 \be 
 | P_\asym  \rangle = ( 1 + \invo ) \Scal{Ê1 + \frac{1}{2}Ê\omega_{ij} a^i a^j } 
\Scal{ \w + Z^\prime_{ij} a^{i} a^{j} } | 0 \rangle
 \label{BPSrep} 
 \ee
where  $\w = M+ i k $ and $Z_{ij}^\prime$ is the component of the central charge 
 $Z_{ij} $ at $\tau = 0$ orthogonal to $\omega_{ij}$ (\ie $I_i^k  Z_{jk}^\prime = 0$).
The data $\omega_{ij} \subset \scal{ÊSU(2) \times SU(6) } \backslash SU(8)$, $\w \in \mathds{C}$ and $Z^\prime_{ij} \in \bigwedge^2 \mathds{C}^6$ provide a complete parametrization of the $Spin^*(16)$ orbit (\ref{BPSorbit}). To see this, note that the only generators of $\so^*(16)$ that act non-trivially on a representative of the grade two component ${\bf 32}^\ord{2}$ associated to a $\gl_1$ generator (\ref{HBPS}) are in the subalgebra of negative grade
\be {\bf 1}^\ord{-4} \oplus \scal{ {\bf 2} \otimes {\bf 12} }^\ord{-2}_\mathds{R} \oplus  \scal{\gl_1 \oplus Ê\so^*(12) }^\ord{0} \subset \so^*(16) \ .
\ee
By construction, the component of grade zero leaves invariant ${\bf h}_\frac{1}{8}$ and acts transitively on the components $\w, \, Z^\prime_{ij}$ associated to regular black holes. The generators of strictly negative grade can be redefined through the addition of generators of strictly positive grade such that they correspond precisely to the generators of $ \su(8)$ that act non-trivially on $\omega_{ij}$ ($\su(8) \cong \su(2) \oplus \su(6)\oplus {\bf 1} \oplus ( {\bf 2} \otimes {\bf 6})$). 

Comparing the general form of $|P\rangle$ (\ref{Pspinor}) to the general representative of the relevant nilpotent orbit (\ref{BPSrep}), we conclude that in the static case (\ie $k=0$), the field dependent ${\bf h}(\phi)$ generator can be defined at any value of $\tau$ uniquely from the  the central charge. Using the decomposition \eqref{Zdecompose} of the  central charge $Z_{ij}$,
\bea 
Z_{ij} &=& \frac{1}{2}Êe^{i \varphi(\phi)} R_i{}^k(\phi) R_j{}^l (\phi)\Scal{Ê\rho_\zero(\phi) \omega^\zero_{ij} + \rho_\un(\phi) \, \omega^\un_{ij} +  \rho_\deux(\phi) \, \omega^\deux_{ij} +   \rho_\trois(\phi) \, \omega^\trois_{ij} }\nn \\
&=& \frac{1}{2}Êe^{i \varphi(\phi)} R_i{}^k(\phi) R_j{}^l (\phi) \rho_\zero(\phi) \omega^\zero_{ij} 
+ Z'_{ij}
 \label{BPSZpara} 
\eea
where $\rho_\zero(\phi) \ge \rho_\n(\phi)$, 
the semi-simple generator  ${\bf h}(\phi)$ can be written 
\be 
\label{omom}
{\bf h}_{\frac{1}{8}}(\phi) \equiv  \omega_{ij}(\phi)  a^i a^j 
-  \omega^{ij}(\phi) a_i a_j  \ ,\qquad \omega_{ij}(\phi) \equiv Êe^{i \varphi(\phi)} 
R_i{}^k(\phi) R_j{}^l(\phi) \, Ê\omega^\zero_{ij} \ ,
\ee
while
\be 
\framebox{$| P\rangle   =   ( 1 + \invo )  \Scal{Ê1 + \frac{1}{2}Ê \omega_{ij}(\phi) a^i a^j } 
e^{U} \Scal{  \rho_\zero(\phi) + Z^\prime_{ij}(\phi) a^{i} a^{j} } | 0 \rangle\ .$}
 \label{BPSrep2} 
 \ee
In the remainder, we shall refrain from writing the dependence on $\phi$ explicitely.

Comparing \eqref{BPSrep2} and \eqref{Pspinor}, we see that  the scalar fields $U$ and $v$ 
satisfy the first order equations 
\bea
\label{BPSlinear}
 \dot{U} &=& -  Êe^{U}  \rho_\zero \\*
Ê{u_{ij}}{}^{IJ} \dot{v}_{klIJ} - v_{ijIJ} {\dot{u}_{kl}}{}^{IJ}   &=& -  6 Ê e^U \Scal{ÊÊ\omega_{[ij} Z_{kl]} + \frac{1}{24} \varepsilon_{ijklmnpq} Ê\omega^{mn} Z^{pq}  } \nonumber 
 \eea
The first equation in (\ref{BPSlinear}) identifies the fake superpotential for BPS black holes as
\be 
\label{WBPS}
\framebox{$W = \rho_\zero \ .$}
\ee
In order to verify that \eqref{BPSlinear} is indeed the gradient flow of $W$, 
one may use the second equation in  \eqref{omom} to write $W=\omega^{ij} Z_{ij} $.  
One then computes (in symmetric gauge, 
but the proof is easily generalized to any other parametrisation of $\M_4$)
\bea d W &=&  \omega^{ ij} \Scal{Êd {u_{ij}}^{IJ} \, Q_{IJ} + d v_{ijIJ} \, Q^{IJ} }  + \omega^{\zero\, kl} d \Scal{Êe^{-\I \varphi} R^i{}_k R^j{}_l } Z_{ij}  \CR
 &=&  \omega^{ ij} \Scal{ÊÊÊ{u_{ij}}{}^{IJ} d v_{klIJ} - v_{ijIJ} d {u_{kl}}{}^{IJ}   } Z^{kl} +  \omega^{ ij} \Scal{ÊÊu^{kl}{}_{IJ} d u_{ij}{}^{IJ} - v^{klIJ} d v_{ijIJ} } Z_{kl}\CR
 && \hspace{20mm} - \I \, \omega^{ij} Z_{ij} d \varphi - 2 \omega^{ik} \, R^j{}_l d R_i{}^l \, Z_{jk} \CR 
&=&  \Esh_\phi (d \phi)_{ijkl}\,  \omega^{ ij} Z^{kl} - \I W d \varphi \CR
&& \hspace{10mm}  +\omega^{ ik} \left[Ê \frac{2}{3} \,   \Scal{ÊÊu^{jl}{}_{IJ} d u_{il}{}^{IJ} - v^{jlIJ} d v_{ilIJ} } - 2 R^j{}_l d R_i{}^l  \right]Ê Z_{jk}\ ,
 \label{dWun}
\eea
where we used the fact that the right-hand-side in the second line is an element of $\su(8)$. 
$\omega_{ij}$ is invariant under an $\su(2) \oplus \su(6)\subset \su(8)$. Moreover, the action of any generator of $\su(8)$ on $\omega_{ij}$ gives an antisymmetric tensor with at least one index in the image of the projector $I_i^j \equiv \omega_{ik} \omega^{jk}$. Using the explicit form of  $Z_{ij}$ (\ref{BPSZpara}), it then follows that the very last line of \eqref{dWun} can be rewritten as
\begin{multline} 
\omega^{ ik} \left[Ê \frac{2}{3} \,   \Scal{ÊÊu^{jl}{}_{IJ} d u_{il}{}^{IJ} - v^{jlIJ} d v_{ilIJ} } - 2 R^j{}_l d R_i{}^l  \right]Ê Z_{jk} \\
=  I^i_j   \left[Ê \frac{1}{3} \,  \Scal{ÊÊu^{jk}{}_{IJ} d u_{ik}{}^{IJ} - v^{jkIJ} d v_{ikIJ} } - R^j{}_k d R_i{}^k \right]Ê  W
\end{multline}
Besides, using the fact that $W$ is real, we compute in the same way that
\bea d W &=& \omega_{ij}  \Scal{Êd {u^{ij}}_{IJ} \, Q^{IJ} + d v^{ijIJ} \, Q_{IJ} }  +   \omega_{kl} d \Scal{Ê
e^{\I \varphi} R_i{}^k R_j{}^l } Z_{ij}  \CR
&=&  \Esh_\phi (d \phi)^{ijkl} \, \omega_{ij} Z_{kl}  + \I W d \varphi \CR
& & \hspace{10mm} + I_i^j \left[Ê \frac{1}{3}   \Scal{ÊÊu_{jk}{}^{IJ} d u^{ik}{}_{IJ} - v_{jkIJ} d v^{ikIJ}  } - R_j{}^k d R^i{}_k \right]  W
\eea
Adding the two expressions of $dW$ and canceling terms, we finally obtain
\be d W = \frac{1}{2}Ê\Esh_\phi (d \phi)_{ijkl} \Scal{Ê \omega^{ij} Z^{kl}  + \frac{1}{24} \varepsilon^{ijklmnpq}  \omega_{mn} Z_{pq} }\ .
\ee
It follows that the equations  (\ref{BPSlinear}) can be expressed as
\be  
\label{bpsflow}
\dot{U} = -  Êe^{U} W \, , \hspace{10mm} \dot{\phi}_{ijkl} =  - 12 e^{U} \,  {\Esh_\phi}^{-2} \left( \frac{ \partial W}{\partial \phi}\right){}_{ijkl} \ , \ee
as expected. Thus, the radial evolution of the radius $U$ and scalars $v$
for $1/8$-BPS solutions follows the gradient flow of $W$ given in \eqref{WBPS}.
We have therefore reproduced the result of \cite{Andrianopoli:2007gt} using the technology
of nilpotent orbits.

Substituting \eqref{bpsflow} back into (\ref{Pspinor}), one can rewrite the momentum vector
as
\be  | P \rangle =e^{U} (1 + \invo) \left(  WÊÊ +  Z_{ij} a^i a^j  + \frac{1}{2}  {\Esh_\phi}^{-1}\left( \frac{ \partial W}{\partial \phi}\right){}_{ijkl}  \,   a^i a^j a^k a^l \right) | 0 \rangle \eeÊ
By virtue of the attractor mechanism, the scalar fields at the horizon 
lie at an extremum of $W$. Thus, $P$ reduces to 
\be 
| P  \rangle \sim  e^{U} (1+ \invo) \Scal{Ê ÊW_* +  Z_{ij\, *} a^i a^j } | 0 \rangle \ee
as $\tau$ goes to infinity. From (\ref{BPSrep})  we conclude that 
$Z^\prime_{ij\, *} = 0$ at the horizon, 
and therefore
\be \rho_\un(Z_{ij\, *}) =  \rho_\deux(Z_{ij\, *}) = \rho_\trois(Z_{ij\, *})  = 0 \ .
\ee
Since the $E_{7(7)}$ invariant $\lozenge(Z)$ depends only on the electromagnetic charges,
it is constant along the flow, and equals to $\rho_\zero(Z_{ij\, *})^4$ at the horizon. Therefore,
the value of $W$ at the horizon can be rewritten as  
\be 
W_* = \sqrt[4]{ \lozenge( Q_{IJ} )} \ .
\ee
Using \eqref{massent}, one recovers the famous 
formula for the entropy of BPS black holes in $\N=8$ supergravity \cite{Kallosh:1996uy}.

Finally, we note that the two-form $\omega_{ij}$ associated to the central charge is  invariant under
 $SU(2) \times SU(6) \subset SU(8)$ and  $E_{6(2)} \subset E_{7(7)}$, respectively. 
Therefore the ``fake superpotential" $W$ admits flat
directions homeomorphic to the symmetric space 
\be 
\cM_{\rm \scriptscriptstyle BPS} \cong \scal{ÊSU(2) \times SU(6)}   
\big\backslash E_{6(2)} \subset \cM_4 \ , 
\ee
which are not determined by the attractor mechanism, but depend on the asymptotic
value of the scalars \cite{Andrianopoli:1997wi,Bellucci:2006xz}.
In order to see this in our formalism, note from (\ref{BPSrep}) that in terms of representations of the $SU(2) \times SU(6)$ isotropy subgroup of $\omega_{ij}$, the components of $Ê{u_{ij}}{}^{IJ} d v_{klIJ} - v_{ijIJ} d {u_{kl}}{}^{IJ}$ are only non-zero along the component $ {\bf 15}_\mathds{C} \subset {\bf 70}$, and are always zero along $({\bf 2} \otimes {\bf 20})_\mathds{R} \subset Ê {\bf 70}$. One therefore understands from the decomposition 
\be \begin{array}{rccccc} \e_{7(7)} \cong &\mathfrak{u}(1)& \oplus& \e_{6(2)} &\oplus& {\bf 27}  \\
\cong & \mathfrak{u}(1)& \oplus& \su(2) \oplus \su(6) \oplus \scal{Ê{\bf 2} \otimes {\bf 20}}_\IR &\oplus& {\bf 2}\otimes {\bf 6} \oplus {\bf 15} \end{array}
\ee
that the $40$ flat directions generate a $\cM_{\rm \scriptscriptstyle BPS} $ subspace.

\subsection{Non-BPS extremal black holes}
Non-supersymmetric extremal spherically symmetric black holes can be treated in much the
same way. In this case, the Noether charge $Q$ (and therefore the momentum $P$) 
lies in a  $\Spin$ orbit \DSOXVI00000020 of nilpotent elements of $\e_{8(8)} \ominus \so^*(16)$, 
which is itself a Lagrangian submanifold  of   the  $E_{8(8)}$ orbit \DEVIII00000002 \cite{Bossard:2009at}
\be 
\frac{Spin^*(16)}{USp(8) \ltimes {\bf 27}} \subset \frac{E_{8(8)}}{E_{6(6)}\ltimes \scal{Ê( {\bf 1} \oplus {\bf 27} \oplus \overline{\bf 27} )^\ord{2} \oplus {\bf 1}^\ord{4} }}\ .
 \label{NBorbit}  
\ee
The associated graded decompositions of $\e_{8(8)}$ and $\so^*(16)$
can be read off the weighted Dynkin diagram,
\be \label{vingthuit}
 \begin{array}{ccccccccccc}
\so^*(16) &\cong& & & {\bf 28}^\ord{-2}& \oplus &\scal{ÊÊ\gl_1 \oplus \su^*(8)}^\ord{0} &
 \oplus &\overline{\bf 28}^\ord{2} &&  \\
 \e_{8(8)} \ominus \so^*(16) 
& \cong& {\bf 1}^\ord{-4}Ê&\oplus &\overline{\bf 28}^\ord{-2} &\oplus &{\bf 70}^\ord{0} &\oplus & {\bf 28}^\ord{2}& \oplus& {\bf 1}^\ord{4} \cong {\bf 128}_+ \end{array}
\ee
This is the same 5-grading as the one relevant for the minimal nilpotent orbit 
with weighted Dynkin diagrams (\DSOXVI00000010, \DEVIII00000001) associated to  
$1/2$-BPS black holes, after rescaling the generator ${\bf h}$ by two (see Appendix A). 
An orbit representative is a generic element of $ {\bf 28}^\ord{2}$, whose  stabilizer defines 
a $USp(8) \subset SU^*(8)^\ord{0}$ subgroup (equivalently, a generic element of the ${\bf 56}^\ord{2}$ whose stabilizer defines an $E_{6(6)} \subset E_{7(7)}^\ord{0}$ subgroup). 
Its associated semi-simple generator takes the form
\be 
{\bf h}_\asym \equiv \frac{1}{2}\Scal{Êe^{i \alpha}  \Omega_{ij} a^i a^j - e^{-i\alpha}Ê 
\Omega^{ij} a_i a_j} \ ,
\ee
where the complex two-form $\Omega_{ij}$ satisfies
\footnote{Equivalently, $\rho_\zero(\Omega_{ij}) = \rho_\un(\Omega_{ij}) =\rho_\deux(\Omega_{ij}) =\rho_\trois(\Omega_{ij}) = 1$ and $\varphi(\Omega_{ij})  = \frac{\pi}{4}$.}
\beÊ
\Omega_{[ij} \Omega_{kl]} + \frac{1}{24} \varepsilon_{ijklmnpq}Ê\Omega^{mn} \Omega^{pq} =0  
\ ,\qquad \Omega_{ik} \Omega^{jk} = \delta_i^j\ .
\ee
Indeed, using the identities 
\be [Ê{\bf h}_\asym , a^i \pm e^{-i\alpha}Ê\Omega^{ij} a_j ] = \pm \scal{Êa^i \pm e^{-i\alpha}Ê\Omega^{ij} a_j } \ee
one obtains the generators of degree $\pm2$ of $\so^*(16)$ as the `squares' of $a^i \pm e^{-i\alpha}Ê\Omega^{ij} a_j$, respectively, and the $\su^*(8)$ generators as their commutators. One can then check that the elements of ${\bf 1}^\ord{4} $ are of the form
\be  | \C^\ord{4}_\asym\rangle   = \I\, N \, e^{-2i \alpha }Ê \, e^{ \frac{1}{2}Êe^{i\alpha}Ê\Omega_{ij} a^i a^j } 
\, | 0\rangle \label{NUT} 
\ee
where $N$ is real. 
The elements of grade two ${\bf h}_\asym |\C^\ord{2}\rangle = 2 | \C^\ord{2} \rangle$ can be computed to be of the form \cite{Bossard:2009my}
\be
 | \C^\ord{2}_\asym \rangle  = ( 1 + \invo )   \Scal{Ê1 + \frac{1}{4} e^{i\alpha}Ê\Omega_{ij} a^i a^j }Ê\scal{Ê e^{ - 2i \alpha}  M  + e^{-i\alpha} \Xi_{ij} a^i a^j } |0\rangle \ ,
 \label{nonBPSzero}
\ee
where $\Xi_{ij}$ satisfies
\be 
\Xi_{ij} =   \Omega_{ik} \Omega_{jl} \Xi^{kl} \ ,\qquad 
\Omega^{ij}  \Xi_{ij} = 0\ .
 \label{real27} 
\ee
These conditions state that $\Xi_{ij}$ is an element of the real ${\bf 27}$ representation of the $USp(8)$ subgroup of $SU(8)$ that leaves invariant $\Omega_{ij}$. Such elements $|\C^\ord{2}_\asym\rangle$ correspond to special values of the central charges in the asymptotic region, whose 
overall phase $\varphi(Z)$ is determined by the value of the NUT charge. 
The most general charge with strictly negative $E_{7(7)}$ invariant $\lozenge(Z)<0$
can be obtained with a linear combination of $|\C^\ord{2}_\asym\rangle $ 
and $|\C^\ord{4}_\asym\rangle$,  associated to the $\gl_1$ generator 
\be 
{\bf h}  \equiv \frac{1}{2}\Scal{Êe^{i \alpha} ( 1 - i \lambda )   \Omega_{ij} a^i a^j - e^{-i\alpha}Ê  ( 1 + i \lambda ) \Omega^{ij} a_i a_j}  + i \lambda \scal{Êa^i a_i - 4 }\ .
\ee
In this case the general element of grade two $|\C^\ord{2}\rangle$ reads
\begin{multline} 
 | \C^\ord{2} \rangle  = N\, ( 1 + \invo )   
 \left( Êe^{-2i\alpha} \Scal{Ê 1 + i\lambda } + e^{-i\alpha} \frac{ 1  + 2 i \lambda}{4}   \Omega_{ij} a^i a^j  + \frac{i \lambda}{16}  \Omega_{ij} \Omega_{kl} a^i a^j a^k a^l \right . \\ \left .  
 + N^{-1}\, \Scal{Ê1 + \frac{1}{4} e^{i\alpha}Ê\Omega_{ij} a^i a^j }Ê\,  e^{-i\alpha} \Xi_{ij} a^i a^j  \right) \, | 0\rangle \ ,
\label{nonBPSZero1}
\end{multline}
with NUT charge 
\be 
\label{nut}
k = N \Scal{Ê\lambda \cos 2\alpha  - \sin 2\alpha } \ .
\ee
The parameters $\Omega_{ij} \in USp(8) \backslash SU(8)$, $\alpha,\, \lambda,\, N$ and $\Xi_{ij} \in {\bf 27}$ give a complete parametrization of the $Spin^*(16)$ orbit (\ref{NBorbit}). To see this, we note that the elements of $\so^*(16)$ that act non-trivially on a generic element of grade two of (\ref{vingthuit}) combine into representations of its $USp(8)\subset SU^*(8)$ isotropy subgroup as
\be 
\scal{Ê{\bf 1} \oplus {\bf 27}}^\ord{-2} \oplus \scal{ÊÊ\gl_1 \oplus  {\bf 27}}^\ord{0} \oplus {\bf 1}^\ord{2} \subset \so^*(16)  \ .
\ee
The component of grade zero leaves ${\bf h}$ invariant by definition, and acts transitively on the set of parameters $N, \, \Xi_{ij}$ defining regular black holes. The generators of $\su(8)$ that act non-trivially on $\Omega_{ij}$ are given by linear combinations in ${\bf 27}^\ord{-2} \oplus {\bf 27}^\ord{2}$, 
while the remaining 
nilpotent generators ${\bf 1}^\ord{-2} \oplus {\bf 1}^\ord{2}$ act by shifting the two parameters $\alpha$ and $\lambda$. 

Since we are interested in static solutions only, we must cancel the NUT charge \eqref{nut} 
by choosing $\lambda =  \tan( 2 \alpha )$ (with $- \frac{\pi}{2} < \alpha < \frac{\pi}{2}$). The
asymptotic value of $P$ thus takes the form
\be 
\framebox{$| P_\asym  \rangle  = ( 1 + \invo )    \Scal{Ê1 + \frac{1}{4} e^{i\alpha}Ê\Omega_{ij} a^i a^j }Ê\left( M \Scal{Ê1 + \frac{i}{4} e^{-i\alpha} \sin(2\alpha)  \Omega_{kl} a^k a^l } + e^{-i\alpha}  \Xi_{ij} a^i a^j  \right)  |0\rangle$}  \label{nonBPS}
\ee
where $M=\cos^{-1}(2\alpha) N $ is the physical mass. 
The asymptotic central charge $Z_{ij}$ 
is readily obtained from the part of \eqref{nonBPS} bilinear in the oscillators,
\be 
\label{Zdec}
Z_{ij} = \frac{1}{2}Ê\Scal{Êe^{i\alpha} + i e^{-i\alpha} \sin 2\alpha  } 
\varrho \, \Omega_{ij} + e^{-i\alpha} \Xi_{ij} \ ,
\ee
where $\varrho\equiv M/2$.
Due to the conditions (\ref{real27}), the complex two-form $Z_{ij}$  
can brought by an $SU(8)$ rotation into
the skew-diagonal form
\begin{multline} 
\tilde R^k{}_i \tilde R^l{}_j Z_{kl} \, \, \hat{=}  \, \, \frac{e^\frac{i\pi}{4}}{2}Ê\, \left( \begin{array}{cc}0 & \, \, 1  \\- 1 & \, \, 0 \end{array}\right) \otimes  \left[   Ê\Scal{Êe^{i\alpha} + i e^{-i\alpha}  \sin 2\alpha }  \left( \begin{array}{cccc} \, \, \varrho \, \, Ê& 0 & 0 & 0 \\0& \, \, \varrho \, \,  & 0 & 0 \\  0 & 0 &\, \,  \varrho \, \, & 0 \\ 0 & 0 & 0 & \varrho \end{array} \right)  \right . \\*
\left . +  e^{-i\alpha} \, \left( \begin{array}{cccc} \, \, {\scriptstyle \xi_\un +\xi_\deux +\xi_\trois} \, \, Ê& 0 & 0 & 0 \\ 0& \, \, -\xi_\un \, \,  & 0 & 0 \\  0 & 0 &\, \, - \xi_\deux \, \, & 0 \\ 0 & 0 & 0 &- \xi_\trois \end{array} \right) \right] \label{Zdiag}\ ,
\end{multline}
where $\varrho>0$ and $\xi_\un, \xi_\deux, \xi_\trois \in \IR$ can be chosen such that 
\be 
 \xi_\un + \xi_\deux + \xi_\trois \, \ge \,  -\xi_\un\, \ge \,  -\xi_\deux Ê\, \ge \, - \xi_\trois \label{tetrah0} \ .
\ee
The condition of non-saturation of the BPS bound $M> \rho_\n$ further requires that 
\be
\varrho\, \cos 2\alpha  \,  > \,  \xi_\un + \xi_\deux + \xi_\trois \, \ge \,  -\xi_\un\, \ge \,  -\xi_\deux Ê\, \ge \, - \xi_\trois \label{tetrah} \ .
\ee
In the parametrization \eqref{Zdiag}, the $E_{7(7)}$ quartic invariant \eqref{defloz}
factorizes into 
\be
\label{lozengexi}
\lozenge(Z) = - 16 \cos^2 (2 \alpha) \,   \Scal{Ê\varrho  \cos(2\alpha) - \xi_\un-\xi_\deux-\xi_\trois} \prod_{\n=\un}^\trois\Scal{Ê\varrho\cos(2\alpha)  + \xi_\n}  \ ,
\ee
and is therefore strictly negative inside the tetrahedron (\ref{tetrah}). It vanishes whenever
$M=\rho_\m$ for any $\m$, i.e. when the black hole becomes BPS with vanishing horizon area. 
The quantities $\alpha,\, \varrho,\, \xi_\un,\, \xi_\deux$ and $\xi_\trois$ define a basis of the $SU(8)$ invariant functions of $Z_{ij}$ well suited for non-BPS solutions, 
alternative to the basis of functions $\rho_\n,\varphi$ adapted to the BPS case.
The diagonalization problem \eqref{Zdiag} is central to the determination 
of the fake superpotential for non-BPS solutions, and is reduced in Appendix \ref{AppW}
to the solution of a sextic polynomial.

Comparing the general form of $|P\rangle$ (\ref{Pspinor}) to the general representative of the relevant nilpotent orbit (\ref{nonBPS}), we conclude that in the static case (\ie $k=0$), the field dependent  ${\bf h}$ generator is determined in terms of the central charges $Z_{ij}$. Using the decomposition  (\ref{Zdec}),
\be
 Z_{ij} = \frac{1}{2} e^{-i\alpha(\phi)}Ê\tilde{R}_i{}^k(\phi) \tilde{R}_j{}^l(\phi) \biggl( Ê\Scal{Êe^{2 i\alpha(\phi)} + i \sin\scal{Ê2\alpha(\phi)}  } \varrho(\phi) \, \Omega_{kl} \biggr)  + e^{-i\alpha(\phi)}  \Xi_{ij}(\phi) \label{ZNB}  \ee
where
\be   \Xi_{ij}(\phi) = \frac{1}{2}Ê\tilde{R}_i{}^k(\phi) \tilde{R}_j{}^l(\phi) \biggl(  \xi_\un(\phi) \varpi^\un_{kl} + \xi_\deux(\phi) \varpi^\deux_{kl} + \xi_\trois(\phi) \varpi^\trois_{kl} \biggr)  \ee 
with $\xi_\n(\phi)$ taking values in the tetrahedron (\ref{tetrah}), the ${\bf h}$ generator is defined by 
\be  Ê{\bf h }Ê=  \frac{1}{2} \Scal{Ê  {\hat{\Omega}}_{ij}(\phi)  a^i a^j -  {\hat{\Omega}}^{ij}(\phi) a_i a_j}   + i  \tan\scal{Ê2\alpha(\phi)}   \scal{Êa^i a_i - 4 } \ee
with
 \be  
{\hat{\Omega}}_{ij}(\phi) = \frac{e^{-i \alpha(\phi) }}{\cos\scal{ 2\alpha(\phi)}}    \tilde{R}_i{}^k(\phi) \tilde{R}_j{}^l(\phi) \Omega_{kl} 
\ee
Indeed, one computes that the general solution to the equation ${\bf h} | P \rangle = 2  | P \rangle$ is
\begin{multline} | P \rangle = ( 1 + \invo )    \Scal{Ê1 + \frac{1}{4} e^{2i\alpha} \cos(2\alpha) Ê{\hat{\Omega}}_{ij} a^i a^j }Êe^{U} \left(  2 \varrho  \Scal{Ê1 + \frac{i}{8} \sin(4\alpha)  \hat{\Omega}_{kl}  a^k a^l } \right . \\* \biggl .  + e^{-i\alpha}  \Xi_{ij}  a^i a^j  \biggr)  |0\rangle  \label{PnonBPS}
\end{multline}

The non-BPS nilpotent orbit therefore determines the scalar fields $U$ and $v$ through the first order equations 
\be
\begin{split}
 \dot{U} & = -  Ê2\, e^{U} \varrho  \\
 {u_{ij}}{}^{IJ} \dot{v}_{klIJ}  - v_{ijIJ} {\dot{u}_{kl}}{}^{IJ} &= - 6 e^U 
 \Scal{Ê {\hat{\Omega}}_{[ij} Z_{kl]} + \frac{1}{24} \varepsilon_{ijklmnpq} 
 Ê{\hat{\Omega}}^{mn} Z^{pq}  }   
 \label{NBlinear}
 \end{split}
 \ee
The first equation identifies the fake superpotential  for extremal non-BPS black holes as 
\be 
\framebox{$W = 2 \varrho  \ , $}
\label{WNB}
\ee
where $\varrho$ is the diagonal component of the central charge $Z_{ij}$
in the  non-standard diagonalization problem \eqref{Zdiag}. 
In particular, $W$ is larger than the modulus
of any of the eigenvalues $\rho_\m$ of the central charge matrix, as required for
a non-BPS solution. Equation \eqref{WNB} is one of the main results in this paper.

In order to check that the second line in \eqref{NBlinear} describes the gradient flow of $W$,
we rewrite $W =Ê\frac{1}{2} \Re\bigl[ {\hat{\Omega}}^{ij} Z_{ij} \bigr]$, 
and compute  $dW$ following the same steps as in \eqref{dWun}. In this way we obtain
\begin{multline} 
d W = \frac{1}{4}Ê\Esh_\phi (d \phi)_{ijkl} \Scal{Ê {\hat{\Omega}}^{ij} Z^{kl}  + \frac{1}{24} \varepsilon^{ijklmnpq}  {\hat{\Omega}}_{mn} Z_{pq} } \\*
+\frac{1}{2}  \Re\biggl(     \tilde{R}^i{}_k  \tilde{R}^j{}_l \Omega^{kl} \, Z_{ij} \, d  \frac{e^{i \alpha }}{\cos2\alpha} \biggr) \\* 
+    \Re\left(  {\hat{\Omega}}^{ik} \left[  \frac{1}{3} \Scal{ÊÊu^{jl}{}_{IJ} d u_{il}{}^{IJ} - v^{jlIJ} d v_{ilIJ} } - \tilde{R}^j{}_l d \tilde{R}_i{}^l   \right] Z_{jk}\right)   \label{dWdeux}  \end{multline}
One computes easily that 
\be \tilde{R}^i{}_k  \tilde{R}^j{}_l \Omega^{kl} \,  d  \frac{e^{i \alpha }}{\cos 2\alpha} = i \Scal{Ê1 - 2 i \tan 2\alpha}  {\hat{\Omega}}^{ij} d \alpha \ee
and using the explicit form of the central charge (\ref{ZNB}) one computes that
\be \frac{1}{2}\, {\hat{\Omega}}^{ij} Z_{ij} = 2 \Scal{Ê1 + 2 i \tan\scal{Ê2 \alpha} } \varrho \ee
such that the second line in (\ref{dWun}) cancels. 
The anti-hermiticity of the $\su(8)$ generators implies that 
\begin{multline}   {\hat{\Omega}}^{i[k}  \left[  \frac{1}{3} Ê\Scal{ÊÊu^{j]l}{}_{IJ} d u_{il}{}^{IJ} - v^{j]lIJ} d v_{ilIJ} } - \tilde{R}^j{}_l d \tilde{R}_i{}^l   \right]  \\* = - \mathring{\Omega}^{kp} \mathring{\Omega}^{jq} \hat{\Omega}_{i[p} \left[  \frac{1}{3}   \Scal{ÊÊu_{q]l}{}^{IJ} d u^{il}{}_{IJ} - v_{q]lIJ} d v^{ilIJ} }- \tilde{R}_q{}^l d \tilde{R}^i{}_l   \right]  \label{ReaIma} \end{multline}
where $\mathring{\Omega}_{ij} \equiv \cos(2 \alpha) {\hat{\Omega}}_{ij}$ is normalized such that $\mathring{\Omega}_{ik} \mathring{\Omega}^{jk} = \delta_i^j $.  With respect to  ${\hat{\Omega}}_{ij}$, the generators of $\su^*(8)$ decompose into the generators of the $\mathfrak{usp}(8)$ subalgebra that leave ${\hat{\Omega}}_{ij}$ invariant, and the remaining $27$ generators that transform as a traceless antisymmetric tensor of rank two with respect to $USp(8)$. It follows that
\begin{multline}  {\hat{\Omega}}^{ik} \left[  \frac{1}{3} \Scal{ÊÊu^{jl}{}_{IJ} d u_{il}{}^{IJ} - v^{jlIJ} d v_{ilIJ} } - \tilde{R}^j{}_l d \tilde{R}_i{}^l   \right] Z_{jk} \\
=  {\hat{\Omega}}^{ik} \left[  \frac{1}{3} \Scal{ÊÊu^{jl}{}_{IJ} d u_{il}{}^{IJ} - v^{jlIJ} d v_{ilIJ} } - \tilde{R}^j{}_l d \tilde{R}_i{}^l   \right]  e^{-i\alpha}Ê\Xi_{jk}  \label{Imaginary} \end{multline}
And because 
\be e^{-i\alpha}Ê\Xi_{ij}  = \mathring{\Omega}_{ik}  \mathring{\Omega}_{jl} e^{i\alpha}Ê\Xi^{kl} \label{ReaRe}  \ee
by definition (\ref{real27}), (\ref{Imaginary}) is pure imaginary and does not contribute to (\ref{dWun}). We finally obtain that 
\be d W = \frac{1}{4}Ê\Esh_\phi (d \phi)_{ijkl} \Scal{Ê \hat{\Omega}^{ij} Z^{kl}  + \frac{1}{24} \varepsilon^{ijklmnpq}  \hat{\Omega}_{mn} Z_{pq} } \ee
such that the first order equations (\ref{NBlinear}) can be rewritten as
\be 
\dot{U} =  - e^{U} W \, , \hspace{10mm}\dot{\phi}_{ijkl} = - 12  e^{U}  \,  {\Esh_\phi}^{-2} \left( \frac{ \partial W}{\partial \phi}\right){}_{ijkl}  \label{gradEqua} 
\ee
This confirms that \eqref{WNB} is indeed a fake superpotential for the non-BPS extremal solutions.

In Appendix \ref{AppW} we discuss how $W$ \eqref{WNB} can be evaluated in practice, and show 
that $W^2$ can be obtained in general as a specific root of a polynomial (\ref{sixDegree}) of degree six,
whose coefficients are polynomial in the central charges $Z_{ij}$. At some particular
loci in the $\rho_\n, \varphi$ space however, this polynomial becomes reducible.
This happens  in particular when $\varphi=0 \mod \pi/4$, $\alpha=0$, where
the sextic polynomial (\ref{sixDegree})
becomes fully reducible. For  $\varphi=\pi/4 \mod \pi/2$,
the physical root is $W_0$ in \eqref{sixroots}, \ie 
\be 
\framebox{
$\varphi=\pi/4 : \ \qquad 
W = \frac{1}{2} \Scal{Ê\rho_\zero + \rho_\un+\rho_\deux+\rho_\trois} \ .
$}
\ee
This result is in agreement with \cite{Andrianopoli:2007gt}. Our formula \eqref{WNB} 
is however more general, and does not assume any restriction on the moduli nor on 
the charges. Similarly, at $\varphi=0 \mod \pi/2$, the physical root is the largest 
of $W_\n$ in \eqref{sixroots2}, i.e. 
\be
\label{physr2t}
\framebox{$
\varphi=0:\ \qquad
W = \frac12 \Scal{ \rho_\zero  + \rho_\un + \rho_\deux + \rho_\trois }  - {\rm min}(\varrho_\m)\ .$}
\ee

Having obtained the fake superpotential for non-BPS black holes, one may 
substitute (\ref{gradEqua}) into \eqref{PnonBPS} and rewrite the momentum as 
\be  
| P  \rangle =e^{U} (1 + \invo) \left(  W ÊÊ +  Z_{ij} a^i a^j  + \frac{1}{2}    {\Esh_\phi}^{-1} \left( \frac{ \partial W}{\partial \phi}\right){}_{ijkl}  a^i a^j a^k a^l \right) | 0 \rangle
 \ee
By virtue of the attractor mechanism, the scalar fields at the horizon  
again lie at an extremum value of $W$. It follows that 
\be 
| P \rangle \sim e^{U} (1+ \invo) \Scal{Ê ÊW_* +  Z_{ij\,*} a^i a^j } | 0 \rangle \ee
as $\tau$ goes to infinity. 
Using (\ref{PnonBPS}), one concludes that at the horizon, the central charges satisfy
\be 
\xi_\un(Z_{ij\, *}) =  \xi_\deux(Z_{ij\, *})=  \xi_\trois(Z_{ij\, *}) = 0 \ ,\qquad
\alpha(Z_{ij\, *}) = 0 
 \ee
In terms of the $SU(8)$ invariant functions appearing in the standard diagonalization
(\ref{Zdecompose}), this can be expressed  as
\be \rho_\zero(Z_{ij\, *}) =  \rho_\un(Z_{ij\, *})= \rho_\deux(Z_{ij\, *})= \rho_\trois(Z_{ij\, *}) = \frac{1}{2}\sqrt[4]{ - \lozenge(Z)} Ê\ ,\qquad
 \varphi(Z_{ij\, *}) = \frac{\pi}{4} \ .
 \ee
In particular, at the horizon 
\be W_* = \sqrt[4]{ - \lozenge(Q_{IJ})}  \ ,
\ee
where we again noted that $\lozenge(Z)$ depends only on the the conserved charges.
Using \eqref{massent} we recover the known entropy formula for non-BPS
extremal black holes in $\N=8$ supergravity \cite{Ferrara:2006em}.

Similarly to the BPS case, the symplectic form $\Omega_{ij}$ is invariant under 
$USp(8) \subset SU(8)$ and  $E_{6(6)} \subset E_{7(7)}$, respectively. This implies 
that fake superpotential $W$ exhibits flat directions homeomorphic to the symmetric space
\be 
\label{mnonbps}
\cM_{\rm non-BPS} \cong \Sp  \big\backslash E_{6(6)} \ .
\ee
along which the attractor mechanism is inactive \cite{Ferrara:2007pc,Ferrara:2007tu,Bellucci:2006xz}
(here $\Sp$ is the quotient of $USp(8)$ by its $\mathds{Z}_2$ centre leaving invariant the representations of even rank). Indeed, the symplectic form $\hat{\Omega}_{ij}$ associated to the central charge is left invariant by a $USp(8) \subset SU(8)$ subgroup, with respect to which the components of $Ê{u_{ij}}{}^{IJ} d v_{klIJ} - v_{ijIJ} d {u_{kl}}{}^{IJ}$ are only non-zero along the components $\mathds{R} \oplus {\bf 27} \subset {\bf 70}$, and are always zero along ${\bf 42} \subset Ê {\bf 70}$ as can be seen from (\ref{PnonBPS}). From the graded decomposition
\be \begin{array}{rccccc} \mathfrak{e}_{7(7)} \cong &\overline{\bf 27}^\ord{-2} &\oplus &\scal{Ê\gl_1 \oplus \mathfrak{e}_{6(6)} }^\ord{0}  &\oplus& {\bf 27}^\ord{2}\\
\cong& {\bf 27}^\ord{-2}Ê&\oplus & \scal{Ê\gl_1 \oplus \mathfrak{usp}(8) \oplus {\bf 42}}^\ord{0} &\oplus & {\bf 27}^\ord{2} \end{array} \ , \ee
it is evident that the 42 flat directions at the horizon parametrize 
the symmetric space \eqref{mnonbps}. It should be noticed that 
this moduli space of flat directions is isomorphic to the moduli space  $\cM_5$
of the five-dimensional supergravity model obtained after 
decompactification \cite{Ferrara:2007tu,Bellucci:2006xz}.

We conclude this section by discussing the relation between the nilpotent orbits of the moment  $P$ under 
$G_3$, and the orbits of the electromagnetic charges $Q_{IJ}$ under $G_4$,  
as classified in \cite{Ferrara:1997uz,Bellucci:2006xz}. To this aim, notice that, as an $\e_{8(8)}$ 
element, $P$ is proportional to 
\be
 {\bf e}Ê= {\bf H} + \hat{\bf Z}_* \ ,
 \qquad
\hat{\bf Z}_* \equiv \frac{1}{2}ÊW_*^{-1}Ê\Scal{ÊZ_{*\, ij} \scal{Ê{\bf F}^{ij} - {\bf E}^{ij} } - Z_*^{ij}Ê\scal{Ê{\bf F}_{ij} - {\bf E}_{ij} }} \ ,\ee
at $\tau \rightarrow +\infty$. One can then use the Cartan involution $\dagger$ defining $Spin_{\scriptscriptstyle \rm c}(16)\subset E_{8(8)}$ to complete ${\bf e}$ into a triplet $({\bf e},\, {\bf f},\, {\bf h})$,
\be {\bf f} = {\bf e}^\dagger = {\bf H} - \hat{\bf Z}_* \quad , \quad \quad {\bf h} = [ {\bf e}Ê, {\bf f} ]Ê= - 2 [Ê{\bf H}Ê,  \hat{\bf Z}_* ]Ê\ . \ee
Indeed, it follows from  
\be [Ê{\bf H} , [ {\bf H}Ê, \hat{\bf Z}_* ] ]Ê= \hat{\bf Z}_* \quad , \quad \quad [Ê\hat{\bf Z}_* , [Ê{\bf H} , \hat{\bf Z}_* ] ] = {\bf H}  \ee
that $[ {\bf h} , {\bf e}]Ê= 2 {\bf e}Ê$.
The maximal reductive subgroup $J^{\scriptscriptstyle \rm R}_{\bf e}$ of the stabilizer of ${\bf e}$ in $E_{8(8)}$ is the intersection of the stabilizers of ${\bf e}$ and ${\bf h}$ in $E_{8(8)}$.  The stabilizer of 
 ${\bf h}$ in $E_{8(8)}$ is a $GL(1,\IR) \times E_{7(7)}$ subgroup, such that the $E_{7(7)}$ factor is conjugate but  distinct  from the four-dimensional duality group $G_4=E_{7(7)}$. Because  any element ${\bf x}$ of a reductive algebra $\mathfrak{j}_{\bf e}$ decomposes into two elements ${\bf x}_\pm\in \mathfrak{j}_{\bf e}$, such that  
\be {\bf x} = {\bf x}_+ + {\bf x}_- \ , \qquad {\bf x}_\pm^\dagger = \pm {\bf x}_\pm \ , \ee
$J^{\scriptscriptstyle \rm R}_{\bf e}$ must stabilize ${\bf H}$ and $\hat{\bf Z}_*$ separately. On the other hand, the stabilizer of ${\bf H}$ is $GL(1,\mathds{R})\times G_4$. At the horizon (\ie at $\tau \rightarrow + \infty$), $J^{\scriptscriptstyle \rm R}_{\bf e}$ therefore coincides
with the stabilizer of the central charge $Z_{*\, ij}$ inside $E_{7(7)}$. $J^{\scriptscriptstyle \rm R}_{\bf e}$ is then the intersection of the two $E_{7(7)}$ subgroups defined by ${\bf H}$ and ${\bf h}$, respectively. Away from the horizon however (at finite $\tau$), these two subgroups differ by a similarity transformation. 

The above discussion is in fact completely general and applies for any extremal black hole with non-vanishing horizon area, in any theory admitting a symmetric scalar manifold whose isometry group acts faithfully on the electromagnetic charges. The stabilizers of the central charges computed in \cite{Ferrara:1997uz,Bellucci:2006xz}, indeed match perfectly the reductive stabilizers of the  corresponding nilpotent orbits given in \cite{Levi}. We have already seen in the case of maximal supergravity that we have the following correspondence 
\bea J_{P}Ê\cong E_{6(2)} \ltimes \scal{Ê {\bf 27}^\ord{2} \oplus {\IR}^\ord{4}}  \times \IR &   \quad \Leftrightarrow \quad  &J_{Q} \cong E_{6(2)} \ , \CR
  J_{P}Ê\cong E_{6(6)} \ltimes \scal{Ê(  {\bf 27} \oplus \overline{\bf 27} )^\ord{2} \oplus {\IR}^\ord{4}} \times \IR & \quad    \Leftrightarrow \quad &J_{Q}  \cong E_{6(6)} \ , 
  \eea
  between the stabilizer $J_P$ of $P$ in $E_{8(8)}$ and the stabilizer $J_Q$ of $Q_{IJ}$ in $E_{7(7)}$.  In the example of the exceptional $\cN=2$ supergravity,  to be discussed in Section 3.1 below, 
   $G_4 \cong E_{7(-25)}$ and $G_3 \cong E_{8(-24)}$, and there are three nilpotent orbits of $E_{8(-24)}$ associated to extremal black holes,  characterized  by the stabilizers 
\bea J_{P}Ê\cong E_{6(-78)} \ltimes \scal{Ê {\bf 27}^\ord{2} \oplus {\IR}^\ord{4}}  \times \IR &   \quad \Leftrightarrow \quad  &J_{Q} \cong E_{6(-78)} \ , \CR
J_{P}Ê\cong E_{6(-14)} \ltimes \scal{Ê {\bf 27}^\ord{2} \oplus {\IR}^\ord{4}}  \times \IR   &   \quad \Leftrightarrow \quad  &J_{Q} \cong E_{6(-14)} \ , \CR
  J_{P}Ê\cong E_{6(-26)} \ltimes \scal{Ê(  {\bf 27} \oplus \overline{\bf 27} )^\ord{2} \oplus {\IR}^\ord{4}} \times \IR & \quad    \Leftrightarrow \quad &J_{Q}  \cong E_{6(-26)} \ .
  \eea

\section{Truncations of maximal supergravity and their extensions}
\label{sectrunc}
Our method  for deriving the fake superpotential of static extremal black holes 
applies to any symmetric model of 4D gravity coupled to abelian vector fields and scalar fields 
invariant under a continuous duality  group $G_4$, provided $G_4$ acts faithfully on  electromagnetic charges. Many such models can be obtained by consistent truncations of $\N=8$ supergravity, 
and all others follow by  ``covariantisation". 
In this Section we shall restrict ourselves to models  that can be obtained by reduction on a circle 
of 5D supergravity theories. Indeed, these are the only ones which admit 
non-BPS black holes with no central charge being saturated \cite{Bossard:2009my}, for which our formalism is most fruitful. All other symmetric models  have only two types of generic extremal black holes, and the fake superpotential is always of the form $W = \rho_\n$.  

\subsection{Magic $\N=2$ supergravity}
The maximal $\N=2$  truncation of $\N=8$ supergravity 
is the `magic' supergravity theory associated to the quaternions \cite{Gunaydin:1983rk}. Its
bosonic sector is identical to the bosonic sector of $\N=6$ supergravity. Within this truncation, the central charges decompose into the $\N=2$ central charge $Z= 2 Z_{12} $ and the $\N=6$ central charges $Z_{AB}  $ which define an antisymmetric tensor of $U(6)$, with $S\scal{ÊU(2) \times U(6)}  \subset SU(8)$ such that $U(2)$ is the $\N=2$ R-symmetry group and $U(6)$ the $\N=6$ R-symmetry group. The maximal subgroup of $E_{7(7)}$ that preserves this decomposition is $SU(2) \times Spin^*(12)_{\scriptscriptstyle \rm c}$, and the scalar fields take values in
\be 
\label{mso12}
q) \qquad \M_4 \cong U(6) \backslash  Spin^*(12)_{\scriptscriptstyle \rm c} \ ,
\qquad
\M_3^* \cong  \scal{SL(2)\times Spin^*(12)_{\scriptscriptstyle \rm c}} \backslash E_{7(-5)}\ .
 \ee
In particular, only the components (and the ones obtained by antisymmetric permutations of the indices)
\be \phi_{12AB} \equiv \phi_{AB} \, , \hspace{10mm} \phi_{ABCD} = \frac{1}{2} \varepsilon_{ABCDEF}Ê\phi^{EF} \ee
are non-zero in the symmetric gauge, where $\phi_{AB}$ is an antisymmetric tensor of $U(6)$.

With respect to $S\scal{ U(2) \times U(6)} \subset SU(8)$, the representations of the fermions decompose as follows
\be
 {\bf 8} \cong {\bf 2} \oplus {\bf 6} \ ,\qquadÊ
{\bf 56}Ê\cong {\bf 2} \otimes {\bf 15} \oplus \bar {\bf 6} \oplus {\bf 20}\ ,
 \ee 
such that the fermionic fields of the $\N=2$ truncation are 2 gravitini and symplectic-Majorana spinors in the ${\bf 2}Ê\otimes {\bf 15}$ representation, and the fermionic fields of the $\N=6$ truncation are 6 gravitini and symplectic-Majorana spinors in the $\bar {\bf 6} \oplus {\bf 20}$.

The $SU(8)$ invariant functions of the central charges $Z_{ij}$, $\rho_\zero,\, \rho_\un,\, \rho_\deux,\, \rho_\trois $ and $\varphi$  then become $U(6)$ invariant functions of $Z$ and $Z_{AB}$ 
defined by\footnote{Note that after identifying $\rho_\zero=|Z|$, as we shall do in all $\N=2$ models,
we can no longer assume that the canonical $\N=8$ ordering  $\rho_\zero \ge \rho_\un \ge \rho_\deux \ge \rho_\trois$ is satisfied, although we can still assume that $\rho_\un \ge \rho_\deux \ge \rho_\trois$.}
\begin{multline}  
\hspace{-3mm} \lambda^3 - 2 Z_{AB} Z^{AB} \lambda^2 + \Scal{ 2 \scal{ÊZ_{AB} Z^{AB} }^2 - 4 Z_{AB} Z^{BC} Z_{CD} Z^{DA}Ê} \lambda  - \frac{1}{36} \bigl| \varepsilon_{ABCDEF}ÊZ^{AB} Z^{CD} Z^{EF} \bigr|^2   \\*
= \scal{Ê\lambda - {\rho_\un}^2 } \scal{Ê\lambda - {\rho_\deux}^2 } \scal{Ê\lambda - {\rho_\trois}^2 } 
\end{multline}
and 
\be \rho_\zero = |Z| \, , \hspace{10mm}Ê4 \varphi = \arg\Bigl[ ÊZ \varepsilon^{ABCDEF} \, Z_{AB}ÊZ_{CD} Z_{EF}  \Bigr] \ .
\ee
There are three classes of generic extremal static black holes in this model. They all 
correspond to the same complex  nilpotent orbit of $E_7(\IC)$, with weighted Dynkin diagram \DEVII2000000. We shall label them by their $Spin^*(12)\times SL(2,\IR)$ weighted Dynkin diagrams, which define both the real nilpotent orbit of $E_{7(-5)}$ in which their Noether charge lies, and the  $Spin^*(12)\times SL(2,\IR)$ orbit in which the coset component of their Maurer--Cartan form $P$
lies. $1/2$-BPS black holes correspond to the weighted Dynkin diagram  \DSOXII0000004, and are
controlled by the fake superpotential  $W = |Z| $. The non-BPS ones with $Z_* = 0$ \DSOXII0200000 (which are $1/6$-BPS in the context of $\N=6$ supergravity) are controlled by the fake superpotential $W = \rho_\un(Z_{AB})$, and the non-BPS black holes with $Z_* \ne  0$ \DSOXII0000202, by the fake superpotential $W= 2 \varrho(Z,Z_{AB})$, which is the same function of $\rho_\zero,\, \rho_\un,\, \rho_\deux,\, \rho_\trois $ and $\varphi$ as for the non-BPS black holes of maximal supergravity.

In addition to \eqref{mso12},there are three additional $\N=2$ `magic' supergravity theories
in $D=4$,  associated to the octonions,  complex and real numbers respectively. 
Their  moduli spaces are given by \cite{Gunaydin:1983rk}
 \bea 
o) \quad & \M_4 \cong \scal{ÊU(1) \times E_{6(-78)} } \backslash  E_{7(-25)} \  ,&
 \qquad \M_3^*  \cong \scal{ÊSL(2) \times E_{7(-25)}  } \backslash  E_{8(-24)}\ , \nn\\
c) \quad  & \M_4 \cong S\scal{U(3) \times U(3)}Ê\backslash SU(3,3) \ ,& 
\qquad   \M_3^*  \cong \scal{ÊSL(2) \times SU(3,3)  } \backslash  E_{6(2)}\ , \nn\\
r)\quad  &  \M_4 \cong U(3) \backslash Sp(6,\IR) \ ,&
\qquad    \M_3^*  \cong \scal{ÊSL(2) \times Sp(6,\IR)  } \backslash  F_{4(4)} \ ,
 \eea 
respectively. The truncations from q) to c) and r) amount to restricting $Z_{AB}$ to the ${\bf 3} \otimes {\bf 3}$ of $S\scal{ÊU(3) \times U(3)} \subset SU(6)$ and the symmetric tensor representation ${\bf 6}$ of its diagonal subgroup $U(3) \subset S\scal{ÊU(3) \times U(3)}$, respectively. Case o)
cannot be obtained by truncation from q) , and needs to be discussed separately.

In this case, the electromagnetic charges transform  in the $\bf 56$ of $E_{7(-25)}$, which decomposes as $\IC \oplus {\bf 27}$ with respect to $U(1) \times E_{6(-78)}$,
 with ${\bf 27}$ being the complex fundamental representation of $ E_{6(-78)}$. With respect to  the $SU(2)\times SU(6) \subset E_{6(-78)}$ subgroup associated to the `truncation to the quaternions', the complex ${\bf 27}$ representation decomposes as ${\bf 15} \oplus {\bf 2} \otimes {\bf 6} $, and the complex charge $Z_a$ splits this way into $Z_{AB}$ and $ Z_\alpha^A$. The action of the remaining generators of $\e_{6(-78)}$ in the $({\bf 2} \otimes {\bf 20})_\IR$ of $SU(2) \times SU(6)$, is defined as follows 
 \be \delta Z_{AB}Ê= \Lambda^{\alpha}_{ABC}ÊZ^C_\alpha \, , \hspace{10mm}Ê\delta Z^A_\alpha = - \Lambda_\alpha^{ABC}ÊZ_{BC} \ee
 in term of the complex self-dual parameters $\Lambda^\alpha_{ABC} = \frac{1}{6} \varepsilon^{\alpha\beta} \varepsilon_{ABCDEF}Ê\Lambda_\beta^{DEF} $. Using these generators, one computes the quadratic and the cubic invariants\footnote{Do not confuse  the 
$E_{6(-78)}$-invariant  tensor $t^{abc}$ appearing in \eqref{E6inv}
with the $E_{6(-26)}$-invariant tensor appearing in the standard cubic prepotential of $\cM_4$.}
of $E_{6(-78)}$ as
 \bea Z_a Z^a &=& Z_{AB} Z^{AB} + Z^A_\alpha Z^\alpha_A \CR
 t^{abc} Z_a Z_b Z_c &= & 3 \, \varepsilon^{\alpha \beta} Z_{AB} Z_\alpha^A Z_\beta^B + \frac{1}{4} \varepsilon^{ABCDEF}ÊZ_{AB} Z_{CD} Z_{EF} 
 \label{E6inv}
 \eea  
A generic element $Z_a$ can always be rotated to a basis in which its component $Z_\alpha^A = 0$, and $Z_{AB}$ takes a block diagonal form such that a generic charge $Z_a$ is parametrized by 50 angles of $Spin(8) \backslash E_{6(-78)}$ and four invariants. The $SU(6)$ invariant functions $\rho_\un,\, \rho_\deux,\, \rho_\trois $ lift in a unique way to the roots of the $E_{6(-78)}$ invariant polynomial
\begin{multline}  
\lambda^3 - 2 \, Z_a Z^a\, Ê\lambda^2+  2\,  t_{abe} t^{cde}\,  Z_c Z_d Z^a Z^b  \, \lambda - \frac{4}{9} \bigl| \, t^{abc}ÊZ_a Z_b Z_c  \, \bigr|^2   \\*
= \scal{Ê\lambda - {\rho_\un}^2 } \scal{Ê\lambda - {\rho_\deux}^2 } \scal{Ê\lambda - {\rho_\trois}^2 } \ ,
\end{multline}
while $\rho_\zero$ and $\varphi$ are the $U(1) \times E_{6(-78)}$ invariants
\be 
\rho_\zero = [Z]\ ,\qquad 
4 \varphi = \arg\left[ Z \, t^{abc} Z_a Z_b Z_c  \right]Ê \ .
\ee
In terms of these invariants, the $E_{7(-25)}$ quartic invariant is defined similarly to (\ref{Lozrho}), 
\be  \lozenge(Z) = \Scal{Ê|Z|^2 - 2  Z_a Z^a }^2   - 8 \,  t_{abe} t^{cde}\,  Z_c Z_d Z^a Z^b  + \frac{16}{3}  \, \Re \bigl[Ê Z \, t^{abc} Z_a Z_b Z_c \bigr] 
\ee

As in magic supergravity q) there are three classes  of generic extremal black holes in this model, all of which correspond to the complex  nilpotent orbit of weighted Dynkin diagram \DEVIII00000002 in $\e_{8\, \IC}$. We shall label them by their $E_{7(-25)} \times SL(2,\IR)$ weighted Dynkin diagram. 
$1/2$-BPS black holes, with weighted Dynkin diagram \DEVIISL00000004, are controlled by the fake superpotential $W = |Z|$; non-BPS $Z_* = 0$ black holes,
with weighted Dynkin diagram  \DEVIISL20000000,  by   the fake superpotential $W = \rho_\un(Z_a)$; and the non-BPS $Z_* \ne  0$ black holes, with weighted Dynkin diagram \DEVIISL00000022, by the fake superpotential $W= 2 \varrho(Z,Z_a)$. The latter 
is again the same function of $\rho_\zero,\, \rho_\un,\, \rho_\deux,\, \rho_\trois $ and $\varphi$ as in
$\N=8$ supergravity.

We will now explain how the fake superpotential controlling the non-BPS $Z_* \ne 0$  black holes can be derived from the parametrization of the nilpotent orbit of weighted Dynkin diagram \DEVIISL00000022. The latter tells us that the generator ${\bf h} \in \sl_2 \oplus \e_{7(-25)} $ of the triplet associated to the nilpotent orbit defines the following five-graded decomposition of  $ \sl_2 \oplus \e_{7(-25)} $:
\be \sl_2 \oplus \e_{7(-25)}  \cong \scal{Ê{\bf 1} \oplus {\bf 27} }^\ord{-2}  \oplus \scal{Ê\gl_1 \oplus \gl_1 \oplus \e_{6(-26)} }^\ord{0} \oplus  \scal{Ê{\bf 1} \oplus \overline{\bf 27} }^\ord{2} \ee
The corresponding decomposition of the coset component $\e_{8(-24)}Ê\ominus  \scal{Ê\sl_2 \oplus \e_{7(-25)}} $ is 
\be {\bf 2}Ê\otimes {\bf 56} \cong {\bf 1}^\ord{-4} \oplus  \scal{Ê{\bf 1} \oplus \overline{\bf 27} }^\ord{-2} \oplus \scal{Ê{\bf 27} \oplus \overline{\bf 27} }^\ord{0}Ê\oplus \scal{Ê{\bf 1} \oplus {\bf 27} }^\ord{2} \oplus {\bf 1}^\ord{4} \ ,\ee
while 
the orbit of a representative ${\bf e} \in  \scal{Ê{\bf 1} \oplus {\bf 27} }^\ord{2}$ associated to a regular black holes is\footnote{The representative of non-compact stabilizer in $E_{6(-26)}$ such as $F_{4(-20)}$ or  $Spin(9,1) \times SO(1,1)$ are associated to black holes with naked singularities.} 
\begin{multline}  
\scal{Ê F_{4(-52)} \ltimes ( {\bf 1} \oplus {\bf 26} )^\ord{2}  }  \big\backslash  \scal{ÊSL(2) \times E_{7(-25)} } \\* \subset  \scal{Ê E_{6(-26)}  \ltimes \scal{Ê( {\bf 1} \oplus {\bf 26}  \oplus \overline {\bf 26})^\ord{2}\oplus{\bf 1}^\ord{4}   }} \big\backslash E_{8(-24)}\ .
 \end{multline}
The generator ${\bf h}$ is parametrized in this case by two phases $\alpha$ and $\beta$ and an element $\Omega_a \in F_{4(-52)} \backslash E_{6(-78)} $ satisfying\footnote{which generalizes the symplectic form $\Omega_{AB} \in USp(6) \backslash SU(6)$ satisfying $\Omega_{AC} \Omega^{BC} = \delta_A^B$.}
\be \Omega_a = - \frac{1}{2} \, t_{abc}\,  \Omega^b \Omega^c\ ,\qquad
Ê\Omega_a \Omega^a = 6\ .
\ee
Let us consider for instance general charges $Z$ and $Z_a$, which we parametrize as 
\be 
Z  =  e^{-i(\alpha -3 \beta) }Ê\Scal{Ê\scal{Êe^{2i\alpha}Ê+ i \sin2\alpha }Ê\varrho - \Omega^a \Xi_a }\ , \quad 
Z_a = e^{-i(\alpha + \beta)} \left( \frac{1}{2}Ê\scal{Êe^{2i\alpha} + i \sin2\alpha} \varrho\,  \Omega_a + \Xi_a \right)  
\ee
where $\Xi_a$ lies in the ${\bf 1} \oplus {\bf 26}$ of the $F_{4(-52)}\subset E_{6(-78)}$ stabilizer of $\Omega_a$ defined as
\be \Xi_a = \Scal{Êt_{abc}\,  \Omega^c + \frac{1}{2}Ê\Omega_a \Omega_b } \Xi^b \ .
\ee
As in maximal supergravity, such decomposition is unique as long as the eigenvalues of $\Xi_a$ 
lie in the tetrahedron (\ref{tetrah}). To such electromagnetic charges, one associates a generator ${\bf h} \in Ê\sl_2 \oplus \e_{7(-25)}$ that acts on the elements of $P_\asym$ as follows 
\bea {\bf h}\,  \w &=& \frac{1}{\cos2\alpha} \Scal{Êe^{i(\alpha-3\beta)}ÊZ + e^{i(\alpha+\beta)} \Omega^a Z_a } - 4 i \tan2\alpha \, \w \CR
  {\bf h}\,  Z &=& \frac{1}{\cos2\alpha} \Scal{Êe^{-i(\alpha-3\beta)}Ê\w + e^{i(\alpha+ \beta)} \Omega^a \Sigma_a } - 2 i \tan2\alpha \, Z  \CR
 {\bf h}\,  Z_a &=& \frac{1}{\cos2\alpha} \Scal{Êe^{i(\alpha-3\beta)}Ê\Sigma_a  + \frac{1}{2} e^{-i(\alpha+\beta)} \Omega_a \w + e^{i (\alpha+ \beta)} t_{abc}\, \Omega^b \Sigma^c  } - 2 i \tan2\alpha \, Z_a \CR
 {\bf h}\,  \Sigma_a &=& \frac{1}{\cos2\alpha} \Scal{Êe^{-i(\alpha-3\beta)}ÊZ_a  + \frac{1}{2} e^{-i(\alpha+\beta)} \Omega_a Z + e^{i (\alpha+ \beta)} t_{abc}\, \Omega^b Z^c  } 
 \eea
The solution to the equation $[ {\bf h} , P_\asym ]Ê= 2 P_\asym$ is then 
\be \w = 2 \varrho \, , \hspace{10mm} \Sigma_a =  e^{2i\beta} \Scal{Êi \sin2\alpha \varrho\, \Omega_a + \Xi_a - \frac{1}{2} \Omega_a \Omega^b \Xi_b  } \label{OctoSol}  \ee
As for the maximally supersymmetric case one shows that the equation $[ {\bf h}  , P ]Ê= 2 P$ is equivalent to a gradient flow with respect to the fake superpotential $W = 2 \varrho(Z,Z_a)$. The only non-trivial step compared to (\ref{dWun}, \ref{dWdeux}) is to check that the $\e_{6(-78)}$ generators acting on $\Omega_a$, give an element satisfying 
\be \Lambda_a{}^b \Omega_b = - \Scal{Êt_{abc}\,  \Omega^c + \frac{1}{2}Ê\Omega_a \Omega_b } \Lambda^b{}_d \Omega^d \ , \ee
generalizing (\ref{ReaIma}, \ref{Imaginary}, \ref{ReaRe}), such that only the coset component of the scalar derivative contributes. 

From (\ref{OctoSol}), it is easily seen that  the flat directions of the fake superpotential 
span the ${\bf 26}$ real representation of $F_{4(-52)}$ inside the complex ${\bf 27}$ of $E_{6(-78)}$. The maximum of $W$  correspond to $\alpha = \Xi_a = 0$, which gives rise to a
 manifold of flat directions
\be 
\M_{\rm non-BPS}  \cong  F_{4(-52)} \backslash E_{6(-26)} \ ,
\ee
isomorphic to the moduli space after decompactification to five 
dimensions. 
At the attractor point, 
$\lozenge(Z_*,Z_{*\, a}) = - {W_*}^4$ and one recovers the familiar expression of the entropy for non-BPS black holes with $Z_* \ne 0$,
\be 
S_{\scriptscriptstyle \rm BH} = \pi \sqrt{Ê- \lozenge(p,q)} \ .
\ee

It is rather remarkable that the nilpotent orbits of the Lie algebras $\mathfrak{f}_{4(4)},\, \e_{6(2)},\, \e_{7(-5)}$ and $\e_{8(-24)}$ 
associated  to the real numbers, complex numbers, quaternions and octonions, respectively,
satisfying the nilpotency conditions
\be [{\bf e}_{|{\bf 26}}]^3 = 0 ,\quad  [{\bf e}_{|{\bf 27}}]^3 = 0   ,\quad  [{\bf e}_{|{\bf 56}}]^3 = 0   ,\quad  [{\bf e}_{|{\bf 3875}}]^5 = 0 , 
\ee
are in one-to-one correspondence  \cite{magicstratVIII,magicstratVII,magicstratVI,magicstratIV}, 
\be
\begin{split}
 F_{4(4)} & \backslash \bigl\{ \, {\bf e} \in \mathfrak{f}_{4(4)} \ | \  [{\bf e}_{|{\bf 26}}]^3 = 0 \, \bigr\} \, 
 \cong\,   E_{6(2)} \backslash \bigl\{\,  {\bf e} \in \e_{6(2)}\  | \  [{\bf e}_{|{\bf 27}}]^3 = 0 \,   \bigr\}  \\
&\cong \,  E_{7(-5)} \backslash \bigl\{\,  {\bf e} \in \e_{7(-5)}\  | \  [{\bf e}_{|{\bf 56}}]^3 = 0 \,   \bigr\} \,   
\cong\,   E_{8(-24)} \backslash \bigl\{\,  {\bf e} \in \e_{8(-24)}\  | \  [{\bf e}_{|{\bf 3875}}]^5 = 0 \,   \bigr\} \ ,
\end{split}
\ee
so that the corresponding moduli space of extremal black holes have the same stratified structure in term of $K_3^*$ orbits, which is displayed in Figure \ref{magicHasse}.


\FIGURE{\includegraphics[height=7cm]{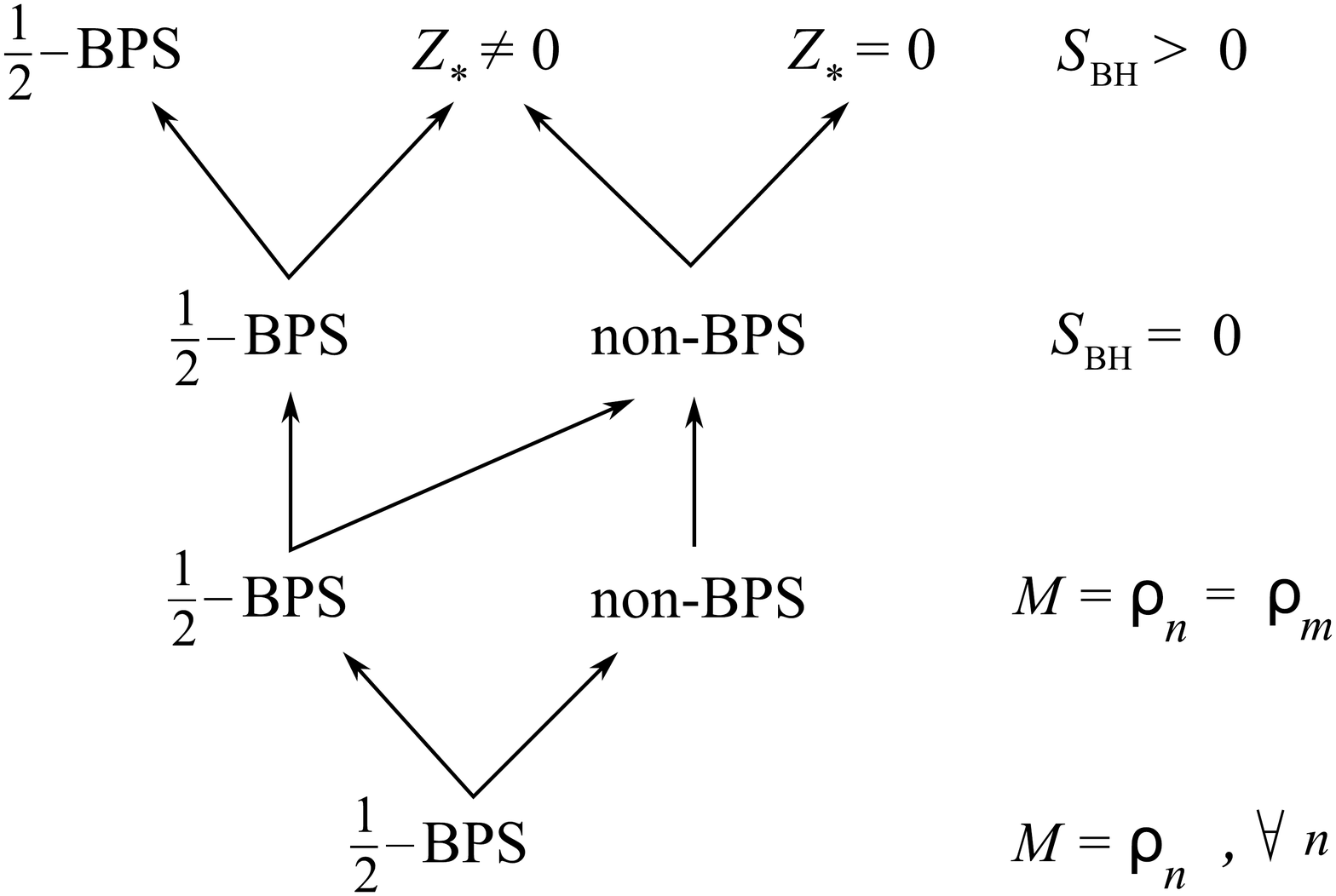}\caption{Stratification of the moduli space of extremal solutions for magic supergravity models.   \label{magicHasse}}}


\subsection{Axion-dilaton $\N=2$ supergravity}  

The quaternionic magic supergravity \eqref{mso12} can also be further truncated  by singling out 
an additional
 complex charge $Z_\un = 2 Z_{34}$ from $Z_{AB}$, such that the internal symmetry group is restricted to $U(2) \times U(4) \subset U(6)$. The remaining charges transform as a complex vector of $SO(6) \cong {\mathds{Z}_2}Ê\backslash SU(4)$ and the scalar fields parametrize the symmetric spaces
\be  
\begin{split}
\M_4 \cong U(1) \backslash  SL(2,\mathds{R})\,  \times \, \scal{ÊSO(2) \times SO(6)}  \backslash  SO(2,6) 
\ ,\qquad\\
\M_3^* \cong \scal{SO(2,2) \times SO(2,6)} \backslash  SO(4,8)\ . 
\end{split}
\ee
More generally, one obtains this way all $\N=2$ supergravity theories with $n+1$ 
vector multiplets ($ 0 <  n \le 6$) coupled to scalar fields parametrising the symmetric space
\be 
\label{mso4}
\begin{split}
 \M_4 \cong U(1) \backslash  SL(2,\mathds{R})\,  \times \, 
 \scal{ÊSO(2) \times SO(n)}  \backslash  SO(2,n)
\ ,\qquad\\
\M_3^* \cong \scal{SO(2,2) \times SO(2,n)} \backslash  SO(4,n+2)\ . 
\end{split}
\ee
Denoting  the $n$ charges transforming in the vector representation of $SO(n)$ as $Z_{\rm i}$, 
one can identify
\be 
\rho_\zero = |Z| ,\quad \rho_\un = |Z_\un|Ê \ ,\qquad 4 \varphi = Ê\arg\bigl[ÊZ \bar Z_\un \bar Z^{\rm i} \bar Z_{\rm i} \bigr]\ ,
\ee
while $\rho_\deux$ and $\rho_\trois$ follow from the roots of the polynomial\footnote{Note that raising the $SO(n)$ index ${\rm i}$ does not involve complex conjugation.} 
\be \lambda^2 - Z^{\rm i} \bar Z_{\rm i} \lambda + \frac{1}{4}Ê| Z^{\rm i} Z_{\rm i} |^2 = \scal{Ê\lambda -{ \rho_\deux}^2  }Ê\scal{Ê\lambda -{ \rho_\trois}^2  }\ .
\ee

For $n>2$ there are four $SL(2,\IR) \times SL(2,\IR) \times SO(2,n)$ orbits of generic extremal black holes associated to four distinct nilpotent orbits of $\so(4,2+n)$ of degree
 three in the spinor representation (\ie such that ${\bf e}^3 = 0$ and ${\ad_{\bf e}}^4 \ne 0$). For 
  $1/2$-BPS black holes, $W = |Z|$; for non-BPS black holes
  with $Z_* = Z_*^{\rm i} = 0 $, $W= |Z_\un|$; 
  for the non-BPS black holes with  $Z_* = Z_{\un\, *} = 0 $,
\be W = \sqrt{Ê\frac{1}{2}ÊZ^{\rm i} \bar Z_{\rm i} +\frac{1}{2}Ê \sqrt{Ê\scal{ÊZ^{\rm i} \bar Z_{\rm i}}^2 - \bigl| Z^{\rm i} Z_{\rm i} \bigr|^2 } }\ ; \ee
and finally, for non-BPS black holes with $Z_* \ne 0$, 
$W= 2 \varrho$ with the same function $\varrho$ as in $\N=8$ supergravity. 
Of course these results extend straightforwardly to any value of $n$ larger than six. 

The STU model, to be expanded upon in Section \ref{secstu}, 
corresponds to $n=2$. The STU truncation of maximal supergravity is defined by a basis such that  
\be 
Z_{ij} \, \, \hat{=}  \, \, \frac{1}{2}ÊÊ\, \left( \begin{array}{cc}0 & \, \, 1  \\- 1 & \, \, 0 \end{array}\right) \otimes  \left( \begin{array}{cccc} \, \, Z \, \, Ê& 0 & 0 & 0 \\0& \, \, \bar Z_\un \, \,  & 0 & 0 \\ \, \, 0 & 0 &\, \,  \bar Z_\deux \, \, & 0 \\ 0 & 0 & 0 & \bar Z_\trois \end{array} \right)  \ee
which is preserved by a subgroup $S\scal{ U(2) \times U(2) \times U(2) \times U(2)}\subset SU(8)$ as well as 
\be [SL(2,\mathds{R})]^3  \times [SU(2)]^4 \subset E_{7(7)}  \ee
The $SU(2)$ subgroups leave invariant the bosonic fields  and will be disregarded. Considering the whole field content of the STU model, only one $SU(2)$ factor acts non-trivially on the 
fermionic fields, and corresponds to the R-symmetry group. There are five different types of extremal static black holes in this model. Four of them, including the $1/2$-BPS ones, correspond to $1/8$-BPS black holes within $\N=8$ supergravity. Their fake superpotential  is then 
$W = |Z| ,\, |Z_\un| ,\,  |Z_\deux| ,\,  |Z_\trois|$, respectively. The last type corresponds to the non-BPS black holes of $\N=8$ supergravity;  its fake superpotential  is obtained from $W = 2 \varrho$ by 
substituting $|Z| ,\, |Z_\un| ,\,  |Z_\deux| ,\,  |Z_\trois|$ and $\frac{1}{4} \arg\scal{ÊZ \bar Z_\un\bar Z_\deux\bar Z_\trois}$ to $\rho_\zero,\, \rho_\un,\, \rho_\deux,\, \rho_\trois $ and $\varphi$.

There are two interesting further truncations, which are obtained by restriction to the diagonal subgroup of either two $SL(2,\mathds{R})$ factors or the three of them. In the first case one obtains the $S^2T$ model (\ie (\ref{mso4}) with $n=1$). It admits three types of extremal static black holes, the $1/2$-BPS ones, for which $W = |Z|$; the non-BPS ones with $Z_* = 0$, for which $W = |Z_\un|$; and the non-BPS ones with  $Z_* \ne 0$, for which $W = 2\varrho(Z,Z_\un,Z_\deux)$. As explained in Section \ref{secs2t},  the degree six polynomial (\ref{sixDegree}) factorizes and an explicit expression of $W$ in terms 
of $Z,\, Z_\un$ and $Z_\deux$ can be obtained by solving the quartic polynomial (\ref{fourDegree}). 

The subsequent truncation is the $S^3$ model, which is the circle compactification  of the minimal $\N=2$ supergravity theory in five dimensions. 
\be \M_4 \cong U(1) \backslash SL(2,\mathds{R}) \, , \hspace{10mm} \M_3^* \cong SO(2,2) \backslash G_{2(2)} \ee
This theory admits two types of extremal black holes, the $1/2$-BPS ones for which $W = |Z|$, and the non-BPS ones with $Z_* \ne 0$,  corresponding to 
the two nilpotent orbits of $\fg_{2(2)}$ in which $[{\bf e}_{|\bf 7}]^3= 0$ \cite{Gaiotto:2007ag}.  
As explained in \ref{secs3},   the degree six polynomial (\ref{sixDegree}) 
reduces to a cubic polynomial \eqref{threeDegree}.
Identifying $\rho_\zero=|Z|$ and $\rho_\un=|Z_1|$ where $Z_\un = -2 \I S_2 DZ$, 
one obtains an algebraic expression for the fake superpotential,
\be
\framebox{$
W = \frac12 \sqrt{
| Z|^2 + 3  \scal{ | Z_\un|^2 +  L_+  + L_- }   } \ ,$}
\label{W1mod}
\ee
where $L_\pm$ are given by \eqref{eqL} below, which we rewrite for convenience, 
\be
L_\pm^3 =  |Z_\un|^4 \Scal{Ê| Z_\un|^2 + 3 | Z|^2 } - \frac{1}{2}Ê\Scal{ÊZ {\bar Z_\un}^3 
+ \bar Z {Z_\un}^3 } \Scal{Ê|Z|^2 + 3 |Z_\un|^2 } \pm
 \frac{\sqrt{-\lozenge}}{2}Ê \bigl| Z {\bar Z_\un}{}^3 - \bar Z { Z_\un}^3 \bigr| \ ,
\ee
and $\lozenge$ is the moduli-independent quartic $SL(2,\IR)$ invariant
\be \lozenge(Z,Z_\un) =  \Scal{Ê|Z|^2 - |Z_\un|^2}^2 - 4 \bigl | Z \bar Z_\un - {Z_\un}^2 \bigr |^2 \ .
\ee
The fake superpotential $W$ in \eqref{W1mod} is plotted as a function of $\varphi$ 
in Figure \ref{fakefig}. Contrary to the superpotential for BPS black holes, which is $\varphi$ independent, $W$ is maximal at $\varphi=\pi/2(n+\frac12)$, $n\in \IZ$. Note that 
\eqref{W1mod} is only valid in the region where $ \lozenge<0$, such that $W>|Z|$.
In the region where $ \lozenge>0$ it must be replaced by the BPS superpotential 
$W=|Z|$.


\FIGURE{\includegraphics[height=9cm]{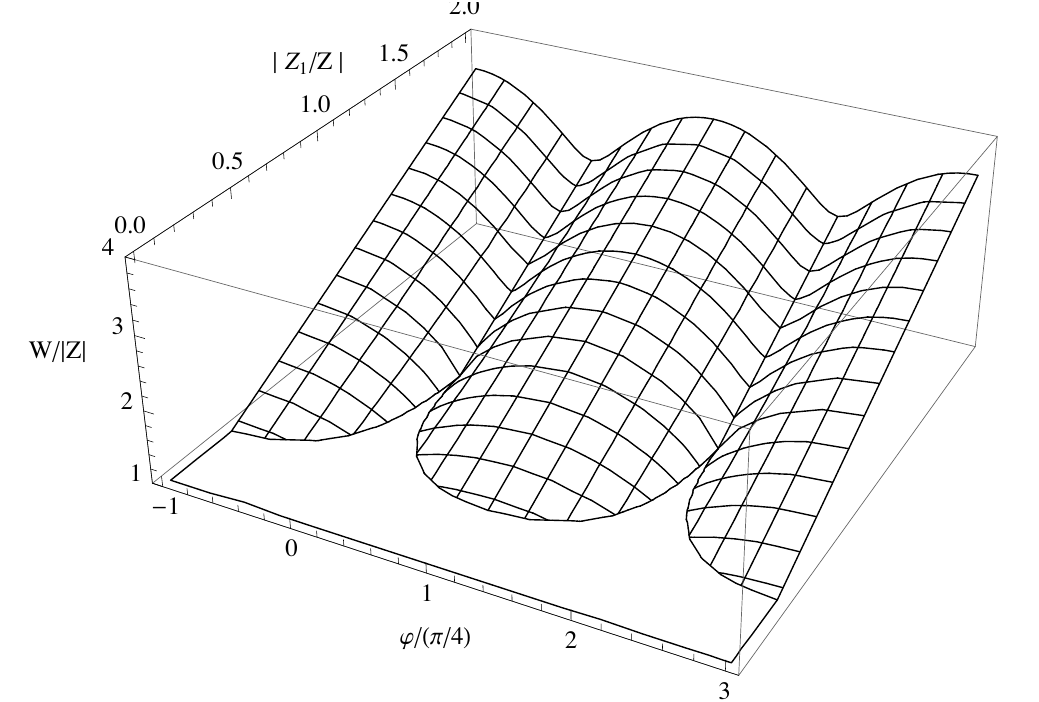}
\caption{Normalized fake superpotential $W/|Z|$ 
as a function of $|Z_1/Z|$ and $\varphi/(\pi/4)$.  \label{fakefig}}}

\subsection{$\N=4$ supergravity}
$\N=8$ supergravity can be truncated to $\N=4$ supergravity coupled to $n=6$ vector multiplets
 by restricting the central charges to two antisymmetric tensors of $SU(4)\times SU(4)$ 
 such that $S\scal{ÊU(4) \times U(4)} \subset SU(8)$.  The central charges $Z_{AB}$ are then defined as an antisymmetric tensor of the first $SU(4)$ that we shall identify as the R-symmetry group of the theory. For further generalization, it will be more convenient to represent the other six complex electromagnetic charges as a complex vector $Z^{\rm i}$ of $SO(n) \cong \mathds{Z}_2 \backslash SU(4)$
 for $n=6$. Then, the $SU(8)$ invariant functions become  $S\scal{ÊU(4) \times U(4)}$ invariant functions 
 defined by the two polynomials
\bea \scal{Ê\lambda - {\rho_\zero}^2  }Ê\scal{Ê\lambda - {\rho_\un}^2 } &=& Ê\lambda^2 - 2 Z_{AB} Z^{AB}Ê\lambda + 2  \Bigl(  \scal{ÊZ_{AB} Z^{AB} }^2 - 2 Z_{AB} Z^{BC} Z_{CD} Z^{DA}  \Bigr) \ , \CR
 \scal{Ê\lambda - {\rho_\deux}^2  }Ê\scal{Ê\lambda - {\rho_\trois }^2 }
 &=&  \lambda^2 - Z^{\rm i} \bar Z_{\rm i}\,  \lambda + \frac{1}{4}Ê| Z^{\rm i} Z_{\rm i} |^2 
 \ ,\eea
and 
\be 4 \varphi  = \arg\bigl[Ê\varepsilon^{ABCD}ÊZ_{AB} Z_{CD} \bar Z_{\rm i} \bar Z^{\rm i} \bigr] \ .
\ee
This model admits three types of generic extremal black holes, which are in one-to-one correspondence with the nilpotent orbits of $\so(8,8)$ of degree three in the spinor representation \cite{Bossard:2009at} (\ie such that ${\bf e}^3 = 0$ and ${\ad_{\bf e}}^4 \ne 0$). They are: the $1/4$-BPS black holes, 
for which 
\be W = \sqrt{ÊZ_{AB} Z^{AB} + \sqrt{ÊÊ4 Z_{AB} Z^{BC} Z_{CD} Z^{DA}  - \scal{ÊZ_{AB} Z^{AB} }^2} } \, ;\ee
the non-BPS black holes with $Z_{*\, AB} = 0$, for which 
\be W = \sqrt{Ê\frac{1}{2}ÊZ^{\rm i} \bar Z_{\rm i} +\frac{1}{2}Ê \sqrt{Ê\scal{ÊZ^{\rm i} \bar Z_{\rm i}}^2 - \bigl| Z^{\rm i} Z_{\rm i} \bigr|^2 } } \, ;
\ee
and the non-BPS ones with $Z_{*\, AB} \ne 0$, for which the fake superpotential is the 
same function $2 \varrho(Z_{AB} , Z_{\rm i})$ as in $\N=8$ supergravity. These results extend straightforwardly to $\N=4$ supergravity coupled to any number $n >2$ of vector multiplets,
with moduli space
\be  
\label{mso8}
\begin{split}
\M_4 \cong U(1) \backslash  SL(2,\mathds{R})\,  \times \, \scal{ÊSO(6) \times SO(n)}  
\backslash  SO(6,n) 
\ ,\qquad\\
\M_3^* \cong \scal{SO(6,2) \times SO(2,n)} \backslash  SO(8,n+2)\ . 
\end{split}
\ee
Non-BPS multi-black hole solutions with $Z_{*\, AB} \ne 0$ have
been discussed in \cite{Bossard:2009mz}.  

For $n=2$, the scalar moduli space \eqref{mso8} of the $\cN=4$ model 
becomes identical to that of the $\N=2$ model
\eqref{mso4} with $n=6$. The nilpotent orbit associated to the non-BPS black holes with $Z_{*\, AB}= 0$ splits into two inequivalent orbits of the connected component of $SO(8,4)$, giving rise to two inequivalent $SO(6,2) \times SL(2,\IR) \times SL(2,\IR)$ orbits . With only one vector multiplet, there is no non-BPS black holes with   $Z_{*\, AB}= 0$.  The degree six polynomial defining the fake superpotential $W = 2 \varrho$ associated to non-BPS black holes with $Z_{*\, AB}\ne  0$ reduces to the degree four polynomial (\ref{fourDegree}) in such a way that the explicit form of $W$ can then be derived straightforwardly.

\subsection{Axion-dilaton gravity}
The non-standard diagonalization problem defining $\varrho(Z)$ can in fact be formulated more generally for the case of Einstein gravity coupled to $m+n$ abelian vector fields and to scalars  valued in
the symmetric spaces (with $m\ge 1,\, n\ge1$)
\be  
\label{mso8}
\begin{split}
\M_4 \cong U(1) \backslash  SL(2,\mathds{R})\,  \times \, \scal{ÊSO(m) \times SO(n)}  
\backslash  SO(m,n) 
\ ,\qquad\\
\M_3^* \cong \scal{SO(m,2) \times SO(2,n)} \backslash  SO(m+2,n+2)\ .
\end{split}
\ee
For $m=2$ and $m=6$, respectively  one recovers 
the cases of axion-dilaton $\cN=2$ supergravities 
and $\cN=4$ supergravities discussed previously. Denoting the $SO(m)$ and $SO(n)$ vectors of charges by $Z_{\rm a}$ and $Z_{\rm i}$, respectively,\footnote{For $\N=2$, $Z_{\rm a}$ is $\frac{Z\pm i\bar Z_\un}{\sqrt{2}}$ for a $=1$ and $2$, respectively. For $\N=4$, $\frac{i}{ \sqrt{2}} Z_{\rm a} [\gamma^{\rm a}]_{AB} =  Z_{AB} $, where the $\gamma^{\rm a}$'s are Dirac matrices of $Spin(6)$.} the non-standard diagonalization problem can be formulated in term of two unit-norm vectors $\Omega_{\rm a}$ and $\Omega_{\rm i}$ as follows,
\bea Z_{\rm a} &=& e^{- i(\alpha- \beta) }Ê\Scal{Ê\scal{Êe^{2i\alpha} + i \sin2\alpha }Ê\, \varrho \, \Omega_{\rm a} + \xi \, \Omega_{\rm a} + i \Xi_{\rm a } } \quad \ , \CR
\bar Z_{\rm i} &=& e^{- i(\alpha+\beta) }Ê\Scal{Ê- i \scal{Êe^{2i\alpha} + i \sin2\alpha }Ê\, \varrho \, \Omega_{\rm i} + i \xi \, \Omega_{\rm i} +  \Xi_{\rm i } } \ , \label{OrthoDiag}
\eea
where $\Xi_{\rm a}$ and $\Xi_{\rm i}$ are real vectors orthogonal to $\Omega_{\rm a}$ and $\Omega_{\rm i}$, respectively.

We use similar notations  as we did in exceptional $\N=2$ supergravity, \ie $\w  = M + i k $,  $\Sigma$ denotes the complex scalar charge associated to $\sl_2$, and $\sigma_{\rm a\,  i }$ denote the real scalar charges associated to $\so(m,n)$.  The nilpotent orbit is characterized by the generator ${\bf h}$ of $\so(m,2) \oplus \so(2,n)$, which acts as follows 
\bea {\bf h}\,  \w &=& \frac{2}{\cos2\alpha} \Scal{Êe^{i(\alpha-\beta)}Ê\Omega^{\rm a} Z_{\rm a}  + i  e^{i(\alpha+\beta)} \Omega^{\rm i} \bar Z_{\rm i}  } - 4 i \tan2\alpha \, \w \CR
  {\bf h}\,  Z_{\rm a}Ê &=& \frac{1}{\cos2\alpha} \Scal{Êe^{-i(\alpha-\beta)}Ê\Omega_{\rm a}Ê \w + e^{i(\alpha- \beta)} \Omega_{\rm a} \Sigma + 2 i e^{i(\alpha+\beta) } \Omega^{\rm i}  \sigma_{\rm a\,  i } } - 2 i \tan2\alpha \, Z_{\rm a}  \CR
 {\bf h}\,  \bar Z_{\rm i} &=& \frac{1}{\cos2\alpha} \Scal{Ê i e^{i(\alpha+\beta)}Ê\Omega_{\rm i }Ê\bar \Sigma  - i  e^{-i(\alpha+\beta)} \Omega_{\rm i}  \w + 2 e^{i (\alpha- \beta)} \Omega^{\rm a}  \sigma_{\rm a\, i}  } - 2 i \tan2\alpha \, \bar Z_{\rm i} \CR
 {\bf h}\,  \Sigma  &=& \frac{2}{\cos2\alpha} \Scal{Êe^{-i(\alpha-\beta)}Ê\Omega^{\rm a}ÊZ_{\rm a}  + i e^{i(\alpha+\beta)} \Omega^{\rm i} Z_{\rm i}   } \CR
  {\bf h}\,  \sigma_{\rm a\, i} &=& \frac{2}{\cos2\alpha} \Re \Scal{Êe^{-i(\alpha-\beta)}Ê\Omega_{\rm a}Ê\bar Z_{\rm i}  - i e^{-i(\alpha+\beta)} \Omega_{\rm i} Z_{\rm a}  } 
 \eea
The solution to the equation ${\bf h}ÊP_\asym = 2 P_\asym$ is then
\be
 \w = 2 \varrho \ , \qquad \Sigma = 2 e^{2i\beta}Ê\scal{Ê\xi + i \sin(2\alpha) \varrho } \ , \qquad
\sigma_{\rm a \, i} = \Omega_{\rm a } \Xi_{\rm i } + \Omega_{\rm i } \Xi_{\rm a } + 2 \sin(2\alpha) \varrho \, \Omega_{\rm a}Ê\Omega_{\rm i}Ê\ .\label{OrthoSol} 
\ee
As before,  the linear equation ${\bf h}  P = 2 P$ is equivalent to a gradient flow with respect to the fake superpotential $W = 2 \varrho(Z_{\rm a}Ê, Z_{\rm i})$. For $m\le6,\, n\le6$, $W$ is necessarily the same function of the five $SO(2) \times SO(m) \times SO(n)$ invariants 
\bea 
\scal{Ê\lambda - {\rho_\zero}^2  }Ê\scal{Ê\lambda - {\rho_\un}^2 } &=&   \lambda^2 - Z^{\rm a} \bar Z_{\rm a}\,  \lambda + \frac{1}{4}Ê| Z^{\rm a} Z_{\rm a} |^2 \ , \CR
 \scal{Ê\lambda - {\rho_\deux}^2  }Ê\scal{Ê\lambda - {\rho_\trois }^2 }
 &=&  \lambda^2 - Z^{\rm i} \bar Z_{\rm i}\,  \lambda + \frac{1}{4}Ê| Z^{\rm i} Z_{\rm i} |^2 
 \ ,\eea
and 
\be 4 \varphi  = \arg\bigl[ÊZ_{\rm a} Z^{\rm a}  \bar Z_{\rm i} \bar Z^{\rm i} \bigr] 
\ee
as in $\cN=8$ supergravity, as it can be obtained by consistent truncation, and therefore remains so for any values $m$ and $n$. 

From (\ref{OrthoSol}) it is apparent that the flat directions of $W$ are the components of $\sigma_{\rm a \, i}$ orthogonal to both $\Omega_{\rm a}$ and $\Omega_{\rm i}$, as well as the component $\Omega^{\rm a}Ê\Omega^{\rm i }Ê\sigma_{\rm a\, i }Ê- \Im (  e^{-2i\beta}Ê\Sigma ) $, which altogether generate  the
expected  symmetric space 
\be \cM_5 \cong GL(1,\IR) \times \scal{ÊSO(m-1) \times SO(n-1) }Ê\backslash SO(m-1,n-1) \ .
\ee
The subgroup $GL(1,\IR) \times  SO(m-1,n-1) \subset SL(2,\IR) \times SO(m,n)$ is also the stabilizer of the  charges at the horizon (at $\tau \rightarrow + \infty$), where they are of the form
\be Z_{*\, {\rm a}} = \frac{1}{2}Êe^{ i\beta_* }Ê W_* \, \Omega_{*\, {\rm a}} \ , \qquad Z_{*\, {\rm i}} =  \frac{i}{2} e^{ i\beta_* }W_*\,  \Omega_{*\, {\rm i}} \ .
\ee


\section{Extremal black holes in the $STU$ model \label{secstu}}

In this section, we expand the previous results in the case of the $STU$
model of $\N=2$ supergravity, which generates all other symmetric 
models by truncation or covariantization. In particular, we interpret
the BPS and non-BPS, $Z_*=0$ solutions in terms of the para-quaternionic
structure on $\M_3^*$, and confirm our identification of nilpotent orbits
on explicit solutions.

\subsection{Moduli spaces in $D=4$ and $D=3$\label{STUmodsec}}

The STU model of $\cN=2$ supergravity is governed by the prepotential
\be
F = - X^1 X^2 X^3 / X^0 \ ,\quad
S=\frac{X^1}{X^0}\ ,\quad
T=\frac{X^2}{X^0}\ ,\quad
U=\frac{X^3}{X^0}\ ,\quad
\ee
The  4D moduli space consists of 3 copies of the Poincar\'e upper half plane, with 
metric\footnote{In this section, to avoid notational conflict, we rename the
variable  $U$ appearing in the metric ansatz \eqref{4dmetric} into $\phi/2$.}
\be
ds^2_{\cM_4} = \frac{dS_1^2+dS_2^2}{S_2^2} +  \frac{dT_1^2+dT_2^2}{T_2^2}
+ \frac{dU_1^2+dU_2^2}{U_2^2}\ ,
\ee
and so is the symmetric space $\cM_4=[U(1)\backslash SL(2,\IR)]^3$.
For conciseness we shall denote $(S,T,U)=(S^{(1)},S^{(2)},S^{(3)})$, while
$S^{(i)}_{1}$ and $S^{(i)}_{2}$ will denote the real and imaginary part of $S^{(1)}$.

Upon dimensional reduction along a space-like direction, the resulting 
moduli space in 3 dimensions is the c-map of $\cM_4$,
\be
\label{cmap}
\begin{split}
ds^2_{\cM_3} =& d\phi^2 +  ds^2_{\cM_4}
+\frac14 e^{-2\phi} (d\sigma + \tzeta^\Lambda d\zeta_\Lambda -  \zeta^\Lambda d\tzeta_\Lambda)^2\\
&-e^{-\phi} \left[ (d\tzeta_\Lambda + \Re\cN_{\Lambda\Sigma} \zeta^\Sigma)
[\Im\cN]^{\Lambda \Lambda'}
(d\tzeta_{\Lambda'} + \Re\cN_{\Lambda'\Sigma} \zeta^\Sigma)
+ d\zeta^\Lambda [\Im\cN]_{\Lambda \Lambda'}d\zeta^{\Lambda'}\right]
\end{split}
\ee
(recall that $\Im\cN$ is definite negative, so the metric on $\cM_3$ is definite positive).
Upon dimensional reduction along a time-like direction, the resulting 
moduli space in 3 dimensions is instead the pseudo-Riemannian space
\be
\label{cmapst}
\begin{split}
ds^2_{\cM_3^*} =& d\phi^2 +  ds^2_{\cM_4}
+\frac14 e^{-2\phi}  (d\sigma + \tzeta^\Lambda d\zeta_\Lambda -  \zeta^\Lambda d\tzeta_\Lambda)^2\\
&+\frac12 e^{-\phi} \left[ (d\tzeta_\Lambda + \Re\cN_{\Lambda\Sigma} \zeta^\Sigma)
[\Im\cN]^{\Lambda \Lambda'}
(d\tzeta_{\Lambda'} + \Re\cN_{\Lambda'\Sigma} \zeta^\Sigma)
+ d\zeta^\Lambda [\Im\cN]_{\Lambda \Lambda'}d\zeta^{\Lambda'}\right]\ ,
\end{split}
\ee
obtained from the Riemannian space \eqref{cmap} by analytic continuation 
$\zeta^\Lambda\to \I \zeta^\Lambda, \tzeta_\Lambda \to \I \tzeta_\Lambda,\sigma\to-\sigma$.

Both $\cM_3$ and $\cM_3^*$ are symmetric spaces for the group $G_3=SO(4,4)$,
\be
\label{M3sym}
\cM_3 = [SO(4)]^2 \backslash SO(4,4) \ ,\qquad
\cM_3^*= [SO(2,2)]^2 \backslash SO(4,4)\ .
\ee
Note that in the second case, the denominator is not a maximal compact subgroup of $G_3$,
which accounts for the signature $(8,8)$ of the metric on $\cM_3^*$. To check that the metrics 
\eqref{cmap} and \eqref{cmapst} are the right-invariant metrics on the two symmetric 
spaces \eqref{M3sym},
it is convenient to choose an explicit parametrization of the Lie algebra 
$\fg_3=\mathfrak{so}(4,4)\ni X=\sum {\underline E}_a E_a$, where $E_a$ are the 28 generators 
and ${\underline E}_a$ are the 28 dual coordinates,
\be
\label{Xso44}
X=
\left(
\begin{array}{cccccccc}
 {\underline H}_{2}+{\underline H}_{3} & -{\underline E}_{3} & -{{\underline F}_{q_1}}& {{\underline F}_{q_0}}& 0 & -{\underline E}_{2} &
   {{\underline E}_{p^0}}& {{\underline E}_{p^1}}\\
 -{\underline F}_{3} & {\underline H}_{2}-{\underline H}_{3} & -{{\underline F}_{p^2}}& {{\underline F}_{q_3}}& {\underline E}_{2} & 0 &
   {{\underline E}_{p^3}}& -{{\underline E}_{q_2}}\\
 -{{\underline E}_{q_1}}& -{{\underline E}_{p^2}}& {\underline H}+{\underline H}_{1} & -{\underline E}_{1} & -{{\underline E}_{p^0}}&
   -{{\underline E}_{p^3}}& 0 & -{\underline E}_0 \\
 {{\underline E}_{q_0}}& {{\underline E}_{q_3}}& -{\underline F}_{1} & {\underline H}-{\underline H}_{1} & -{{\underline E}_{p^1}}&
   {{\underline E}_{q_2}}& {\underline E}_0 & 0 \\
 0 & {\underline F}_{2} & -{{\underline F}_{p^0}}& -{\underline F}_{p^1}
 & -{\underline H}_{2}-{\underline H}_{3} & {\underline F}_{3} &
   {{\underline E}_{q_1}}& -{{\underline E}_{q_0}}\\
 -{\underline F}_{2} & 0 & -{{\underline F}_{p^3}}& {{\underline F}_{q_2}}& {\underline E}_{3} & {\underline H}_{3}-{\underline H}_{2} &
   {{\underline E}_{p^2}}& -{{\underline E}_{q_3}}\\
 {{\underline F}_{p^0}}& {{\underline F}_{p^3}}& 0 & {\underline F}_{0} & {{\underline F}_{q_1}}&
   {{\underline F}_{p^2}}& -{\underline H}-{\underline H}_{1} & {\underline F}_{1} \\
 {\underline F}_{p^1}& -{{\underline F}_{q_2}}& -{\underline F}_{0} & 0 & -{{\underline F}_{q_0}}&
   -{{\underline F}_{q_3}}& {\underline E}_{1} & {\underline H}_{1}-{\underline H}
\end{array}
\right)
\ee
which preserve the  $SO(4,4)$ quadratic form is $\eta=
{\scriptsize \begin{pmatrix} 0 & 1_4\\ 1_4 & 0 \end{pmatrix}}$,
$X^t \eta + \eta X=0$.  This basis is adapted to the branching
\be
\mathfrak{so}(4,4) =[\sl(2,\IR)]^4 \oplus (2,2,2,2)\ , 
\ee
where  the four commuting $SL(2,\IR)$ subgroups are generated by 
\be
[ E_i, F_i ] = H_i\ ,\qquad [H_i, E_i] = 2 E_i\ ,\qquad [H_i, F_i] = -2 F_i\ \qquad i=0,1,2,3
\ee
while the 16 real generators in the coset fit in an hypercube $E_{a_0,a_1,a_2,a_3}$
with $a_i=1,2$ (see Figure \ref{hypercube}). Decomposing with respect to the Cartan generator
$H_0$ leads to the real 5-grading
\be
\label{5greal}
F_0\vert_{-2} \oplus \{ F_{p^\Lambda}, F_{q_\Lambda} \}\vert_{-1}
\oplus \left( H \oplus \{ E_i, F_i, H_i\}_{i=1,2,3} \right)\vert_0
\oplus \{ E_{p^\Lambda}, E_{q_\Lambda} \}\vert_{1}\oplus E_0\vert_{2} \ .
\ee
A canonical basis of Cartan generators 
is $\bigl( H_1,H_2,H_3;\frac12(H-H_1-H_2-H_3)\bigr)$, where the last generator is the 
one attached to the middle node of the Dynkin diagram.

\FIGURE{\includegraphics[height=9cm]{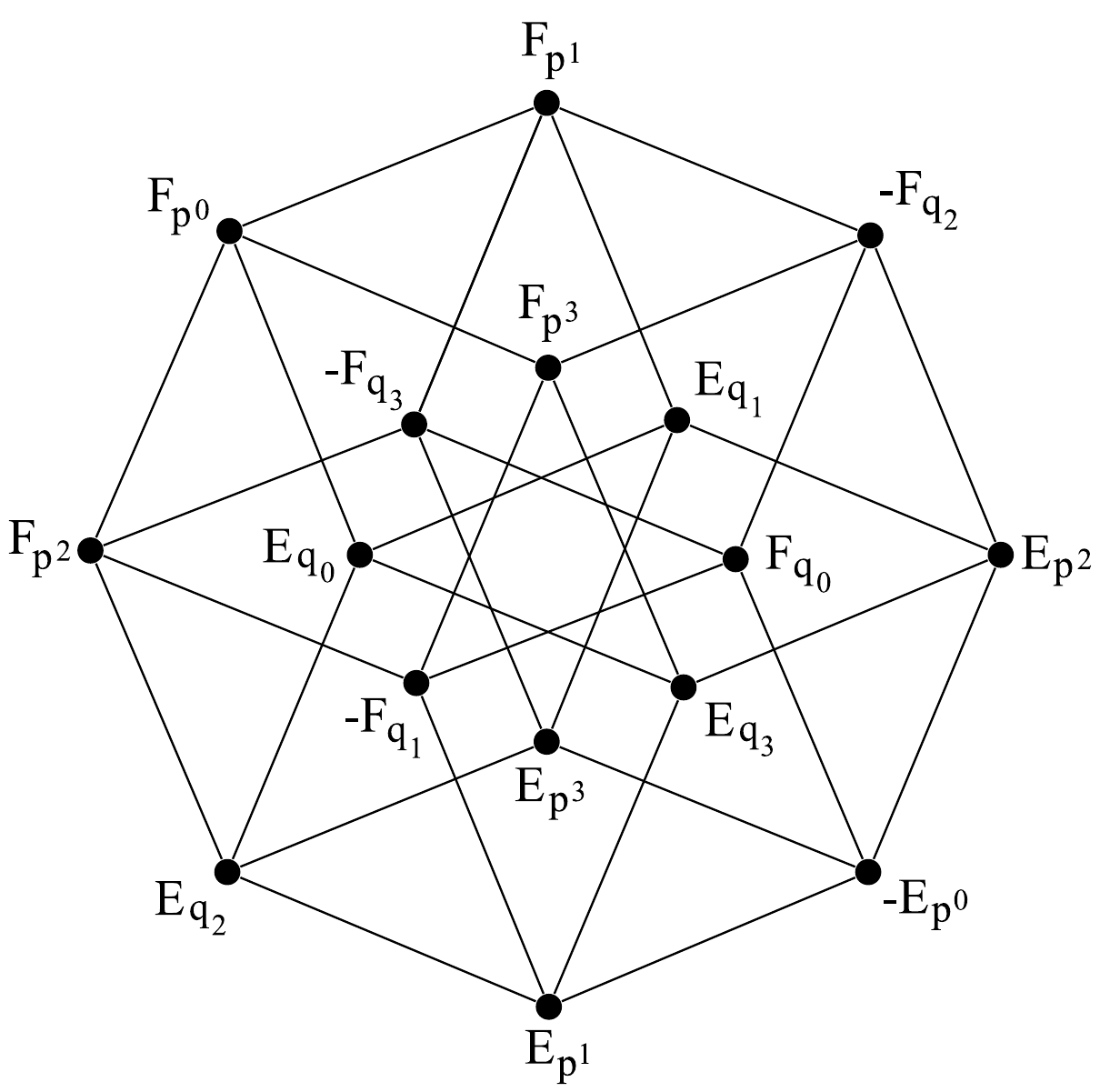}
\caption{(2,2,2,2) roots in $SO_\asym(4,4)$  \label{hypercube}}}


The symmetric space $\cM_3$ can then be parametrized in the Iwasawa gauge by the
coset element
\be
\label{iwa}
\V = e^{-\frac12 \phi\, H_0} \cdot \left( 
\prod_{i=1,2,3}
e^{-\frac12 (\log S^{(i)}_2) H_i} \cdot e^{ - S^{(i)}_1 E_i} \right) \cdot
e^{-\zeta^\Lambda E_{q_\Lambda}-  \tzeta_\Lambda E_{p^\Lambda}}\cdot
e^{-\frac12\sigma E_0}\ .
 \ee
The metric \eqref{cmap} on $\cM_3$ is then the right-invariant metric 
obtained from the Maurer-Cartan one-form $\theta= d\V \cdot \V^{-1}$,
\be
ds^2_{\cM_3} = \Tr( P^2 )\ ,\quad
P = \frac12 ( \theta + \theta^T ) \ ,\quad
 \label{dsp}
\ee
The Iwasawa parametrization \eqref{iwa} can also be used for the pseudo-Riemannian
space $\cM_3^*$, although it is suitable only on an open subset of  the full  $\cM_3$.
The metric \eqref{cmapst} is obtained as in \eqref{dsp} upon replacing $P$ by
\be
P_* = \frac12 ( \theta + \eta'\, \theta^T {\eta'}^{-1})\ ,\quad
\eta' = {\rm diag}(-1,-1,1,1,-1,-1,1,1)\ .
\ee
where $\eta'$ is the quadratic form preserved 
by $SL(2,\IR)^4=SO(2,2)\times SO(2,2)\subset G_3$.

The Hamiltonian associated to  geodesic motion on $\cM_3^*$ with 
Lagrangian $\mathcal{L} = ds^2_{\cM_3^*}/d\rho$ is given by 
\be
\label{ham}
\begin{split}
H =& \frac14 \left[ p^2_\phi + \sum_{i=1,2,3} (S_2^{(i)})^2 \left(p^2_{S_1^{(i)}} + p^2_{S_2^{(i)}} \right) \right]
+4 e^{2\phi} p_\sigma^2 \\
&+e^{\phi}
 \left[ (P_{\zeta^\Lambda} - \Re\cN_{\Lambda\Sigma} P_{\tzeta_\Sigma})
[\Im\cN]^{\Lambda \Lambda'}
(P_{\zeta^{\Lambda'}} - \Re\cN_{\Lambda'\Sigma'} P_{\tzeta_{\Sigma'}})
+P_{\tzeta_\Lambda} [\Im\cN]_{\Lambda \Lambda'} P_{\tzeta_{\Lambda'}}
\right]  ,
\end{split}
\ee
where
\be
P_{\zeta^\Lambda}  = p_{\zeta^\Lambda} -  \tzeta_\Lambda p_\sigma\ ,\qquad
P_{\tzeta_\Lambda}  = p_{\tzeta_\Lambda} + \zeta^\Lambda p_\sigma \ ,\qquad
\ee
and $p_a=\pa\mathcal{L}/\pa\dot\phi^a$ are the momenta conjugate to $\phi_a$.
The conserved Noether charges are given by
\be
Q = \V^{-1} P_* \V \ .
\label{Qp}
\ee
The electric and magnetic charges $q_\Lambda, p^\Lambda$, NUT charge $k$ 
are proportional to the components $E_{q_\Lambda}, E_{p^\Lambda}, E_k$ of 
the Noether charge:
\be
\label{defpq}
\left\{ 
\begin{matrix}
p^\Lambda  &=  \sqrt2 E_{p^\Lambda} = -\frac{\sqrt2}{2}  
( p_{\tzeta_\Lambda}-\zeta^\Lambda \, p_\sigma )\\
q_\Lambda &=\sqrt2 E_{q_\Lambda} = -\frac{\sqrt2}{2} 
( p_{\zeta^\Lambda}+\tzeta_\Lambda \, p_\sigma )
\end{matrix}
\right. \ ,\qquad
k = 2 E_0 = - 2 p_\sigma\ .
\ee
Note that the charges $(p^\Lambda,q_\Lambda)$ transform as $(2,2,2)$ of $G_4$,
so can be fit into a cube $E_{2,a_1,a_2,a_3}$, corresponding to the front face
of the hypercube in Figure \ref{hypercube}. The Cayley hyperdeterminant of this cube,
\be
\begin{split}
\lozenge &=  \frac1{8} 
\epsilon^{a_1b_1}\epsilon^{a_2b_2}\epsilon^{a_3d_3}
\epsilon^{c_1d_1}\epsilon^{c_2d_2}\epsilon^{b_3c_3}\,
E_{2,a_1,a_2,a_3} E_{2,b_1,b_2,b_3} E_{2,c_1,c_2,c_3} E_{2,d_1,d_2,d_3}\\
& = -4 q_0 p^1 p^2 p^3 + 4 p^0 q_1 q_2 q_3 + 4 ( p^1 p^2 q_1 q_2 + p^2 p^3 q_2 q_3 + p^3 p^1
q_3 q_1 )\\
&\quad - (p^0 q_0 + p^1 q_1 + p^2 q_2 + p^3 q_3)^2   
\end{split}
\ee
provides a quartic polynomial invariant under $[SL(2,\IR)]^3$. This is in fact the
truncation of the $E_{7(7)}$ quartic invariant \eqref{defloz} in $\N=8$ supergravity.

For static black holes with $p_\sigma=0$,
the Hamiltonian \eqref{ham} reduces to
\be
\label{ham2}
H = \dot\phi^2 +  \sum_{i=1,2,3} \frac{(\dot S_1^{(i)})^2+(\dot S_2^{(i)})^2}{(S^{(i)}_2)^2} 
- e^\phi V_{\rm \scriptscriptstyle BH}\ ,  
\ee
where $V_{\rm \scriptscriptstyle BH}$ is the ``black hole potential"
\be
\label{VBH}
V_{\rm \scriptscriptstyle BH} = - \frac12 \left[ 
 (q_\Lambda - \Re\cN_{\Lambda\Sigma} p^{\Sigma})
[\Im\cN]^{\Lambda \Lambda'}
(q_{\Lambda'} - \Re\cN_{\Lambda'\Sigma'} p^{\Sigma'})
+p^{\Lambda} [\Im\cN]_{\Lambda \Lambda'} p^{\Lambda'} \right] \ .
\ee
Note that $V_{\rm \scriptscriptstyle BH}>0$, but the actual potential $- e^\phi V_{\rm \scriptscriptstyle BH}$ is unbounded from below.

\FIGURE{\begin{picture}(0,0)%
\includegraphics{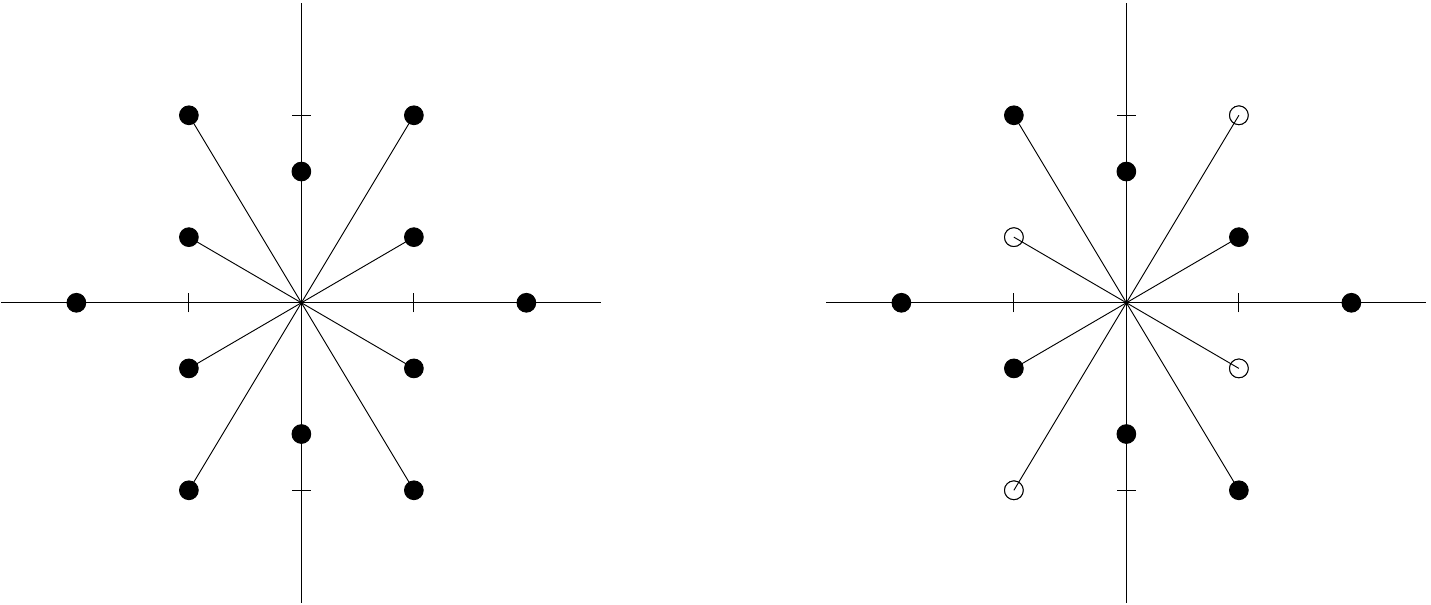}%
\end{picture}%
\setlength{\unitlength}{2368sp}%
\begingroup\makeatletter\ifx\SetFigFont\undefined%
\gdef\SetFigFont#1#2#3#4#5{%
  \reset@font\fontsize{#1}{#2pt}%
  \fontfamily{#3}\fontseries{#4}\fontshape{#5}%
  \selectfont}%
\fi\endgroup%
\begin{picture}(11424,4824)(1189,-6373)
\put(4801,-5761){\makebox(0,0)[lb]{\smash{{\SetFigFont{11}{13.2}{\rmdefault}{\mddefault}{\updefault}{\color[rgb]{0,0,0}$E_{q_0}$}%
}}}}
\put(12651,-3661){\makebox(0,0)[lb]{\smash{{\SetFigFont{11}{13.2}{\rmdefault}{\mddefault}{\updefault}{\color[rgb]{0,0,0}$L_0$}%
}}}}
\put(8326,-3811){\makebox(0,0)[lb]{\smash{{\SetFigFont{7}{8.4}{\rmdefault}{\mddefault}{\updefault}{\color[rgb]{0,0,0}$-i$}%
}}}}
\put(11926,-3811){\makebox(0,0)[lb]{\smash{{\SetFigFont{7}{8.4}{\rmdefault}{\mddefault}{\updefault}{\color[rgb]{0,0,0}$+i$}%
}}}}
\put(9226,-3811){\makebox(0,0)[lb]{\smash{{\SetFigFont{7}{8.4}{\rmdefault}{\mddefault}{\updefault}{\color[rgb]{0,0,0}$-i/2$}%
}}}}
\put(11026,-3811){\makebox(0,0)[lb]{\smash{{\SetFigFont{7}{8.4}{\rmdefault}{\mddefault}{\updefault}{\color[rgb]{0,0,0}$+i/2$}%
}}}}
\put(10426,-1861){\makebox(0,0)[lb]{\smash{{\SetFigFont{11}{13.2}{\rmdefault}{\mddefault}{\updefault}{\color[rgb]{0,0,0}$R_0$}%
}}}}
\put(10351,-2536){\makebox(0,0)[lb]{\smash{{\SetFigFont{7}{8.4}{\rmdefault}{\mddefault}{\updefault}{\color[rgb]{0,0,0}$+3/2i$}%
}}}}
\put(10351,-5536){\makebox(0,0)[lb]{\smash{{\SetFigFont{7}{8.4}{\rmdefault}{\mddefault}{\updefault}{\color[rgb]{0,0,0}$-3i/2$}%
}}}}
\put(11401,-2461){\makebox(0,0)[lb]{\smash{{\SetFigFont{11}{13.2}{\rmdefault}{\mddefault}{\updefault}{\color[rgb]{0,0,0}$K_{++++}$}%
}}}}
\put(11401,-5761){\makebox(0,0)[lb]{\smash{{\SetFigFont{11}{13.2}{\rmdefault}{\mddefault}{\updefault}{\color[rgb]{0,0,0}$K_{+---}$}%
}}}}
\put(11401,-4711){\makebox(0,0)[lb]{\smash{{\SetFigFont{11}{13.2}{\rmdefault}{\mddefault}{\updefault}{\color[rgb]{0,0,0}$K_{+(+--)}$}%
}}}}
\put(11401,-3436){\makebox(0,0)[lb]{\smash{{\SetFigFont{11}{13.2}{\rmdefault}{\mddefault}{\updefault}{\color[rgb]{0,0,0}$K_{+(++-)}$}%
}}}}
\put(8026,-3436){\makebox(0,0)[lb]{\smash{{\SetFigFont{11}{13.2}{\rmdefault}{\mddefault}{\updefault}{\color[rgb]{0,0,0}$K_{-(++-)}$}%
}}}}
\put(8026,-4711){\makebox(0,0)[lb]{\smash{{\SetFigFont{11}{13.2}{\rmdefault}{\mddefault}{\updefault}{\color[rgb]{0,0,0}$K_{-(+--)}$}%
}}}}
\put(8026,-5761){\makebox(0,0)[lb]{\smash{{\SetFigFont{11}{13.2}{\rmdefault}{\mddefault}{\updefault}{\color[rgb]{0,0,0}$K_{----}$}%
}}}}
\put(8026,-2461){\makebox(0,0)[lb]{\smash{{\SetFigFont{11}{13.2}{\rmdefault}{\mddefault}{\updefault}{\color[rgb]{0,0,0}$K_{-+++}$}%
}}}}
\put(12151,-4261){\makebox(0,0)[lb]{\smash{{\SetFigFont{11}{13.2}{\rmdefault}{\mddefault}{\updefault}{\color[rgb]{0,0,0}$J_0^+$}%
}}}}
\put(8101,-4261){\makebox(0,0)[lb]{\smash{{\SetFigFont{11}{13.2}{\rmdefault}{\mddefault}{\updefault}{\color[rgb]{0,0,0}$J_0-$}%
}}}}
\put(9751,-2986){\makebox(0,0)[lb]{\smash{{\SetFigFont{11}{13.2}{\rmdefault}{\mddefault}{\updefault}{\color[rgb]{0,0,0}$J_i^+$}%
}}}}
\put(9676,-5161){\makebox(0,0)[lb]{\smash{{\SetFigFont{11}{13.2}{\rmdefault}{\mddefault}{\updefault}{\color[rgb]{0,0,0}$J_i^-$}%
}}}}
\put(5851,-3661){\makebox(0,0)[lb]{\smash{{\SetFigFont{11}{13.2}{\rmdefault}{\mddefault}{\updefault}{\color[rgb]{0,0,0}$H$}%
}}}}
\put(1726,-3811){\makebox(0,0)[lb]{\smash{{\SetFigFont{7}{8.4}{\rmdefault}{\mddefault}{\updefault}{\color[rgb]{0,0,0}$-2$}%
}}}}
\put(5326,-3811){\makebox(0,0)[lb]{\smash{{\SetFigFont{7}{8.4}{\rmdefault}{\mddefault}{\updefault}{\color[rgb]{0,0,0}$+2$}%
}}}}
\put(2626,-3811){\makebox(0,0)[lb]{\smash{{\SetFigFont{7}{8.4}{\rmdefault}{\mddefault}{\updefault}{\color[rgb]{0,0,0}$-1$}%
}}}}
\put(4426,-3811){\makebox(0,0)[lb]{\smash{{\SetFigFont{7}{8.4}{\rmdefault}{\mddefault}{\updefault}{\color[rgb]{0,0,0}$+1$}%
}}}}
\put(3751,-2536){\makebox(0,0)[lb]{\smash{{\SetFigFont{7}{8.4}{\rmdefault}{\mddefault}{\updefault}{\color[rgb]{0,0,0}$+3/2$}%
}}}}
\put(3751,-5536){\makebox(0,0)[lb]{\smash{{\SetFigFont{7}{8.4}{\rmdefault}{\mddefault}{\updefault}{\color[rgb]{0,0,0}$-3/2$}%
}}}}
\put(4801,-2461){\makebox(0,0)[lb]{\smash{{\SetFigFont{11}{13.2}{\rmdefault}{\mddefault}{\updefault}{\color[rgb]{0,0,0}$E_{p^0}$}%
}}}}
\put(4801,-4711){\makebox(0,0)[lb]{\smash{{\SetFigFont{11}{13.2}{\rmdefault}{\mddefault}{\updefault}{\color[rgb]{0,0,0}$E_{q_i}$}%
}}}}
\put(4801,-3436){\makebox(0,0)[lb]{\smash{{\SetFigFont{11}{13.2}{\rmdefault}{\mddefault}{\updefault}{\color[rgb]{0,0,0}$E_{p^i}$}%
}}}}
\put(2026,-3436){\makebox(0,0)[lb]{\smash{{\SetFigFont{11}{13.2}{\rmdefault}{\mddefault}{\updefault}{\color[rgb]{0,0,0}$F_{q_i}$}%
}}}}
\put(2026,-4711){\makebox(0,0)[lb]{\smash{{\SetFigFont{11}{13.2}{\rmdefault}{\mddefault}{\updefault}{\color[rgb]{0,0,0}$F_{p^i}$}%
}}}}
\put(2026,-5761){\makebox(0,0)[lb]{\smash{{\SetFigFont{11}{13.2}{\rmdefault}{\mddefault}{\updefault}{\color[rgb]{0,0,0}$F_{p^0}$}%
}}}}
\put(5551,-4261){\makebox(0,0)[lb]{\smash{{\SetFigFont{11}{13.2}{\rmdefault}{\mddefault}{\updefault}{\color[rgb]{0,0,0}$E_0$}%
}}}}
\put(1501,-4261){\makebox(0,0)[lb]{\smash{{\SetFigFont{11}{13.2}{\rmdefault}{\mddefault}{\updefault}{\color[rgb]{0,0,0}$F_0$}%
}}}}
\put(3151,-5086){\makebox(0,0)[lb]{\smash{{\SetFigFont{11}{13.2}{\rmdefault}{\mddefault}{\updefault}{\color[rgb]{0,0,0}$E_i$}%
}}}}
\put(3151,-2986){\makebox(0,0)[lb]{\smash{{\SetFigFont{11}{13.2}{\rmdefault}{\mddefault}{\updefault}{\color[rgb]{0,0,0}$F_i$}%
}}}}
\put(3826,-1861){\makebox(0,0)[lb]{\smash{{\SetFigFont{11}{13.2}{\rmdefault}{\mddefault}{\updefault}{\color[rgb]{0,0,0}$\sum H_i$}%
}}}}
\put(2026,-2461){\makebox(0,0)[lb]{\smash{{\SetFigFont{11}{13.2}{\rmdefault}{\mddefault}{\updefault}{\color[rgb]{0,0,0}$F_{q_0}$}%
}}}}
\end{picture}%

\caption{Two-dimensional projection of the root diagram with respect to the split
Cartan torus $H_0,H_1+H_2+H_3$ (left) and the compact Cartan torus $L_0=J_0^3,R_0=J_1^3+J_2^3+J_3^3$ (right).
The compact (resp. non-compact) roots are indicated by a white
(resp. black) dot. The long roots generate $SL(3,\IR)$ (left)
and $SU(2,1)$ (right) subgroups, respectively.\label{g2root}}}

Since $\cM_3^*$ has restricted holonomy group $SO(2,2)^2$, it is possible to construct
a  vielbein $V_{A A' A'' A'''}$ transforming as $(2,2,2,2)$, covariantly constant under
the $\sl(2,\IR)^4$-valued spin connection. In order to make contact with the
standard construction of the quaternionic vielbein for the $c$-map geometry
\eqref{cmap} \cite{Ferrara:1989ik}, we adapt the construction of \cite{Gunaydin:2007qq}
to the pseudo-Riemannian case. 

For this purpose,  we perform a Cayley rotation within each $SL(2)$ factor, and define 
\be
J_i^+ = \frac12 ( E_i + F_ i + \I H_i )\ ,\quad
J_i^3=E_i-F_i\ ,\quad 
J_i^- = \frac12 ( E_i + F_ i - \I H_i ) \ , \quad 
\ee
for $i=0,1,2,3$. This defines a $[SL(2,\IC)]^4$ subgroup of $SO(8,\IC)$.
Under the branching
\be
\mathfrak{so}(4,4) \cong [\su(2,\IR)]^4 \oplus (2,2,2,2)\ ,
\ee
the 16 generators in the pseudo-real coset $(2,2,2,2)$ can now be fit in a complex hypercube
$K_{A_0,A_1,A_2,A_3}$ with $A_i=\pm 1$, where $\I A_i$ is the eigenvalue under $J_i^3$ 
and $K$ satisfies the reality condition
\be
K_{-A_0,-A_1,-A_2,-A_3}=(K_{A_0,A_1,A_2,A_3})^*\ .
\ee 
The entries in $K$ can be obtained from 
the real coset $E_{a_0,a_1,a_2,a_3}$ by means of a Cayley rotation
$C=e^{-\frac{\I\pi}{4} \sum_{i=0\dots 3} (E_i + F_i)}$, 
for example,
\be
\begin{split}
K_{++++} &= C\cdot F_{p^0}\cdot C^{-1} \\
&= \frac{1}{4} (-E_{p^0}-i E_{p^1}-i E_{p^2}-i E_{p^3}+i
   E_{q_0}-E_{q_1}-E_{q_2}-E_{q_3}\\
   &\quad +F_{p^0} +i
   F_{p^1}+i F_{p^2}+i F_{p^3}-i
   F_{q_0}+F_{q_1}+F_{q_2}+F_{q_3})
   \end{split}
\ee

 The action of $L_0\equiv J_0^3$ defines a complex 5-grading, and the combined action of $L_0$
 and $R_0=J_1^3+J_2^3+J_3^3$ projects the root system onto the $G_2$ root system, 
 with multiplicity 3 for the short roots. We now perform a Weyl reflection with respect to
 $K_{+---}$, in such a way that the four $SU(2)$ subalgebras generated by $J_0^\pm,J_i^\pm$
 are rotated by an angle $\pi/3$ into the four $SU(2)$ subalgebra 
 generated by\footnote{Here the parenthesis indicates that one should include the three possible permutations.} $K_{\pm\pm\pm\pm}$ and $K_{\pm(\pm\mp\mp)}$,  
 with Cartan generators
 \be
 J_0^3+J_1^3+J_2^3+J_3^3, \qquad J_0^3\pm J_1^3\mp J_2^3 \mp J_3^3\ ,
 \ee
 respectively. The algebra $\so(4,4)$ again decomposes as 
 $[\su(2)]^4 \oplus (2,2,2,2)$ with respect to this $SU(2)^4$, and the projection
 of the Maurer-Cartan one-form on the coset yields the desired 
 vielbein. This may be represented in the same Figure \ref{hypercube},
 upon replacing
 \be
 \begin{split}
  E_{p^0}\mapsto -\bar v\ ,\quad E_{p^i}\mapsto i E_{(i)} \ ,\quad 
 E_{q_i}\mapsto e_{(i)}\ ,\quad E_{q_0}\mapsto -i u\ ,\quad \\
 F_{p^0}\mapsto -\bar v\ ,\quad F_{p^i}\mapsto -i \bar E_{(i)} \ ,\quad 
 F_{q_i}\mapsto i \bar e_{(i)}\ ,\quad F_{q_0}\mapsto i \bar u\ ,\quad 
\end{split}
\ee
 where $u,v,e^{(i)},E^{(i)}$ denote the right-invariant one-forms
 \be
 \label{uveE}
\begin{array}{llllllllllll}
v &=& -d\phi +  \frac{i}{2} 
e^{-\phi}  (d\sigma + \tzeta^\Lambda d\zeta_\Lambda -  \zeta^\Lambda d\tzeta_\Lambda)&\ ,\ &
u &=&\sqrt2\, e^{(K-\phi)/2} X^\Lambda ( d\tzeta_\Lambda + \cN_{\Lambda\Sigma} d\zeta^\Sigma )\\
e^{(i)} &=& i (dS^{(i)}_{1} + i \, dS^{(i)}_{2})/S^{(i)}_{2} &\ ,\ &
E^{(i)} &=&2i\sqrt{2} \, S^{(i)}_{2} \, e^{-\phi/2}\, f_i^\Lambda \, 
( d\tzeta_\Lambda + \bar\cN_{\Lambda\Sigma} d\zeta^\Sigma )
\end{array}
\ee
where $f_i^I = e^{K/2} D_i X^I = e^{K/2} [\pa_i + (\pa_i K) ] X^I$. 
It is straightforward to check that the vielbein $V_{A A' A'' A'''}$
satisfies the reality condition 
\be
V_{A A' A'' A'''}  = \epsilon_{AB} \epsilon_{A'B'} \epsilon_{A''B''} \epsilon_{A'''B'''} 
( V_{B B' B'' B'''} )^*
\ee
In term of these one-forms, the metric takes the simple form
\be
ds^2_{\cM_3} = -\frac12 \epsilon_{AB} \epsilon_{A'B'} \epsilon_{A''B''} \epsilon_{A'''B'''} 
V_{A A' A'' A'''} V_{B B' B'' B'''}= 
u \bar u + v\bar v+ \sum_{i=1,2, 3} (e^{(i)} \bar e^{(i)} + E^{(i)} \bar E^{(i)})
\ee
The spin connection for each of the $SU(2)$ factors is given by 
 \be
\begin{pmatrix} {\underline K}_{++++} \\
{\underline J_3} \\
 {\underline K}_{----} 
\end{pmatrix}= 
\begin{pmatrix} u \\
\frac12 \left( (v- \bar v) + (e_1 - \bar e_1) + ( e_2 - \bar e_2) + ( e_3 - \bar e_3)  \right)\\
-\bar u
\end{pmatrix} 
\ee
and
\be
\begin{pmatrix} {\underline K}_{-(++-)} \\
{\underline J_i}' \\
 {\underline K}_{+(+--)} 
\end{pmatrix}= 
\begin{pmatrix} E_i  \\
\frac12 \left( (v- \bar v) + (e_i - \bar e_i) -  (e_j - \bar e_j) -  (e_k - \bar e_k) \right)\\
-\bar E_i
\end{pmatrix} 
\ee
The constant covariance of the vielbein can be checked using  the Maurer-Cartan equation 
$d\theta+\theta\wedge\theta=0$, 
\bea
\label{maurereq}
du &+& \frac14 
\left(  (e_1 - \bar e_1) + ( e_2 - \bar e_2) + ( e_3 - \bar e_3) -(v+ \bar v)  \right) u
= \frac12 e_i \wedge E_i\ ,\nn\\
dE_i &+& \frac14 
\left(  ( e_j - \bar e_j) + ( e_k - \bar e_k) - (e_i - \bar e_i)  -(v+ \bar v)  \right) E_i
 = \frac12 \left( |\epsilon_{ijk}| \bar E_j \wedge e_k + u \wedge \bar e_i  \right)\nn \\
 dv &=& \frac12 ( v\bar v+ u \bar u+ E_i \bar E_i )\nn \\
 de_i &+& \frac12 \bar e_i e_i = 0
\eea
Singling out the first $SU(2)$ as the R-symmetry group, we recover the
quaternionic vielbein for the $c$-map geometry \eqref{cmap} \cite{Ferrara:1989ik},
 \be
\label{quatviel}
V^{\alpha A} = \begin{pmatrix}
\bar u &   v \\ 
\bar  e^{\bar \I} & E_{\I} \\ 
-\bar E_{\bar \I} & e^{\I} \\
 - \bar v & u
\end{pmatrix}
=-i  \begin{pmatrix}
 {\underline  K}_{+---} &    {\underline J}_0^+ \\ 
- {\underline J}_i^- &   {\underline K}_{+(++-)}  \\ 
 {\underline K}_{-(+--)} &  {\underline J}_i^+  \\ 
 {\underline J}_0^- & -  {\underline K}_{-+++}
\end{pmatrix}\ ,
\ee

Finally, in order to describe the pseudo-Riemannian manifold $\cM_3^*$ we conjugate
the generators by $\exp(i\pi H_0/2)$, whose effect is to Wick rotate
\be
E_0\mapsto -E_0\ ,\qquad 
\begin{pmatrix} E_{p^\Lambda} \\ E_{q_\Lambda}\end{pmatrix} \mapsto
i \begin{pmatrix} E_{p^\Lambda} \\ E_{q_\Lambda}\end{pmatrix}\ ,\quad
\begin{pmatrix} F_{p^\Lambda} \\ F_{q_\Lambda}\end{pmatrix}\mapsto
-i \begin{pmatrix} F_{p^\Lambda} \\ F_{q_\Lambda}\end{pmatrix}\ ,\quad
F_0\mapsto -F_0\ .
\ee
The (2,2,2,2) coset $V^*_{A A' A'' A'''} $ is now obtained from  
Figure \ref{hypercube} by replacing
\be
\begin{split}
 E_{p^0}\mapsto -\bar v\ ,\quad E_{p^i}\mapsto E_{(i)} \ ,\quad 
 E_{q_i}\mapsto e_{(i)}\ ,\quad E_{q_0}\mapsto -u\ ,\quad \\
  F_{p^0}\mapsto -v\ ,\quad F_{p^i}\mapsto - \bar E_{(i)} \ ,\quad 
 F_{q_i}\mapsto \bar e_{(i)}\ ,\quad F_{q_0}\mapsto  \bar u\ ,\quad 
\end{split}
 \ee
 In terms of this new vielbein $V^*_{A A' A'' A'''} $, the pseudo-Riemannian metric
 \eqref{cmapst} may be written as
\be
\label{dscmapu}
\begin{split}
ds^2_{\cM_3^*} &= -\frac12 \epsilon_{AB} \epsilon_{A'B'} \epsilon_{A''B''} \epsilon_{A'''B'''} 
V^*_{A A' A'' A'''} V^*_{B B' B'' B'''}\\
&= \left( v\bar v+ \sum_{i=1,2, 3} e^{(i)} \bar e^{(i)}  \right) 
-\left( u \bar u + \sum_{i=1,2, 3}  E^{(i)} \bar E^{(i)} \right)\ .
\end{split}
\ee
As will become clear in Section \ref{secgeobps}, the two terms in bracket correspond to
the kinetic and potential terms in the Hamiltonian \eqref{ham2}.

\subsection{Near horizon solutions}

For simplicity, we focus on solutions with $D4$ and $D0$ brane charges only, and
denote $P^i=p^i, P^0=-p^0, Q_i=q_i, Q_0=-q_0$ to conform with existing literature.
For extremal solutions, the near horizon geometry is $AdS_2\times S_2$ with 
constant values for the complex scalars $S^{(i)}$, which have to extremize the
``black hole potential" $V_{\rm \scriptscriptstyle BH}$. For $D4-D0$, it is consistent to set the axions
$S^{(i)}_1$ to zero. In this case, $V_{\rm \scriptscriptstyle BH}$ simplifies to
\be
V_{\rm \scriptscriptstyle BH}= \frac{Q_0^2 + 
(P^1)^2 T_2^2 U_2^2 + (P^2)^2 S_2^2 U_2^2 + (P^3) ^2 S_2^2 T_2^2}
{2 S_2 T_2 U_2}\ .
\ee
Extremization with respect to $S_2,T_2,U_2$ leads to a unique 
minimum,
\be
\label{genextr}
S_{2,*} = \sqrt{\frac{|Q_0 P^1|}{|P^2 P^3|}}\ ,\quad
T_{2,*} = \sqrt{\frac{|Q_0 P^2|}{|P^1 P^3|}}\ ,\quad
U_{2,*} = \sqrt{\frac{|Q_0 P^3|}{|P^1 P^2|}} \ ,\quad
\ee
corresponding to an entropy
\be
S_{\rm \scriptscriptstyle BH} = \pi V_{{\rm \scriptscriptstyle BH},*}= 2\pi \sqrt{|Q_0 P^1 P^2 P^3|} \ .
\ee
The extremum \eqref{genextr} exists for charges $Q_0, P^i$ of any sign, but the
supersymmetry properties of the solution do depend on the signs of the charges.
The values of the central charges at the horizon are summarized in the table below, where
$z={\rm sgn}(Q_0) \sqrt2 |Q_0 P^1 P^2 P^3|^{1/4}$ (see also \cite{Bellucci:2007zi,Bergshoeff:2008be,Bellucci:2008sv}):
\be 
\begin{array}{|c|ccc|cccc|}  \hline
& P^1 Q_0 &  P^2 Q_0 & P^3 Q_0 & Z_* & Z_{1,*} & Z_{2,*} & Z_{3,*} \\   \hline
(a) &  + & + & + & -z & 0 & 0 & 0 \\
(b) &   + & - & - & 0 & -z & 0 & 0 \\
(c) &   - & + & - & 0 & 0 & -z & 0 \\
 (d) &  - & - & + & 0 & 0 & 0 & -z \\ \hline 
(e) &   - & - & - & \frac{z}{2} & -\frac{z}{2} & -\frac{z}{2} & -\frac{z}{2} \\
(f) &   -& + & + & -\frac{z}{2} & \frac{z}{2} & -\frac{z}{2} & -\frac{z}{2} \\
(g) & + & - & + & -\frac{z}{2} & -\frac{z}{2} & \frac{z}{2} & -\frac{z}{2}  \\
(h) &  + & + & - & -\frac{z}{2} & -\frac{z}{2} & -\frac{z}{2} & \frac{z}{2} \\\hline
\end{array} 
\ee
Case $(a)$ corresponds to BPS black holes of the $STU$ model.
The next three cases $(bcd)$ are the so-called non-BPS, $Z_*=0$
extremal solutions. They are non-BPS in the $STU$ model, but can still be
lifted to 1/8-BPS black holes in maximal supergravity \cite{Ferrara:2007pc}. 
All these cases have $\lozenge>0$. In contrast, the remaining four cases $(efgh)$ have $\lozenge<0$
and are genuinely non-BPS. They are
characterized by the fact that the four central charges $Z,Z_i$ at the horizon 
are non-vanishing,
and in fact equal in modulus. We shall discuss these solutions in more detail below.

\subsection{Extremal $Z_*=0$ solutions \label{secgeobps}} 

In this subsection, we discuss the geodesic flow interpretation of BPS black holes, 
as well as so-called non-BPS, $Z_*=0$ black holes, as they all
appear on the same footing in the context of the $STU$ model.

Our starting point is the $SL(2,\IR)^4$ vielbein $V^*_{AA'A''A'''}$ computed at the end
of Section \ref{STUmodsec} in terms of the invariant one-forms \eqref{uveE}, and the
relation \eqref{defpq} between the electromagnetic and NUT charges 
$q_\Lambda, p^\Lambda,k$ and canonical momenta conjugate to $\zeta^\Lambda,
\tzeta_\Lambda,\sigma$. For a static black hole, $k=-2 p_\sigma=0$, which allows
to express the differentials $d\zeta^\Lambda, d\tzeta_\Lambda, d\sigma$ in terms of the
electromagnetic charges $q_\Lambda, p^\Lambda$. In this way, we find
\be
\label{uveEonshell}
\begin{array}{llllllllllll}
v &=& -d\phi  &\ ,\ &
u &=&-i e^{\phi/2} Z \\
e^{(i)} &=& i (dS^{(i)}_{1} + i dS^{(i)}_{2})/S^{(i)}_{2} &\ ,\ &
E^{(i)} &=& -i e^{\phi/2} Z_i
\end{array}
\ee
where $Z\equiv Z_0$ and $Z_i$ are the central charge and ``scalar charges'', respectively, 
\be
\label{Zcent}
Z = e^{K/2} ( q_\Lambda X^\Lambda - p^\Lambda F_\Lambda)\ ,\qquad
Z_i = -2 i S^{(i)}_{2}  \scal{\pa_i+\frac{1}{2}\pa_i K} Z\ .
\ee
For completeness we list the central charges:
\be
\begin{split}
2 \sqrt{2 S_2 T_2 U_2} Z &=
\scal{q_0 + q_1 S_1 + q_2 T_1 + q_3 U_1
+ p^3 S_1 T_1  + p^2 S_1 U_1 + p^1 T_1 U_1 - 
   p^0 S_1 T_1 U_1} \\&
 + \I S_2 \scal{q_1 + p^3 T_1 + (p^2 - p^0 T_1) U_1}  
 + \I T_2 \scal{q_2 + p^3 S_1 + (p^1 - p^0 S_1) U_1}   \\&
 + \I U_2 \scal{q_3 + p^2 S_1 + (p^1 - p^0 S_1) T_1} 
 -   S_2 T_2 (p^3 - p^0 U_1)   \\&-
 S_2 U_2 (p^2 - p^0 T_1)   - T_2 U_2 (p^1 - p^0 S_1)   + 
 \I p^0 S_2 T_2 U_2 \ ,
 \end{split}
 \ee
 \be
 \begin{split}
 2 \sqrt{2 S_2 T_2 U_2} Z_1 &=
 \scal{q_0 + q_1 S_1 + q_2 T_1 + q_3 U_1+ p^3 S_1 T_1  + p^2 S_1 U_1 + p^1 T_1 U_1 - 
   p^0 S_1 T_1 U_1} \\& 
   - \I S_2 \scal{q_1 + p^3 T_1 + (p^2 - p^0 T_1) U_1} 
 + \I  T_2 \scal{q_2 + p^3 S_1 + (p^1 - p^0 S_1) U_1}  \\&
  + \I U_2 \scal{q_3 + p^2 S_1 + (p^1 - p^0 S_1) T_1} 
  + S_2 T_2 (p^3 - p^0 U_1)  \\&
  + S_2 U_2 (p^2 - p^0 T_1)  
 + T_2 U_2  (-p^1 + p^0 S_1)   - 
 \I p^0 S_2 T_2 U_2 \ ,
 \end{split}
 \ee
 \be
 \begin{split}
 2 \sqrt{2 S_2 T_2 U_2} Z_2 &=
\scal{q_0 + q_1 S_1 + q_2 T_1 + q_3 U_1
 + p^3 S_1 T_1  + p^2 S_1 U_1 + p^1 T_1 U_1 -   p^0 S_1 T_1 U_1} \\&
   + \I S_2 \scal{q_1 + p^3 T_1 + (p^2 - p^0 T_1) U_1}  
   - \I T_2 \scal{q_2 + p^3 S_1 + (p^1 - p^0 S_1) U_1)}  \\&
 +  \I U_2 \scal{q_3 + p^2 S_1 + (p^1 - p^0 S_1) T_1} 
 + S_2 T_2 (p^3 - p^0 U_1)   \\&
 + S_2 U_2  (-p^2 + p^0 T_1)  
 + T_2 U_2(p^1 - p^0 S_1)   - 
 \I p^0 S_2 T_2 U_2 \ ,
 \end{split}
 \ee
 \be
 \begin{split}
 2 \sqrt{2 S_2 T_2 U_2} Z_3 &=
 \scal{q_0 + q_1 S_1 + q_2 T_1 + q_3 U_1+ p^3 S_1 T_1  + p^2 S_1 U_1 + p^1 T_1 U_1 - 
   p^0 S_1 T_1 U_1} \\&
 + \I S_2 \scal{q_1 + p^3 T_1 + (p^2 - p^0 T_1) U_1} 
 + \I T_2 \scal{q_2 + p^3 S_1 + (p^1 - p^0 S_1) U_1}  \\&
 -  \I U_2 \scal{q_3 + p^2 S_1 + (p^1 - p^0 S_1) T_1}   
 +  S_2 T_2 (-p^3 + p^0 U_1)\\&
 +  S_2 U_2 (p^2 - p^0 T_1)   
 + T_2 U_2  (p^1 - p^0 S_1)   - \I p^0 S_2 T_2 U_2 \ .
\end{split}
\ee
It should be noted that $Z_i$ is related to $Z$ by reversal of the sign of one of 
the $S_2^{(i)}$'s followed by complex conjugation, e.g.
\be
Z_1(S,T,U; p^\Lambda, q_\Lambda) = \overline{Z (\bar S,T,U; p^\Lambda, q_\Lambda) }\ .
\ee
Using \eqref{dscmapu}, we see that the black hole potential \eqref{VBH}
can be rewritten as 
\begin{equation}
  V_{\rm \scriptscriptstyle BH}=|Z_0|^2+|Z_1|^2 +|Z_2|^2 +|Z_3|^2 \ .
\end{equation}
Moreover, from standard formulae in special geometry \cite{Ceresole:1995ca} 
\be
\label{dDmc}
D Z = \frac{i}{2S_2^{(i)}} Z_i \, dS^{(i)} \ ,\qquad
D Z_i =  \frac{i}{2S_2^{(k)}}  |\epsilon_{ijk}|  \bar Z_j \wedge  dS^{(k)} -\frac{i}{2S_2^{(i)} } Z d\bar S^{(i)}  
\ee
or from the Maurer-Cartan equations \eqref{maurereq}, one can check that 
\be
 V_{\rm \scriptscriptstyle BH} =W^2 +4 \left( 
 g^{S\bar S} | \pa_S W |^2 
  + g^{T\bar T} | \pa_T W |^2 + g^{U\bar U}| \pa_U W |^2 \right)
 \ee
for any choice of $W$ amongst \cite{Andrianopoli:2007gt}
\be
\label{listW}
W= |Z|\ ,\qquad
W= |Z_1|\ ,\qquad
W= |Z_2|\ ,\qquad
W= |Z_3|\ .
\ee
Therefore,  any $W$ in \eqref{listW}
can be used as a fake superpotential to generate first order equations
\be
\label{eq1}
\dot\phi =- e^{\phi/2} W \ ,\qquad  
\dot S^{(i)} = -e^{\phi/2} g^{S^{(i)} \bar S^{(i)} } \pa_{\bar S^{(i)} } W
\ee
which imply the second order equations from the Hamiltonian \eqref{ham} at zero
energy, and moreover guarantee that the corresponding solution is 
extremal \cite{Ceresole:2007wx}. 

The first choice in \eqref{listW} corresponds to BPS black holes. Its interpretation
as a special kind of geodesic flow on $\cM_3^*$ is well 
known \cite{Gunaydin:2005mx,Gunaydin:2007bg}: indeed, the
BPS attractor flow equations \eqref{eq1} with $W=|Z|$ can be 
written in terms of the invariant forms \eqref{uveEonshell} as
\be
v =- z u\ ,\quad \bar e_{(i)}  = z E_{(i)}\ ,  \quad \bar E_{(i)} = z  e_{(i)}\ ,\quad \bar u= - z \bar v \ ,\quad
z = - i \sqrt\frac{\bar Z}{Z}\ ,
\ee
or equivalently, using the vielbein $V_{AA'A''A'''}$,
\be
V_{AA'A''A'''} \epsilon^{A} = 0\ ,\qquad 
\epsilon^{A} = \begin{pmatrix} 1 \\ z \end{pmatrix}\ .
\ee
Since the index $A$ transforms as a doublet under the R-symmetry $SU(2)$, this is recognized
as the Killing spinor equation for BPS black holes \cite{Gunaydin:2005mx,Gunaydin:2007bg}.
Note that the phase of $Z$ varies according to
\be
d\arg(Z) + \cA = 0
\ee
where $\cA$ is the K\"ahler connection,
\be
\cA = \frac{1}{2i} \left( \pa_i K dt^i - \pa_{\bar i} K d\bar t^{\bar i}  \right)=
\frac12 \left( \frac{dS_1}{S_2} + \frac{dT_1}{T_2} + \frac{dU_1}{U_2} \right)\ . 
\ee

Similarly, the first order equations \eqref{eq1} with $W=|Z_i|$ can be 
written in terms of the invariant forms \eqref{uveEonshell} as
\be
v =- z \bar E_{(i)}\ ,\quad 
e_{(i)}  = z u\ ,\quad
e_{(j)}  = z E_{(k)}\ ,\quad
e_{(k)}  = z E_{(j)}\ ,\quad
z = - i \sqrt\frac{Z_i}{\bar Z_i}\ ,
\ee
or equivalently, using the vielbein $V_{AA'A''A'''}$,
\be
\framebox{$
V_{AA'A''A'''} \epsilon^{A^{(i)}} = 0\ ,\qquad 
\epsilon^{A^{(i)}} = \begin{pmatrix} 1 \\ z \end{pmatrix}\ ,$}
\ee
where we denote $A^{(1)}=A',A^{(2)}=A'',A^{(3)}=A'''$. 
Note that the phase of $Z_i$ varies in this class of solutions according to
\be
d\arg(Z_i) + \cA_{(i)} = 0
\ee
where 
\be
\cA_{(i)} = \frac12 \left( \frac{dS_1^{(i)}}{S_2^{(i)}} - \frac{dS_1^{(j)}}{S_2^{(j)}} 
- \frac{dS_1^{(k)}}{S_2^{(k)}}\right)\ . 
\ee
Thus, the non-BPS, $Z_*=0$ solutions are related to the BPS ones by a permutation 
of the four $SL(2)$ factors.\footnote{This fact has been observed independently in \cite{Bergshoeff:2008be}.}This is in accordance with the fact that these solutions
lift to 1/8-BPS solutions in maximal supergravity \cite{Ferrara:2007pc}. 
Using the explicit form of the solutions
given in Appendix \ref{secsol}, one may check that the Noether charge is nilpotent of degree 3
in the vector representation, and corresponds to the
nilpotent orbits $[(+-+)^2 (-)^2]_{II}$,  $[(-+-)^2 (+)^2]_{I,II}$ in cases (a,b,c,d), respectively.

\subsection{Extremal $Z_*\neq 0$ solutions}

When $\lozenge<0$, the extremal black hole is genuinely non-BPS. The fake superpotential $W$
is given by the same formula \eqref{WNB} as in $\N=8$ supergravity, upon substituting
\be
\rho_\zero = |Z|\ ,\quad
\rho_\un = |Z_1|\ ,\quad
\rho_\deux = |Z_2|\ ,\quad
\rho_\trois = |Z_3|\ ,\quad
4\varphi=\arg(Z \bar Z_1 \bar Z_2 \bar Z_3)\ .
\ee
On the sublocus where $\varphi=\pi/4$ or $\varphi=0$, the simpler forms 
\eqref{physr}, \eqref{physr2t} can be used. In particular, for axion-free solutions
with $Z_*<0$ and $Z_{i,*}>0$, both \eqref{physr} and \eqref{physr2t}
reduce to the formula
proposed in \cite{Ceresole:2007wx,Bellucci:2008sv}
\begin{equation}
\begin{split}
W&= \frac{1}{2} \Scal{Ê- Z + Z_\un+Z_\deux+Z_\trois} \\
&=e^{K/2}
  \left(-Q_0
  +\frac{T\bar{U}+U\bar{T}}{2}P^1
  +\frac{S\bar{U}+U\bar{S}}{2}P^2
  +\frac{S\bar{T}+T\bar{S}}{2}P^3\right)
 \label{WnonBPSZnonnul}
\end{split}
\end{equation}
on the locus $S_1=T_1=U_1=0$. 

Evaluating the Noether charge on the explicit solution given in Appendix \ref{secsol},
it is easy to confirm that such non-BPS, $Z_*\neq 0$ solutions are indeed associated
to the nilpotent orbit  $[(+-+),(-+-),+,-]$ of $SO(4,4)$.

\acknowledgments

The authors thank G. Dall'Agata, H. Nicolai, K. S. Stelle  and S. Trivedi for useful discussions.
B.P. acknowledges ICTS, TIFR and the Monsoon Workshop in String Theory 2008 
for providing a stimulating atmosphere when this project 
was begun. 

\appendix

\section{Nilpotent orbits}

In this appendix, we review some general facts about nilpotent orbits of real Lie groups, 
and discuss in details the nilpotent orbits of $E_{8(8)}$ and $SO(4,4)$ relevant for 
extremal black holes in maximal supergravity and the $STU$ model, respectively.
Useful references for the material of this section are \cite{Collingwood} 
and  \cite{Levi,E8strat,DokovicSO}.

\subsection{Generalities}

Complex nilpotent orbits of $G_\IC$ are classified by conjugacy classes of homomorphisms 
$\mathfrak{sl}(2)\hookrightarrow\fg_\IC$, i.e. by triplets 
$({\bf e},\, {\bf f},\, {\bf h})$ of elements in the Lie algebra $\fg_\IC$ satisfying the $\sl_2$ commutation relations $[{\bf e},{\bf f}]={\bf h}, \, [{\bf h},{\bf e}]=2{\bf e}, [{\bf h},{\bf f}]=-2{\bf f}$. 
Under the adjoint action of ${\bf h}$, $\fg_\IC$ decomposes as a sum of eigenspaces 
$\fg_\IC=\bigoplus \fg^\ord{i}$ with eigenvalues $i\in [-n,n]$, $n\geq 2$, often 
referred to as a $2n+1$-grading (this grading is said to be even if $\fg^\ord{i}=0$ for $i$ odd;
in this case $n$ is even, and $\fg_\IC=\bigoplus \fg^\ord{2i}$ yields a $n+1$-grading). Correspondingly, 
$\fg_\IC$ decomposes into a sum of irreducible representations of $SL(2)$ with 
spin $j\leq n/2$. ${\bf e}$ is an element of $\fg^\ord{2}$, and therefore nilpotent of degree $2n+1$. Its complex adjoint orbit $\cO_\IC({\bf e})$ is $P_\IC\backslash G_\IC$,
where $P_\IC$ is the stabilizer of ${\bf e}$, a non-reductive subgroup of the parabolic subgroup $\bigoplus_{i\ge0} \fg^\ord{i} \subset \fg_\IC$ which contains
 in particular $\fg^\ord{n-1}\oplus
\fg^\ord{n}$. The complex dimension of $\cO_\IC({\bf e})$ is equal to $\dim\fg-\dim\fg^\ord{0}-\dim\fg^\ord{1}$.
Complex orbits can be labelled by their Dynkin indices, which are the coefficients of ${\bf h}$
on the standard generators of the Cartan subalgebra of $\fg$. 
These can take values amongst $0,1,2$ only. The subalgebra $\fg^\ord{0} \subset \fg_\IR$ commuting with ${\bf h}$  is the direct sum of abelian factors associated to the 
nodes with a non-zero label,  and of the semi-simple Lie algebra whose Dynkin diagram coincides with the set of zero nodes. The set of nilpotent orbits admits a partial ordering, with ${\bf e}\leq {\bf e}'$ whenever ${\bf e}'$ lies in the closure of the orbit through ${\bf e}$. It is convenient to display them in a Hasse-type diagram, 
with arrows pointing from ${\bf e}$ to ${\bf e}'$ when ${\bf e}\leq {\bf e}'$. 

We are interested in nilpotent orbits of $\mathfrak{g}_\mathds{R}$, where $\mathfrak{g}_\mathds{R}$
is the Lie algebra of a non-compact real form $G_\IR$ of $G_\IC$, with  maximal compact subgroup 
$K_\mathds{R}$.  The maximal compact Lie algebra $\k_\mathds{R}$ 
of $\fg_\mathds{R}$ is obtained as the $(-1)$-eigenspace of the 
Cartan involution  $^\dagger$ (not to be confused with the adjoint),
so that elements of $\k_\mathds{R}$ are of the form $e-e^\dagger$ where $e\in \fg_\IR$.
The Kostant--Sekiguchi correspondence relates $G_\IR$-orbits in $\fg_\IR$ to 
$K_\IC$-orbits in $\fg_\IC \ominus \k_\IC$. More precisely, the Kostant--Sekiguchi homeomorphism
\be 
{\bf e}  \mapsto \frac{1}{2} \scal{Ê{\bf e} + {\bf e}^\dagger + i [ {\bf e} , {\bf e}^\dagger]  } 
\ee
identifies
\be \frac{\Nil \cap \mathfrak{g}_\mathds{R}}{G_\mathds{R}} \cong \frac{\Nil \cap (\mathfrak{g}_\IC \ominus \k_\IC)}{K_\IC} \ ,
\ee
where $\Nil$ is the variety of nilpotent elements inside $\mathfrak{g}_\IC$. Its inverse can be obtained by use of the complex conjugation $^*$, under 
the convention that the generators of $\mathfrak{g}_\mathds{R}$ are real,
\be 
{\bf e} \mapsto \frac{1}{2} \scal{Ê{\bf e} + {\bf e}^* - i [ {\bf e} , {\bf e}^*]  } \ . 
\ee
The representative of a real orbit $\mathcal{O}_\mathds{R}({\bf e})$ can be chosen 
so that its standard triplet 
$({\bf e},\, {\bf f},\, {\bf h})$ is a Cayley triplet, \ie verifies 
\be
{\bf e}-{\bf f} \in \k_\IR\ ,\qquad {\bf h},\, {\bf e}+{\bf f} \in \mathfrak{g}_\IR \ominus
\k_\IR\ .
\ee
In practice, ${\bf f}= {\bf e}^\dagger$ is obtained from $ {\bf e}$ via  the Cartan involution, after having
normalized ${\bf e}$ properly.
The Kostant--Sekiguchi homeomorphism maps this $\sl_2$ subalgebra of $\mathfrak{g}_\mathds{R}$ to  the following  $\sl_2$ subalgebra of $\mathfrak{g}_\IC$,
\be \bigl\{  {\bf e},\, {\bf f}, \, {\bf h} \bigr\} 
\mapsto 
\bigl\{\sfrac{1}{2} ( {\bf e} + {\bf f} + i{\bf h} ) ,\, \sfrac{1}{2} ( {\bf e} + {\bf f} - i {\bf h} ),\,
 i ({\bf e}-{\bf f} \,  )\bigr\} \ .
\ee
such that the nilpotent element $\sfrac{1}{2} ( {\bf e} + {\bf f} + i{\bf h} )$ lies in $\mathfrak{g}_\IC \ominus \k_\IC$, while the semi-simple element $ i ({\bf e} -{\bf f})$ lies in $\k_\IC$. 

It turns out that the $G_\IR$-orbit of ${\bf e}$ is uniquely characterized
by the $K_\IC$-conjugacy class of $i({\bf e}-{\bf e}^\dagger)$. Moreover, one can 
always choose a representative ${\bf e}_c$ of the  orbit $\mathcal{O}_\IR({\bf e})$ such that  
$i({\bf e}_c-{\bf e}_c^\dagger)$ is in a fixed Cartan subalgebra of $\k_\IC$. The weighed 
Dynkin diagram associated to $\mathcal{O}_\IR({\bf e})$ is then
the Dynkin diagram of $K$ decorated by the eigenvalues of the simple roots of $\k$ 
with respect to the element $i({\bf e}_c-{\bf e}_c^\dagger)$. This diagram  
labels the real nilpotent orbit uniquely if $\k$ is semi-simple,
otherwise it has to be supplemented with  extra labels associated to the Abelian factors. 
In contrast to $G$, the labels of the  weighed 
Dynkin diagram of $K$ are not bounded by 2, although they are still positive.

For applications to black holes,  we are however more interested in the orbit of 
an element ${\bf e}$  lying in
 $\Nil \cap(  \mathfrak{g}_\mathds{R} \ominus \mathfrak{k}^*_\mathds{R})$ 
 under $K^*_\mathds{R}$,
where $K^*_\mathds{R}$ is a non-compact real form of $K_\IC$
defined by the Cartan involution $^\ddagger$. 
When the intersection $\mathcal{O}_\IR({\bf e}) \cap ( \fg_\IR \ominus \k^*_\IR)$ is non-empty,
there exists a  ``starred" Cayley triplet $({\bf e}_{c^*}, {\bf f}_{c^*}= {\bf e}_{c^*}^\ddagger, {\bf h}_{c^*})$ such that 
 ${\bf e}_{c^*}, {\bf f}_{c^*} \in \mathfrak{g}_\mathds{R} \ominus \mathfrak{k}^*_\mathds{R}$
 and  ${\bf h}_{c^*}$ lies in a fixed Cartan subalgebra of $\mathfrak{k}^*_\mathds{R}$. 
The orbit is characterized by the eigenvalues of the simple roots with respect to ${\bf h}_{c^*} $, 
 which furnish a weighted Dynkin diagram for $K^*$. The semi-simple generator
 ${\bf h}_{c^*} $ defines a graded decomposition of the Lie algebra $\k^*$.

 A given  nilpotent $G_\IC$-orbit in general corresponds to  several $G_\mathds{R}$-orbits
 labelled by $K_\mathds{C}$-weighted Dynkin diagrams $d_i$. Similarly, the
 $K_\mathds{C}$-orbit associated to $d_i$ corresponds to several $K^*_\mathds{R}$-orbits,
 labeled by the same \footnote{more properly, by a subset thereof. Indeed, some $G_\IR$-orbits do not 
 intersect with  the coset component $\fg_\IR-\k^*$. 
 This is the case for instance of all the $E_{8(8)}$ nilpotent orbits of dimension strictly greater than 216.} 
 $K_\mathds{C}$-weighted Dynkin diagrams $d_j$. Thus,
 each $K^*_\mathds{R}$-orbit is labelled by a pair $(d_i, d_j)$.
 It turns out, although we do not know a proof of this fact, that extremal black holes
 correspond to choosing $d_i=d_j$. This condition on $P$ complements the nilpotency
 condition on $Q$  mentioned in Section 1.4.

%
\subsection{Nilpotent orbits of $E_{8(8)}$}

\subsubsection*{Minimal orbit, dimension 58}

We start with the minimal orbit of $E_{8(8)}$, with 
weighed Dynkin diagram \DEVIII00000001. It corresponds 
to the graded decomposition
\be 
\mathfrak{e}_{8(8)} \cong {\bf 1}^\ord{-2} \oplus {\bf 56}^\ord{-1} \oplus \scal{Ê\mathfrak{gl}_1 \oplus \mathfrak{e}_{7(7)}}^\ord{0} \oplus {\bf 56}^\ord{1} \oplus  {\bf 1}^\ord{2}  \label{DEVII} 
\ee
of the adjoint representation, familiar from the dimensional reduction from 4 to 3 dimensions.
The ${\bf 3875}$  representation is also five-graded with respect to this decomposition, therefore the elements in the grade $2$ component are nilpotent of degree $3$ in both the adjoint representation and the ${\bf 3875}$. 
To see that such elements can be chosen inside $\mathfrak{e}_{8(8)} \ominus \so^*(16)$, 
we decompose (\ref{DEVII}) further into 
\be \mathfrak{e}_{8(8)} \cong {\bf 1}^\ord{-2} \oplus \scal{Ê{\bf 28}_+ \oplus {\bf 28}_-}^\ord{-1} \oplus \scal{Ê\mathfrak{gl}_1 \oplus \mathfrak{su}^*(8) \oplus {\bf 70}}^\ord{0} \oplus \scal{Ê{\bf 28}_+ \oplus {\bf 28}_-}^\ord{1} \oplus  {\bf 1}^\ord{2} \ .
\ee
This exhibits the embedding $\so^*(16) \subset \mathfrak{e}_{8(8)}$, 
associated to the weighted Dynkin diagram \DSOXVI00000010 of the real $E_{8(8)}$ orbit
\be \so^*(16) \cong {\bf 28}_+^\ord{-1} \oplus \scal{Ê\mathfrak{gl}_1 \oplus \mathfrak{su}^*(8) }^\ord{0}  \oplus  {\bf 28}_+^\ord{1} \ .
\ee
The weighted Dynkin diagram of the real orbit labels the compact element of the Cartan subalgebra of $\so(16)$ that appears in the Cayley triplet associated to the nilpotent element ${\bf e}$, namely ${\bf e}-{\bf e}^\dagger \in {\bf 1}^\ord{2} -  {\bf 1}^\ord{-2}$. As it turns out, it also labels  
a non-compact element of the Cartan subalgebra of $\so^*(16)$ which is ${\bf h} = [Ê{\bf e} , {\bf e}^\ddagger] \in \gl_1^\ord{0}$ for a particular choice of normal triplet. 

The isotropy subgroups inside 
$Spin^*(16)$ and $E_{8(8)}$ follow trivially from the five-graded decomposition 
\be SU^*(8) \ltimes {\bf 28}^\ord{1} \subset E_{7(7)} \ltimes \scal{Ê{\bf 56}^\ord{1} \oplus {\bf 1}^\ord{2} } \ee 

\subsubsection*{Nilpotent orbit of dimension 92}

The next orbit in the Hasse diagram 
is associated to the weighted Dynkin diagram \DEVIII10000000. The corresponding graded decomposition is the five-graded decomposition of 
$\mathfrak{e}_{8(8)}$ in term of irreducible representations of $Spin(7,7)$ 
\be 
\mathfrak{e}_{8(8)} \cong {\bf 14}^\ord{-2} \oplus {\bf 64}^\ord{-1} \oplus \scal{Ê\mathfrak{gl}_1 \oplus \so(7,7)}^\ord{0} \oplus {\bf 64}^\ord{1} \oplus  {\bf 14}^\ord{2} \ .
\label{DSO14} 
\ee
The grade $2$ component is no longer a singlet, and one should further decompose $Spin(7,7)$ irreducible representations with respect to    $Spin(6,7)$ to obtain a singlet in the ${\bf 14} \cong {\bf 1} \oplus {\bf 13} $.\footnote{A time-like vector would lead to a minimal orbit. Note that the choice of an either time-like or space-like vector is equivalent in this case, whereas it would not be in the case of $E_{8(-24)}$, which has a similar five-graded decomposition in irreducible representations of $Spin(3,11)$. Indeed, there are two corresponding $92$ dimensional nilpotent orbits of $E_{8(-24)}$.}    Again it follows from the five-grading that such a nilpotent element satisfy $\C^3 = 0$ in the adjoint representation. The ${\bf 3875}$ is however nine-graded with respect to  the $Spin(7,7)$ decomposition, and only the fifth power of $\C$ vanishes in this representation.
One can easily identify the relevant embeddings of $Spin^*(16) /\mathds{Z}_2 \subset E_{8(8)}$ such that ${\bf 1}^\ord{2} \subset  \mathfrak{e}_{8(8)} \ominus \so^*(16)$ from the decomposition associated to the weighted Dynkin diagram \DSOXVI00010000 of the orbit,
\be \so^*(16) \cong {\bf 6}^\ord{-2} \oplus ( {\bf 4} \otimes {\bf 8} )_\mathds{R}^\ord{-1} \oplus \scal{Ê\mathfrak{gl}_1 \oplus \so(5,1) \oplus \so(2,6) }^\ord{0}  \oplus  ( {\bf 4} \otimes {\bf 8} )_\mathds{R}^\ord{1} \oplus  {\bf 6}^\ord{2} \ .\ee
It turns out that one has two corresponding $Spin^*(16)$ orbits, associated to time-like and space-like vectors ${\bf e} \in {\bf 8}^\ord{2}$, which isotropy subgroup of $SO(2,6)$ are $SO(1,6)$ and $SO(2,5)$, respectively. The corresponding isotropy subgroups of $Spin^*(16)$ and the one of $E_{8(8)}$ follow trivially from the five-graded decomposition 
\be
\left. \begin{matrix}
\Scal{Spin(5,1) \times Spin(1,6) }
\ltimes \scal{Ê( {\bf 4} \otimes {\bf 8} )_\mathds{R}^\ord{1} \oplus  {\bf 6}^\ord{2} }
 \\
\Scal{Spin(1,5) \times Spin(5,2) } \ltimes \scal{Ê( {\bf 4} \otimes {\bf 8} )_\mathds{R}^\ord{1} \oplus  {\bf 6}^\ord{2} }
\end{matrix}
\right\}
\subset Spin(6,7) \ltimes \scal{Ê{\bf 64}^\ord{1} \oplus  \scal{{\bf 1} \oplus Ê{\bf 13}}^\ord{2} } 
\ee
The first line corresponds to regular $1/4$-BPS black holes, while the second corresponds to
solutions with the two non-saturated central charges higher than 
the mass \ie $|z_\un| = |z_\deux | = |\w| < |z_\trois| = |z_\quatre|$.

 \subsubsection*{Nilpotent orbit of dimension 112}
 
The  next nilpotent orbit, with weighted Dynkin diagram \DEVIII00000010,  
is slightly more intricate. Its  graded decomposition is the seven-graded decomposition in $SL(2,\mathds{R}) \times E_{6(6)}$ irreducible representations
\be \mathfrak{e}_{8(8)} \cong {\bf 2}^\ord{-3}\oplus  \overline{\bf 27}^\ord{-2} \oplus ( {\bf 2} \otimes {\bf 27 })^\ord{-1} \oplus \scal{Ê\mathfrak{gl}_1 \oplus \mathfrak{sl}_2 \oplus \mathfrak{e}_{6(6)}}^\ord{0} \oplus( {\bf 2} \otimes \overline{\bf 27 })^\ord{1} \oplus {\bf 27 }^\ord{2}  \oplus {\bf 2}^\ord{3} \ .
\label{DE6} 
\ee
To exhibit a singlet in the grade 2 component, one should decompose further into  $F_{4(4)}$
irreducible representations,
\begin{multline}Ê 
\mathfrak{e}_{8(8)} \cong {\bf 2}^\ord{-3}\oplus \scal{Ê{\bf 1} \oplus {\bf 26}} ^\ord{-2} \oplus \scal{Ê {\bf 2}  \oplus {\bf 2} \otimes {\bf 26 }}^\ord{-1} \oplus \scal{Ê\mathfrak{gl}_1 \oplus \mathfrak{sl}_2 \oplus \mathfrak{f}_{4(4)} \oplus {\bf 26} }^\ord{0} \\*\oplus\scal{Ê {\bf 2}  \oplus {\bf 2} \otimes {\bf 26 }}^\ord{1} \oplus \scal{Ê{\bf 1} \oplus {\bf 26}} ^\ord{2}  \oplus {\bf 2}^\ord{3}  
\end{multline}
It follows from the seven-graded decomposition that such a nilpotent element vanishes to the fourth power in the adjoint representation. In order to identify the existence of a corresponding $Spin^*(16)$ orbit inside $\mathfrak{e}_{8(8)} \ominus \so^*(16)$, one should further decompose\footnote{We added indices to the ${\bf 14}_2$ and the ${\bf 14}_3$ representations of $USp(6)$ to differentiate the real $2$-form representation from the pseudo-real $3$-form representation.} 
\be \mathfrak{f}_{4(4)} \cong \mathfrak{su}(2) \oplus \mathfrak{sp}(3) \oplus ( {\bf 2} \otimes {\bf 14}_3 )_\mathds{R} \ ,\qquad
 {\bf 26} \cong ( {\bf 2} \otimes {\bf 6} )_\mathds{R} \oplus {\bf 14}_2 \ .
 \ee
The embedding of $\so^*(16)$ corresponds to the decomposition
\begin{multline}  \so^*(16) \cong {\bf 1}_+^\ord{-3} \oplus ({\bf 2} \otimes {\bf 6})_\mathds{R}^\ord{-2} \oplus \scal{Ê{\bf 1} \oplus ({\bf 2} \otimes {\bf 6})_\mathds{R} \oplus {\bf 14}_2 }_+^\ord{-1} \oplus \scal{Ê\mathfrak{gl}_1 \oplus \mathfrak{gl}_1 \oplus  \mathfrak{su}(2) \oplus \mathfrak{sp}(3) \oplus {\bf 14}_2 }^\ord{0} \\*   \oplus \scal{Ê{\bf 1} \oplus  ({\bf 2} \otimes {\bf 6})_\mathds{R} \oplus {\bf 14}_2 }_-^\ord{1} \oplus ({\bf 2} \otimes {\bf 6})_\mathds{R}^\ord{2}  \oplus {\bf 1}_-^\ord{3}  \label{SOSp}\end{multline}
where the indices $\pm$ denote the eigenvalue $\pm$ under 
 $\mathfrak{gl}_1 \subset \mathfrak{sl}_2$. This decomposition can be read off
 from the weighted Dynkin diagram  \DSOXVI01000010 . These decompositions enable us
 to compute the  isotropy subgroups of $Spin^*(16)$ and $E_{8(8)}$ as
\begin{multline}  \scal{ÊGL_+(1,\mathds{R}) \times SU(2) \times USp(6) } \ltimes \Scal{Ê\scal{ ({\bf 2} \otimes {\bf 6})_\mathds{R} \oplus {\bf 14}_2 }^\ord{1} \oplus ({\bf 2} \otimes {\bf 6})_\mathds{R}^\ord{2}Ê\oplus {\bf 1}^\ord{3} } \\* \subset \scal{ÊSL(2,\mathds{R}) \times F_{4(4)} } \ltimes \Scal{ ( {\bf 2} \otimes {\bf 26} )^\ord{1} \oplus \scal{Ê{\bf 1} \oplus {\bf 26} }^\ord{2} \oplus {\bf 2}^\ord{3} } \end{multline}Ê
Nevertheless the decomposition (\ref{SOSp}) is not unique, for instance one has a similar decomposition by decomposing $\mathfrak{su}^*(6) \cong \mathfrak{sp}(1,2) \oplus {\bf 14}_2 $ rather than $\mathfrak{su}^*(6) \cong \mathfrak{sp}(3) \oplus {\bf 14}_2 $. The resulting orbit has as isotropy subgroup of $Spin^*(16)$
\be \scal{ÊGL_+(1,\mathds{R}) \times SU(2) \times USp(2,4) } \ltimes \Scal{Ê\scal{ ({\bf 2} \otimes {\bf 6})_\mathds{R} \oplus {\bf 14}_2 }^\ord{1} \oplus ({\bf 2} \otimes {\bf 6})_\mathds{R}^\ord{2}Ê\oplus {\bf 1}^\ord{3} } \ee
and correspond to black holes for which some of the central charges are larger than the mass.

\subsubsection*{Nilpotent orbits of dimension 114}

There are two $114$-dimensional orbits which lie in the same complex orbit of $E_8$ associated to the weighted Dynkin diagram \DEVIII00000002. They are associated to the five-graded decomposition (\ref{DEVII}), however the weight $2$ indicates that the five grading must be considered as an even nine grading, or equivalently that the corresponding nilpotent elements lie inside the grade $1$ component with respect to  the five-grading. Because the ${\bf 3875}$ representation decomposes as well into a five-graded decomposition, the representative of these orbit is nilpotent of degree five in both the adjoint and the ${\bf 3875}$ representation. It turns out that the two real orbits are associated to the 
decomposition of $E_{7(7)}$ under the two real forms $E_{6(2)}$ and $E_{6(6)}$ of $E_6$, 
respectively. In the first case, \footnote{Note that the complex representation ${\bf 27}$ is sometimes written as ${\bf 27} \oplus \overline{\bf 27}$ in the literature.}   
\be \mathfrak{e}_{8(8)} \cong  i \mathds{R}^\ord{-2} \oplus \scal{ \mathds{C} \oplus {\bf 27}}^\ord{-1} \oplus \scal{ \mathds{C} \oplus \mathfrak{e}_{6(2)} \oplus {\bf 27} }^\ord{0} \oplus \scal{ \mathds{C} \oplus {\bf 27}}^\ord{1} \oplus i \mathds{R}^\ord{2} \ . \ee
The nilpotent element can be chosen in the real component of $\mathds{C}^\ord{1}$,
therefore its nilpotency degree  in the adjoint representation is 5.
In order to understand the relevant embedding of $Spin^*(16)$ one should further  decompose 
\be 
\mathfrak{e}_{6(2)} \cong \mathfrak{su}(2) \oplus \mathfrak{su}(6) \oplus ( {\bf 2} \otimes {\bf 20} )_\mathds{R} \ ,\qquad
{\bf 27} \cong ({\bf 2} \otimes {\bf 6}) \oplus {\bf 15} 
\ee
and $\so^*(16)$ into 
\be  \so^*(16) \cong  i \mathds{R}^\ord{-2} \oplus ({\bf 2} \otimes {\bf 6})^\ord{-1} \oplus \scal{ \mathds{C} \oplus \mathfrak{su}(2) \oplus \mathfrak{su}(6)  \oplus {\bf 15} }^\ord{0} \oplus ({\bf 2} \otimes {\bf 6})^\ord{1} \oplus i \mathds{R}^\ord{2} \ , 
\ee
as suggested by the weighted Dynkin diagram \DSOXVI02000000. 
This way one determines the isotropy subgroups inside $Spin^*(16)$ and $E_{8(8)}$ respectively
\be 
\scal{ÊSU(2) \times SU(6)} \ltimes \scal{Ê({\bf 2} \otimes {\bf 6})^\ord{1} \oplus i\mathds{R}^\ord{2}} \subset E_{6(2)} \ltimes \scal{ ( \mathds{R} \oplus {\bf 27})^\ord{1} \oplus i \mathds{R}^\ord{2}} \ .
\ee
It is also interesting to consider the isotropy subgroup inside 
the $E_{7(7)}$ subgroup of $E_{8(8)}$ which intersects with $Spin^*(16)$ on an $SU(8)$ subgroup.
From  the embedding 
\be \mathfrak{e}_{7(7)} \cong i \mathds{R}^\ord{2} - i \mathds{R}^\ord{-2} \oplus \mathfrak{e}_{6(2)} \oplus {\bf 27}^\ord{1} - {\bf 27}^\ord{-1} \ee
one finds that this isotropy subgroup is $E_{6(2)}$. This implies 
that $\lozenge(\w^{-\frac{1}{2}} Z) > 0$ on the corresponding $Spin^*(16)$ orbit. 

The other nilpotent orbit of dimension 114 is associated to the decomposition
\begin{multline} \mathfrak{e}_{8(8)} \cong {\bf 1}^\ord{-2} \oplus \scal{ {\bf 1}^\ord{-3} \oplus {\bf 27}^\ord{-1} \oplus  \overline{\bf 27}^\ord{1} \oplus {\bf 1}^\ord{3} }^\ord{-1} \oplus \scal{ \mathfrak{gl}_1 \oplus \overline{{\bf 27}}^\ord{-2}  \oplus ( \mathfrak{gl}_1  \oplus \mathfrak{e}_{6(6)} )^\ord{0} \oplus  {\bf 27}^\ord{2}  }^\ord{0} \\* \oplus  \scal{ {\bf 1}^\ord{-3} \oplus {\bf 27}^\ord{-1} \oplus  \overline{\bf 27}^\ord{1} \oplus {\bf 1}^\ord{3}}^\ord{1} \oplus {\bf 1}^\ord{2}  \end{multline}
where the internal grading is associated to the three-graded decomposition of $\mathfrak{e}_{7(7)}$ into irreducible representations of $\mathfrak{e}_{6(6)}$. It follows 
 that the nilpotency degree of an element of ${\bf 1}^\gra{1}{3}$  in the adjoint representation is 3.
  In order to get a nilpotent element of degree 5  in the adjoint representation, one could  consider an element in $ {\bf 1}^\gra{1}{-3} + {\bf 1}^\gra{1}{3}$. In order to ensure that
  this element lies  in $\mathfrak{e}_{8(8)} \ominus \so^*(16)$, we have to further  
  decompose $\mathfrak{e}_{6(6)} \cong \mathfrak{sp}(4) \oplus {\bf 42}$ and consider 
  the embedding of $\so^*(16)$ associated to its decomposition \DSOXVI00000020
\begin{multline} \so^*(16) \cong \scal{Ê{\bf 1}^\ord{-3} - {\bf 1}^\ord{3} \oplus {\bf 27}^\ord{-1} - {\bf 27}^\ord{1}}^\ord{-1} \oplus \scal{ \mathfrak{gl}_1 \oplus \mathfrak{sp}(4)^\ord{0} \oplus {\bf 27}^\ord{-2} + {\bf 27}^\ord{2} }^\ord{0} \\* \oplus  \scal{Ê{\bf 1}^\ord{-3} - {\bf 1}^\ord{3} \oplus {\bf 27}^\ord{-1} - {\bf 27}^\ord{1}}^\ord{1} \ .\end{multline}
These decompositions enable us 
to compute the corresponding isotropy subgroups inside $Spin^*(16)$ and $E_{8(8)}$,
\be USp(8)\ltimes {\bf 27}^\ord{1} \subset E_{6(6)} \ltimes \scal{Ê ( { \bf 1}Ê\oplus {\bf 27} \oplus \overline{\bf 27})^\ord{1} \oplus {\bf 1}^\ord{2} } \ee 
The isotropy subgroup inside the $E_{7(7)}$ subgroup of $E_{8(8)}$ which intersects with $Spin^*(16)$ on an $SU(8)$ subgroup is $E_{6(6)}$, as follows from the decomposition 
\be \mathfrak{e}_{7(7)} \cong \overline{\bf 27}^\gra{-1}{1} -  \overline{\bf 27}^\gra{1}{1} \oplus {\bf 1}^\gra{-2}{0} - 2 {\bf 1}^\gra{0}{0} + {\bf 1}^\gra{2}{0} \oplus \mathfrak{e}_{6(6)} \oplus {\bf 27}^\gra{-1}{-1} - {\bf 27}^\gra{1}{-1} \ .\ee
This implies that $\lozenge(\w^{-\frac{1}{2}} Z) < 0$ on the corresponding $Spin^*(16)$ orbit.

Using similar techniques for different decompositions of $\so^*(16)$, one 
gets additional $Spin^*(16)$ orbits with the following isotropy subgroups 
\bea
\left.
\begin{matrix}
USp(2,6)\ltimes {\bf 27}^\ord{1} \\
\scal{ÊSU(2) \times SU(2,4)} \ltimes \scal{Ê({\bf 2} \otimes {\bf 6})^\ord{1} \oplus i\mathds{R}^\ord{2}} 
\end{matrix}\right\} &\subset& E_{6(2)} \ltimes \scal{ ( \mathds{R} \oplus {\bf 27})^\ord{1} \oplus i \mathds{R}^\ord{2}} \CR
\left.
\begin{matrix}
USp(4,4)\ltimes {\bf 27}^\ord{1} \\
\scal{ÊSU(2) \times SU^*(6)} \ltimes \scal{Ê({\bf 2} \otimes {\bf 6})^\ord{1} \oplus i\mathds{R}^\ord{2}} 
\end{matrix}\right\} 
 &\subset& E_{6(6)} \ltimes \scal{Ê ( { \bf 1}Ê\oplus {\bf 27} \oplus \overline{\bf 27})^\ord{1} \oplus {\bf 1}^\ord{2} } \CR
\eea
all of them corresponding to black holes with naked singularities.

\subsection{Nilpotent orbits of $SO(4,4)$}

For $G_\IC=SO(N,\IC)$, complex orbits can be conveniently labeled by Young tableaux,
or partitions of $N$, which summarize the decomposition of the $N$-dimensional vector
representation of $G_\IC$ into finite dimensional representations of $SL(2,\IC)$, or
equivalently the Jordan normal form of the nilpotent element ${\bf e}$ in the vector representation. 
These Young tableaux must be such that lines with even length (equivalently, representations with half integer spin) occur always in pair. Young tableaux with only rows of even length correspond
to two different orbits. 
The closure ordering ${\bf e}\leq {\bf e}'$ holds whenever, for all $p=1\dots N$,
the number of boxes in the first $p$ columns of the Young tableau of ${\bf e}$ is less than
the number of boxes in the first $p$ columns of the Young tableau of ${\bf e}'$.
The nilpotent orbits of $SO(8,\IC)$ are summarized in the 
following Hasse diagram,
\be 
 \includegraphics[width=12cm]{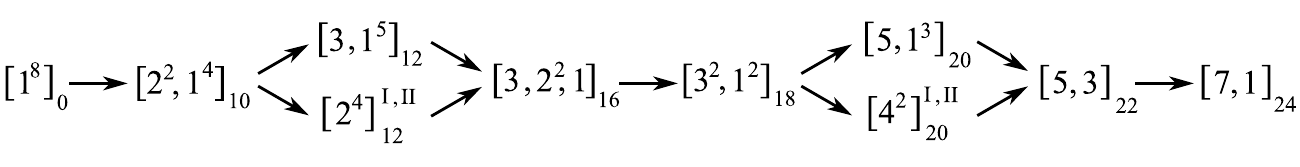}
      \label{hasseSO8}
\ee
where the integers inside the bracket indicate the size of the Jordan blocks 
in the 8-dimensional vector representation, with multiplicity 
given in superscript. The subscript indicates the complex dimension
of the orbit. The triplets of degenerate orbits $[3,1^5],[2^4]^{I,II}$ 
are permuted by triality, as can be easily seen from their
Dynkin labels, \DSOVIII2000, \DSOVIII0200, \DSOVIII0020. Similarly, 
$[5,1^3],[4^2]^{I,II}$, with Dynkin labels \DSOVIII2002, \DSOVIII0202, \DSOVIII0022, are permuted by triality. The remaining orbits,  with Dynkin labels \DSOVIII0001, \DSOVIII1110, \DSOVIII0002, \DSOVIII2220, \DSOVIII2222  are manifestly triality invariant.

The nilpotent orbits $\cO_\IR({\bf e})$ under the real group\footnote{The subscript 0
denotes the component connected to the identity. Recall that $O(p,q,\IR)$ has four connected
components when $pq>0$. The connected component of the identity $SO_\asym(p,q,\IR)$ is homotopy equivalent to $SO(p,\IR)\times SO(q,\IR)$. } $G=SO_\asym(p,q,\IR)$ can be obtained by attaching 
$\pm$ signs to each of the box of the Young tableau, alternating along each row,
such that the total number of $(+,-)$ signs is $(p,q)$ and moreover, such that every
row of even length starts with $+$. A given signed Young tableau
corresponds to 4 different orbits labelled $(I,I), (I,II), (II,I), (II,II)$ when all rows have
even length, 2 different orbits labelled $I,II$ when all rows with odd length have an
even number of $+$ signs, 2 different orbits labelled $I,II$ when all rows with odd length have an
even number of $-$ signs, or else a single orbit. All orbits with the 
same signed Young tableau are related by $O(p,q,\IR)$. The closure ordering 
involves not only the shape of the Young tableaux but also the sign and latin 
letter assignments, and is discussed in detail in \cite{DokovicSO}. For $G=SO_\asym(4,4)$,
the  nilpotent orbits of real dimension 18 and lower are summarized in 
Figure \ref{hasseSO44}.

\FIGURE{\includegraphics[height=18cm]{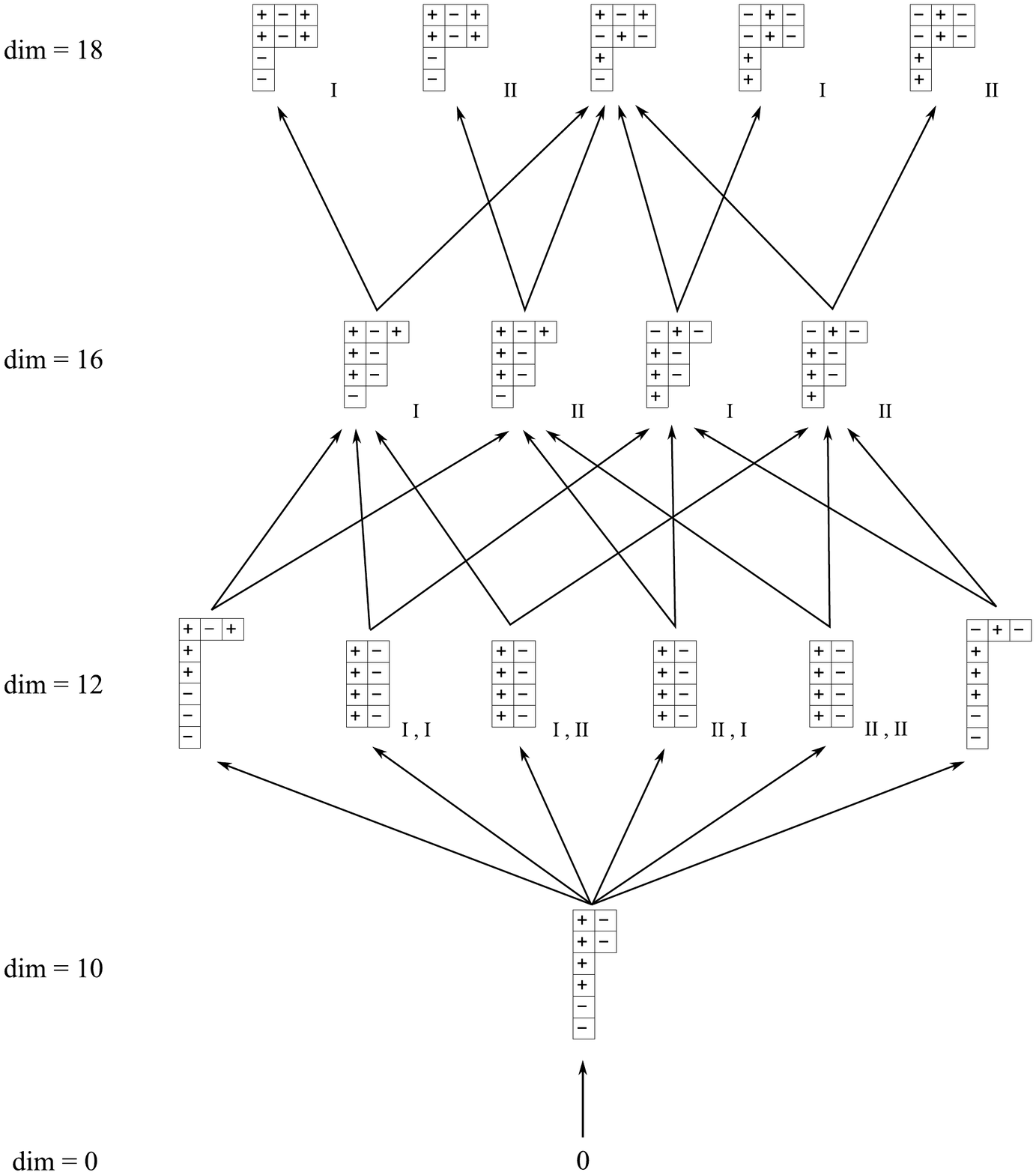}
\caption{Hasse diagram of nilpotent orbits of $SO_\asym(4,4)$ \label{hasseSO44}}}


Starting from the bottom, the smallest (non-zero) orbit is the minimal orbit, 
or real dimension 10, attached to the Young tableau $[(+-)^2]$ (we omit 
the rows of length one, for brevity). It is associated to the 5-grading 
\be
\label{gradmin}
\so(4,4) \cong {\bf 1}^\ord{-2} \oplus \scal{Ê{\bf 2} \otimes {\bf 4} }^\ord{-1} \oplus \scal{Ê\gl_1 \oplus \sl_2 \oplus \so(2,2) }^\ord{0} \oplus  \scal{Ê{\bf 2} \otimes {\bf 4} }^\ord{1} \oplus {\bf 1}^\ord{2} \ ,
\ee
where $\fg^\ord{2}=\mathds{R} {\bf e}, \, \fg^\ord{-2}=\mathds{R} {\bf f}$ and ${\bf h}$ is the singlet in $\fg^\ord{0}$. This 5-grading
originates from the branching $SO(4,4)\supset SO(2,2)\times SO(2,2)
\sim SL(2)\times SL(2)\times SO(2,2)$, where ${\bf h}$ is the non-compact 
Cartan generator of the first factor. The stabilizer of ${\bf e}$ is 
$\so(2,2,\IR)^\ord{0}\oplus \fg^\ord{1}\oplus\fg^\ord{2}$, therefore the 
real minimal orbit is isomorphic to 
\be
\cO_{\rm min} = \frac{ SO_\asym(4,4)}
{(SL(2,\IR)\times SO(2,2,\IR)) \ltimes \bigl[  (Ê{\bf 2} \otimes {\bf 4} )^\ord{1} \oplus {\bf 1}^\ord{2} \bigr]}
\ee
Since the Young tableau $[(+-)^2(+)^2(-)^2]$ corresponds to a nilpotent element of 
order 2 in the vector representation, and is invariant under triality, the minimal orbit 
is characterized by the nilpotency conditions $V^2=S^2=C^2=A^3=0$, where
$V,S,C,A$ are a short-hand notation for the representations of ${\bf e}$ in the 
vector, spinor, conjugate spinor and adjoint representations.

The next set of nilpotent orbits, of real dimension 12, corresponds to three distinct
unsigned Young tableaux, $[3,1^5]$ and $[3^4]$. These are usually of distinct
dimension $2N-4$ and $4N-20$, but the dimensions and stabilizers happen to 
coincide when $N=8$. 
The tableaux $[(\pm\mp\pm)]$ 
correspond to the 3-grading (or more accurately, ``even'' 5-grading)
\be
\so(4,4) \cong {\bf 6}^\ord{-2}Ê\oplus \scal{Ê\gl_1 \oplus \so(3,3) }^\ord{0} \oplus {\bf 6}^\ord{2} 
\ee
which arises from the branching $SO(4,4)\supset SO(1,1)\times SO(3,3)$.
The stabilizer in $SO(3,3)$  of a non-zero vector ${\bf e}\in {\bf 6}^\ord{2}$ is $SO(3,2)$ if $||{\bf e}||^2>0$ 
for the signature $(3,3)$ quadratic norm, $SO(2,3)$ if $||{\bf e}||^2<0$, or $SO(2,2)\ltimes \IR^4$
if $||{\bf e}||^2=0$. The last case returns to the minimal orbit discussed above, while the
first two cases are related by parity. Thus, the 12-dimensional nilpotent 
orbits  $[(+-+)]$ and  $[(-+-)]$ are  isomorphic to
\be
\cO_{12} = \frac{ SO_\asym(4,4)}
{ SO(2,3,\IR) \ltimes [  ({\bf 5}\oplus {\bf 1})^\ord{2} ]} \ .
\ee
On the other hand, the Young tableau $[(+-)^4]$ corresponds to the isomorphic 3-grading
\be
\so(4,4) \cong {\bf 6}^\ord{-2}Ê\oplus \scal{Ê\gl_1 \oplus \sl_4 }^\ord{0} \oplus {\bf 6}^\ord{2} 
\ee
associated to the branching $SO(4,4) \supset SO(1,1) \times SL(4,\IR)$. 
The tableau $[(+-)^4]$ is ``very even", therefore corresponds
to 4 different orbits of $SO_\asym(4,4)$ (but a single orbit of $O(4,4)$).
The stabilizer in $SL(4,\IR)$ of a generic element ${\bf e}\in {\bf 6}^\ord{2}$ is $Sp(4,\IR)$, therefore
the 12-dimensional nilpotent orbits $[(+-)^4]$ are isomorphic to
\be
\cO_{12^\prime} = \frac{ SO_\asym(4,4)}
{ Sp(4,\IR) \ltimes [  ({\bf 5}\oplus {\bf 1})^\ord{2} ]} \ .
\ee
Like the minimal orbit, the dimension 12 nilpotent orbits satisfy $A^3=0$, but have
a higher degree of nilpotency in the 8-dimensional representations. For this purpose,
it is useful to note that the unsigned Young tableaux $[(3)],[(2)^4],[(2)^4]$ are exchanged
by triality. Thus, we conclude that  
$V^2=S^3=C^2=0$ for $[(+-)^4]_{I,I}$ and $[(+-)^4]_{I,II}$, $V^2=S^2=C^3=0$ for $[(+-)^4]_{II,I}$ and $[(+-)^4]_{II,II}$, and $S^2=V^3=C^2=0$ for both $[(+-+)]$ and $[(-+-)]$. Moreover, $[(+-)^4]_{I,I}$, 
$[(+-)^4]_{II,I}$  and $[(+-+)]$ are permuted amongst each other under triality, while 
$[(+-)^4]_{I,II}$, $[(+-)^4]_{II,II}$ and $[(-+-)]$ form a second triality orbit. 
The distinction between the two triplets of orbits corresponds to choosing the representative in the vector representation of $SO(3,3)$ to satisfy either $||{\bf e}||^2>0$ or $||{\bf e}||^2<0$.

The next level of nilpotent orbits, of real dimension 16, corresponds to
two signed Young tableaux $[(+-+)(+-)^2]$ and $[(-+-)(+-)^2]$. They correspond to 
the 7-grading
\begin{multline}
\so(4,4) \cong {\bf 2}^\ord{-3} \oplus \scal{{\bf 1}_\gra{-2}{0}Ê\oplus {\bf 1}_\gra{1}{-1} \oplus {\bf 1}_\gra{1}{1} }^\ord{-2} \oplus \scal{Ê{\bf 2}_\gra{2}{0}Ê\oplus {\bf 2}_\gra{-1}{1} \oplus {\bf 2}_\gra{-1}{-1}}^\ord{-1} \\* \oplus \scal{Ê\gl_1 \oplus \gl_1 \oplus \gl_1 \oplus \sl_2 }^\ord{0} \oplus   \scal{Ê{\bf 2}_\gra{-2}{0}Ê\oplus {\bf 2}_\gra{1}{-1} \oplus {\bf 2}_\gra{1}{1}}^\ord{1}  \oplus \scal{{\bf 1}_\gra{2}{0}Ê\oplus {\bf 1}_\gra{-1}{1}  \oplus {\bf 1}_\gra{-1}{-1}}^\ord{2} \oplus {\bf 2}^\ord{3} 
\end{multline}
where $\bf 2$ denotes the two-dimensional representation of $\mathfrak{sl}_2$, and the subscripts the weights with respect to the two extra $\gl_1$ subalgebras. 
 This originates again from the branching $SO(4,4)\supset [SL(2)]^4$, with 
 a suitable choice of the non-compact Cartan generator ${\bf h}$ inside $ [SL(2)]^3$.
 The stabilizer of a generic element ${\bf e}$ in the grade 2 component (\ie which carries a non-zero element in the three components ${\bf 1}_\gra{2}{0}$, ${\bf 1}_\gra{-1}{1} $ and ${\bf 1}_\gra{-1}{-1}$)  is 
 $ \mathfrak{sl}_2{}^\ord{0}  
\oplus (2\times {\bf 2})^\ord{1} \oplus (3 \times {\bf 1})^\ord{2} \oplus {\bf 2}^\ord{3}$
therefore
 \be
\cO_{16} = \frac{ SO_\asym(4,4)}
{SL(2,\IR)\ltimes \bigl[  (2\times {\bf 2})^\ord{1} \oplus (3 \times {\bf 1})^\ord{2} \oplus {\bf 2}^\ord{3} \bigr]} \ .
\ee
Due to the 7-grading, elements in this orbit satisfy $A^4=0$; moreover, since
the Young tableau $[(+-+)(+-)^2]$ is invariant under triality, they also satisfy
$V^3=S^3=C^3=0$.

Finally, the nilpotent orbits of real dimension 18 correspond three signed Young tableaux,
all associated to the same 5-grading (rather, even 9-grading)
\be
\so(4,4) \cong {\bf 1}^\ord{-4} \oplus \scal{Ê{\bf 2} \otimes {\bf 4} }^\ord{-2} \oplus \scal{Ê\gl_1 \oplus \sl_2 \oplus \so(2,2) }^\ord{0} \oplus  \scal{Ê{\bf 2} \otimes {\bf 4} }^\ord{2} \oplus {\bf 1}^\ord{4} \ ,
\ee
identical to \eqref{gradmin} save for the normalization of the non-compact Cartan 
generator ${\bf h}$. In particular, ${\bf e}\in\fg^\ord{2}$ now transforms  
as a doublet of vectors of $SO(2,2,\IR)$. If the norm of the two vectors is
of the same sign, the stabilizer in  $SO(2,2,\IR)$ is $SO(2)$. The broken
$SO(2)$ can be combined with $SO(2)\subset SL(2)$, leading to 
\be
\cO_{18,+} = \frac{ SO_\asym(4,4)}
{SO(2,\IR)\times SO(2,\IR) \ltimes [ ( {\bf 2} Ê\oplus {\bf 2} \otimes {\bf 2}Ê)^\ord{2} \oplus  {\bf 1}^\ord{4} ]} 
\ee
If instead the two vectors have norms of opposite sign,  the stabilizer in  $SO(2,2,\IR)$ is $SO(1,1)$,
leading instead to 
\be
\cO_{18,-} = \frac{ SO_\asym(4,4)}
{SO(1,1,\IR)\times SO(1,1,\IR) \ltimes [ ( {\bf 2} Ê\oplus {\bf 2} \otimes {\bf 2}Ê)^\ord{2} \oplus  {\bf 1}^\ord{4}  ]} 
\ee
The second one is invariant under parity, and so should correspond to the Young tableau
associated to $[(+-+)(-+-)]$. The two other signed Young tableaux label two different
orbits of $SO_\asym(4,4)$ each, labelled I,II, which are identified under $O(4,4)$.  All 5 nilpotent orbits of dimension 18 satisfy $A^5=0$ (due to the even 9-grading) 
and $V^3=S^3=C^3=0$ (since $V^3=0$, obviously, 
and the Young tableau is invariant under triality). The condition 
$S^3=0$ (or equivalently $C^3=0$) is the condition of interest for extremal black holes.

In fact, from the Hasse diagram
of the complex nilpotent orbits \eqref{hasseSO8}, it follows that the 18 dimensional
orbits covers all nilpotent elements such that $A^5=0$. Higher dimensional orbits correspond
to elements with a higher order of nilpotency in the adjoint representation: from the
Young tableaux in \eqref{hasseSO8}, it is easy to see that the three 
20-dimensional complex orbits satisfy $A^7=V^5=S^4=C^4=0$ and triality
images thereof, that the 22-dimensional orbits satisfies $A^7=V^5=S^5=C^5=0$,
and that the 24-dimensional orbit satisfies $A^{11}=V^7=S^7=C^7=0$.

Let us briefly discuss the fate of these orbits under the two-step restriction 
\be
SO_\asym(4,4)\supset SO_\asym(4,3)\supset G_{2(2)}\ ,
\ee 
which corresponds
to the truncation of the $STU$ model to the $ST^2$ and $S^3$ models.
In the first step $SO_\asym(4,4)\supset SO_\asym(4,3)$, the orbits associated to $[(-+-)^2]$, $[(-+-)(+-)^2]$, 
$[(+-)^4]$ are lost, leaving 3 orbits $[(+-+)^2]_{I,II}$, $[(+-+)(-+-)]$ of dimension 14 and degree 5, 
2 orbits $[(+-+)(+-)^2]_{I,II}$ of 
dimension 12 and degree 4, 2 orbits $[(+-+)],[(-+-)]$ of dimension 10 and degree 3, and 
the minimal orbit $[(+-)^2]$ of dimension 8. In the second step $SO_\asym(4,3)\supset G_{2(2)}$, 
the degree 5 orbits reduce to two inequivalent orbits of dimension 10 (coming
from $[(+-+)^2]_{I,II}$ and $[(-+-)(+-+)]$, respectively), the degree 4 orbits to a single orbit 
of dimension 8, and the only remaining orbit of degree 3 is the minimal orbit
of dimension 6.

\section{Solving the non-BPS diagonalization problem \label{AppW}}

The fake superpotential for extremal non-BPS black holes is given by \eqref{WNB},
where $\varrho$ is proportional to the component of the central charge in the parametrization
\eqref{Zdiag}. In this section, we explain how to compute $W$ from the $SU(8)$ invariants
$\rho_\zero, \rho_\un, \rho_\deux, \rho_\trois$ and $\varphi$. The latter can be computed
from the central charge $Z_{ij}$   
 as explained below \eqref{Zdecompose}.

\subsection{General case}

To compute $\varrho$, let us define four real positive  variables ${\rm x}_\n$ as
\be 
{\rm x}_\n =  \frac12 \left\{  \begin{array}{lcl}  \cos 2 \alpha + \frac{\xi_\n}{\varrho} 
 \hspace{10mm}Ê&\mbox{for}& \hspace{2mm} \n = \un,\,  \deux,\, \trois  \vspace{2mm} \\
 \cos 2 \alpha - \frac{\xi_\un+\xi_\deux+\xi_\trois}{\varrho}   \hspace{10mm}Ê&\mbox{for} & \hspace{2mm} \n= \zero \end{array} \right. \ ,
 \ee
such that $\sum {\rm x}_\m = 2 \cos 2 \alpha$. Here and below, summations or products over $\m$
will always range over $\zero, \un,\,  \deux,\, \trois$.
Using \eqref{lozengexi}, one may rewrite \eqref{WNB} as 
\be W = \frac{1}{2} 
\left( \frac{ \bigl . - \lozenge( Z) \bigr . }{ \cos^2(2\alpha) \cdot
 \prod {{\rm x}_\m}  }\right)^{1/4} \ . \label{Wxn}
\ee
In order to determine the variables ${\rm x}_\n$, we equate
\eqref{Zdecompose} and (\ref{Zdiag}) (keeping in mind that the rotations
$R_i{}^j$ and $\tilde R_i{}^j$ are distinct for the two decompositions), obtaining
\be
{\rho_\n}^2 = W^2 \, \left | {{\rm x}_\n}  - e^{2 i \alpha}   \right |^2 \CR
= W^2  \Scal{Ê\scal{Ê{{\rm x}_\n} - \cos( 2 \alpha)    }^2 +  \sin^2 (2 \alpha) } \ .
\ee
which gives the algebraic equation 
\be  
{\rm x}_\n^2  -2  \,  {{\rm x}_\n}  \cos (2\alpha)  
+ \left( 1 - \frac{{\rho_\n}^2}{W^2} \right)= 0  \ .
\label{polyW} 
\ee
The positive solution of this quadratic equation is 
\be
{\rm x}_\n = \cos(2\alpha) - \sqrt{\scal{  {\rho_\n}/W}Ê^2 - \sin^2(2\alpha)}\ .
\ee
Using moreover (\ref{Wxn}), we obtain that $W$ and $\cos(2\alpha)$ are determined by the two equations
\bea \sum_\n \sqrt{\scal{Ê {\rho_\n}/W }^2 - \sin^2(2\alpha)} &=& \cos^2(2\alpha)\ , \CR
Ê\prod_\n \left( \cos(2\alpha) - \sqrt{ \scal{Ê{\rho_\n}/W }^2 - \sin^2(2\alpha)} \right) &=&  - \frac{\lozenge(Z)}{16 \cos^2(2\alpha) W^4}\ . \label{W2eq} 
\eea

In order to reduce these equations, we denote by $\sigma_n(r)$ 
the elementary symmetric functions of $r_\n\equiv \sqrt{ \scal{Ê{\rho_\n}/W }^2 - \sin^2(2\alpha)}$,
\be
\sigma_1(r) = \sum r_\m\ ,\qquad
\sigma_2(r) = \sum_{\m<\n} r_\m r_\n\ ,\qquad
\sigma_3(r)= \sum_{\m<\n<\p} r_\m r_\n r_\p\ ,\qquad
\sigma_4(r) = \prod r_\m \ ,
\ee
and by $\Sigma(r)$ the elementary symmetric functions of $r_\n^2$, i.e.
$\Sigma_i(r_\n)=\sigma_i(r_\n^2)$. The two sets of symmetric functions are related by
\be
\label{PR}
\Sigma_1 = \sigma_1^2 - 2 \sigma_2\ ,\qquad
 \Sigma_2 = \sigma_2^2 + 2 \sigma_4-2 \sigma_1 \sigma_3\ ,\qquad
 \Sigma_3 = \sigma_3^2 - 2 \sigma_2 \sigma_4\ ,\qquad
 \Sigma_4 = \sigma_4^2 \ .
\ee
This can be inverted by eliminating $\sigma_1$ and $\sigma_3$ between these
equations, leading to a quartic equation in $\sigma_2$,
\be
\Scal{\sigma_2^2 + 2 \sqrt{\Sigma_4} - \Sigma_2}^2 - 4 
\Scal{\Sigma_3+2 \sigma_2 \sqrt{\Sigma_4}} \Scal{\Sigma_1+2\sigma_2} = 0\ .
\ee
Now, since $\sigma_1(r)=2\cos 2\alpha$, the first equation in \eqref{W2eq} implies that
\be
\sigma_2(r) = 2 -\frac12 \sum \frac{\rho_\m^2}{W^2}\ .
\ee
The second equation in \eqref{PR} then leads to
\be
2 \sigma_3(r) \cos2\alpha  - \sigma_4(r) 
= 2 - 3 \sin^4 2\alpha - \scal{ 1 - \frac32 \sin^2 2\alpha }
\sum \frac{\rho_\m^2 }{W^2} + \frac{1}{8W^4} 
\biggl( \sum \rho_\m^4 - 2 \sum_{\n>\m} {\rho_\n}^2 \rho_\m^2 \biggr) 
\ee
Substituting back into \eqref{W2eq}, we obtain
\be
\sigma_4(r) = - \sin^4 2\alpha + \frac12\sin^2 2\alpha \sum_\n \frac{{\rho_\n}^2}{W^2} 
+ \frac{1}{8W^4} \biggl(  \sum {\rho_\m}^4 - 2\sum_{\n>\m} {\rho_\n}^2 {\rho_\m}^2 
- \frac{\lozenge}{\cos^2 2\alpha}   \biggr) \ .\label{Rpoly}
\ee
Squaring both sides, we obtain a polynomial of degree two in $W^2$ 
\begin{multline}
\left[ \lozenge - \biggl( \sum {\rho_\m}^4 - 2 \sum_{\n>\m} {\rho_\n}^2 {\rho_\m}^2 \biggr)
\cos^2 2\alpha \right]^2 
- 8^2 \cos^4 2\alpha 
\prod {\rho_\m}^2 \\*
- 2 W^2  \sin^2 4 \alpha 
 \left(  \biggl(\sum {\rho_\m}^2\biggr) \lozenge 
- \biggl( \sum {\rho_\m}^6 - \sum_{\n\ne \m} {\rho_\n}^4 {\rho_\m}^2 
+ 2 \sum_{\m>\n>\p} {\rho_\m}^2 {\rho_\n}^2 {\rho_\p}^2 \biggr)\cos^2 2\alpha \right)   \\
 + 4 W^4 \lozenge  \sin^2 2 \alpha  \sin^2 4\alpha = 0 \label{polP}
\end{multline}
Substituting (\ref{Rpoly}) inside the third equation in (\ref{PR}), we obtain a second polynomial, whose difference with the former gives a polynomial of degree two in $\cos^2 2\alpha$,
\begin{multline} \label{polS}
\left[ \prod \left(1- \frac{{\rho_\m}^2}{W^2}\right) + 
\frac{\lozenge}{2W^4} \left( 1- \frac14 \sum_\m \frac{{\rho_\m}^2}{ W^2} \right) \right] \cos^4 2\alpha \\
 - \frac{\lozenge}{8W^4} \left( 1 - \frac12 \sum_\m \frac{{\rho_\m}^2}{W^2} - \frac18 W^4 
 \biggl( \sum_{\m} {\rho_\n}^4 - 2 \sum_{\n>\m} {\rho_\n}^2 {\rho_\m}^2 \biggr) \right) \cos^2 2\alpha
 - \frac{3\lozenge^2}{16^2W^8}  = 0
\end{multline}
By eliminating $\cos 2\alpha$ between the two equations, we obtain a polynomial
equation of degree 6 in $W^2$, whose coefficients are symmetric polynomials in ${\rho_\m}^2$
and polynomial in $\lozenge$.  It is convenient to write it in terms of $\lozenge$ and the 
elementary symmetric functions $\Sigma_i(\rho_\m)$, since they can all be expressed in 
terms of the $SU(8)$ invariant functions of the central charges $Z_{ij}$ using \eqref{SUpoly}.
Writing $\Sigma_i\equiv \Sigma_i(\rho_\m)$ for brevity, we obtain
\begin{multline} 
 \label{sixDegree}
64 W^{12} \Bigl[-\lozenge^3+\bigl(3 \Sigma_{1}^2-8 \Sigma_{2}\bigr)
   \lozenge^2+\Bigl(-3 \Sigma_{1}^4+16 \Sigma_{2} \Sigma_{1}^2-16 \Sigma_{3}
   \Sigma_{1}-16 \Sigma_{2}^2+64 \Sigma_{4}\Bigr)
   \lozenge \Bigr.  \\*  \hspace{40mm}Ê\Bigl. +\bigl(\Sigma_{1}^3-4 \Sigma_{2} \Sigma_{1}+8
   \Sigma_{3}\bigr)^2\Bigr]  \\
\hspace{-70mm} +16   W^{10} \Bigl[  
   3 \Sigma_{1}
   \lozenge^3+\bigl(-9 \Sigma_{1}^3+20 \Sigma_{2} \Sigma_{1}+24
   \Sigma_{3}\bigr) \lozenge^2 \Bigr.  \\*  \Bigl.  
   +\Bigl(9 \Sigma_{1}^5-40 \Sigma_{2}
   \Sigma_{1}^3+32 \Sigma_{3} \Sigma_{1}^2+16 \bigl(\Sigma_{2}^2-20 \Sigma_{4}\bigr)
   \Sigma_{1}+128 \Sigma_{2} \Sigma_{3}\Bigr) \lozenge  \Bigr.  \\*  \hspace{10mm} \Bigl. 
   -\bigl(\Sigma_{1}^3-4
   \Sigma_{2} \Sigma_{1}+8 \Sigma_{3}\bigr) \Bigl(3 \Sigma_{1}^4-8 \Sigma_{2}
   \Sigma_{1}^2+32 \Sigma_{3} \Sigma_{1}-16 \Sigma_{2}^2+64 \Sigma_{4}\Bigr)\Bigr]
   \\
\hspace{-20mm} + W^8  \Bigl[-15 \Sigma_{1}^8+240 \Sigma_{2} \Sigma_{1}^6-128 \Sigma_{3}
   \Sigma_{1}^5+32 \bigl(28 \Sigma_{4}-37 \Sigma_{2}^2\bigr) \Sigma_{1}^4+2048
   \Sigma_{2} \Sigma_{3} \Sigma_{1}^3  \Bigr.  \\*  \Bigl. 
   +256 \bigl(7 \Sigma_{2}^3-12 \Sigma_{2}
   \Sigma_{4}\bigr) \Sigma_{1}^2+2048 \Sigma_{3} \bigl(4 \Sigma_{4}-3
   \Sigma_{2}^2\bigr) \Sigma_{1}-27 \lozenge^4+96 \lozenge^3
   \bigl(\Sigma_{1}^2-3 \Sigma_{2}\bigr) \Bigr.  \\*  \Bigl. 
   -2 \lozenge^2 \Bigl(63
   \Sigma_{1}^4-408 \Sigma_{2} \Sigma_{1}^2+384 \Sigma_{3} \Sigma_{1}+496
   \Sigma_{2}^2-960 \Sigma_{4}\Bigr) \Bigr.  \\*  \Bigl. 
   +256 \bigl(\Sigma_{2}^4-8 \Sigma_{4}
   \Sigma_{2}^2+16 \Sigma_{3}^2 \Sigma_{2}+16 \Sigma_{4}^2\bigr) 
+8 \lozenge
   \Bigl(9 \Sigma_{1}^6-96 \Sigma_{2} \Sigma_{1}^4+112 \Sigma_{3} \Sigma_{1}^3
    \Bigr.\Bigr.  \\* \hspace{25mm}Ê \Bigl. \Bigl. +16
   \bigl(17 \Sigma_{2}^2+8 \Sigma_{4}\bigr) \Sigma_{1}^2-576 \Sigma_{2} \Sigma_{3}
   \Sigma_{1}-64 \bigl(2 \Sigma_{2}^3-8 \Sigma_{4} \Sigma_{2}-3
   \Sigma_{3}^2\bigr)\Bigr)\Bigr] \\
\hspace{-33mm}  + W^6\ \Bigl[-\Sigma_{1}^9+32 \Sigma_{2}
   \Sigma_{1}^7-64 \Sigma_{3} \Sigma_{1}^6+32 \bigl(4 \Sigma_{4}-9 \Sigma_{2}^2\bigr)
   \Sigma_{1}^5+640 \Sigma_{2} \Sigma_{3} \Sigma_{1}^4 \Bigr.  \\* 
   +1024 \bigl(\Sigma_{2}^3-2
   \Sigma_{4} \Sigma_{2}-\Sigma_{3}^2\bigr) \Sigma_{1}^3 +2048 \Sigma_{3} \bigl(\Sigma_{2}^3 -
   \Sigma_{2}^2  \Sigma_{1}^2 -4 \Sigma_{4} \Sigma_{2}-2
   \Sigma_{3}^2\bigr)  \\*  \hspace{3mm}  -256 \bigl(5 \Sigma_{2}^4-24 \Sigma_{4} \Sigma_{2}^2-16 \Sigma_{3}^2
   \Sigma_{2}+16 \Sigma_{4}^2\bigr) \Sigma_{1}
   +\lozenge^3
   \bigl(\Sigma_{1}^3-36 \Sigma_{2} \Sigma_{1}+216 \Sigma_{3}\bigr)  \\*  \Bigl.  -\lozenge^2
   \Bigl(3 \Sigma_{1}^5-104 \Sigma_{2} \Sigma_{1}^3+336 \Sigma_{3} \Sigma_{1}^2+368
   \Sigma_{2}^2 \Sigma_{1}+960 \Sigma_{4} \Sigma_{1}   -1728 \Sigma_{2}
   \Sigma_{3}\Bigr)  \Bigr.  \\*  \Bigl. 
   +\lozenge \Bigl(3 \Sigma_{1}^7-100 \Sigma_{2}
   \Sigma_{1}^5+184 \Sigma_{3} \Sigma_{1}^4+16 \bigl(41 \Sigma_{2}^2-28
   \Sigma_{4}\bigr) \Sigma_{1}^3-1728 \Sigma_{2} \Sigma_{3} \Sigma_{1}^2 
   \Bigr.\Bigr.  \\*  \Bigl. \Bigl. \hspace{35mm} 
   +64 \bigl(-19
   \Sigma_{2}^3+12 \Sigma_{4} \Sigma_{2}+12 \Sigma_{3}^2\bigr) \Sigma_{1}+128
   \Sigma_{3} \bigl(31 \Sigma_{2}^2-60 \Sigma_{4}\bigr)\Bigr)\Bigr]
   \\
\hspace{-0mm} + W^4 \Bigl[\Bigl(\Sigma_{2} \Sigma_{1}^4-36 \Sigma_{3} \Sigma_{1}^3-8
   \bigl(\Sigma_{2}^2-42 \Sigma_{4}\bigr) \Sigma_{1}^2+144 \Sigma_{2} \Sigma_{3}
   \Sigma_{1}+16 \bigl(\Sigma_{2}^3-36 \Sigma_{4} \Sigma_{2}-27
   \Sigma_{3}^2\bigr)\Bigr) \lozenge^2 \Bigr.  \\*  \Bigl. 
   \hspace{-25mm}Ê+2 \Bigl(3072 \Sigma_{4}^2-16 \bigl(5
   \Sigma_{1}^4-36 \Sigma_{2} \Sigma_{1}^2-24 \Sigma_{3} \Sigma_{1}+64
   \Sigma_{2}^2\bigr) \Sigma_{4} \Bigr. \Bigr. \\*  \Bigl.\Bigl. 
   \hspace{20mm} +\bigl(\Sigma_{1}^2-4 \Sigma_{2}\bigr)
   \bigl(-\Sigma_{2}  (\Sigma_{1}^2-4 \Sigma_{2} )^2+26 \Sigma_{1} \Sigma_{3}
    (\Sigma_{1}^2-4 \Sigma_{2} )+288 \Sigma_{3}^2\bigr)\Bigr)
   \lozenge \Bigr.  \\*  \Bigl. 
   \hspace{5mm} +4096 \Sigma_{2} \Sigma_{4}^2+\bigl(\Sigma_{1}^2-4
   \Sigma_{2}\bigr)^2 \Bigl(\Sigma_{2} \bigl(\Sigma_{1}^2-4 \Sigma_{2}\bigr)^2-16
   \Sigma_{1} \Sigma_{3} \bigl(\Sigma_{1}^2-4 \Sigma_{2}\bigr)-128
   \Sigma_{3}^2\Bigr) \Bigr.  \\*  \Bigl. 
   \hspace{35mm} +64 \Bigl(\bigl(\Sigma_{1}^2-2 \Sigma_{2}\bigr)
   \bigl(\Sigma_{1}^2-4 \Sigma_{2}\bigr)^2+32 \Sigma_{1} \Sigma_{3}
   \bigl(\Sigma_{1}^2-4 \Sigma_{2}\bigr)+192 \Sigma_{3}^2\Bigr) \Sigma_{4}\Bigr]
   \\
\hspace{-20mm} + W^2 \Bigl[\bigl(-\Sigma_{1}^2+\lozenge+4 \Sigma_{2}\bigr)
   \Sigma_{3} \bigl(\Sigma_{1}^2-4 \Sigma_{2}\bigr)^3+8 \Bigl(\lozenge
   \bigl(-5 \Sigma_{1}^3+20 \Sigma_{2} \Sigma_{1}-72 \Sigma_{3}\bigr) \Bigr.\Bigr.  \\* \Bigl. \Bigl. 
   +2
   \bigl(\Sigma_{1}^2-4 \Sigma_{2}\bigr) \bigl(\Sigma_{1}^3-4 \Sigma_{2}
   \Sigma_{1}+16 \Sigma_{3}\bigr)\Bigr) \Sigma_{4} \bigl(\Sigma_{1}^2-4
   \Sigma_{2}\bigr) \Bigr.  \\*  \Bigl. 
   \hspace{50mm}Ê-512 \bigl(2 \Sigma_{1}^3+3 \lozenge \Sigma_{1}-8
   \Sigma_{2} \Sigma_{1}+24 \Sigma_{3}\bigr) \Sigma_{4}^2\Bigr]
   \\ 
+  \Bigl[\bigl(\Sigma_{1}^2-4 \Sigma_{2}\bigr)^2-64 \Sigma_{4}\Bigr]^2
   \Sigma_{4}   = 0  \hspace{85mm}
\end{multline}

Although still complicated, this polynomial  is much simpler than what
could be expected on the basis of the resultant of two generic 
polynomial equations of the form \eqref{polS} and \eqref{polP}
above, which would generically lead to a polynomial of degree 12 in $W^2$.
In the next subsections, we shall see that the degree is further reduced
at special loci in the parameter space of $(\varphi, \rho_\m)$.

\subsection{Near $\varphi= \pi/4$}

At the point $\varphi= \pi/4$, it is straightforward if tedious to check that the
polynomial \eqref{sixDegree} at $\vareps=0$ admits the six roots
\bea
W_0^2 &=& \frac{1}{4}Ê\Scal{ \rho_\zero  + \rho_\un + \rho_\deux + \rho_\trois }^2 \ ,\qquad 
W_1^2 = \frac{1}{4}Ê\Scal{ \rho_\zero  + \rho_\un - \rho_\deux - \rho_\trois }^2 \ ,\nn\\ 
W_2^2 &=& \frac{1}{4}Ê\Scal{ \rho_\zero  - \rho_\un + \rho_\deux - \rho_\trois }^2 \ ,\qquad 
W_3^2 = \frac{1}{4}Ê\Scal{ \rho_\zero  - \rho_\un - \rho_\deux + \rho_\trois }^2 \ ,\\ 
\label{sixroots}
W^2_{\pm} &=& - \frac{\rho_\zero\rho_\un\rho_\deux\rho_\trois \lozenge}
{4(\rho_\zero\rho_\un+\rho_\deux\rho_\trois)
(\rho_\zero\rho_\deux+\rho_\un\rho_\trois)
(\rho_\zero\rho_\trois+\rho_\un\rho_\deux)}\ .\nn
\eea
The physical root corresponds to $W_0$. Indeed, using (\ref{Lozrho}), one easily checks that
the polynomial \eqref{polP} vanishes at $\alpha=0$ when $\varphi=\pi/4$. Moreover, the quadratic
equation \eqref{polyW} is solved by 
\be
{\rm x}_\n=1-\frac{\rho_\n}{W}= 1 - \frac{ 2 \rho_\n}{\rho_\zero + \rho_\un +\rho_\deux + \rho_\trois}\ ,
\ee 
which satisfies 
$ {\rm x}_\m>0, \sum {\rm x}_\m = 2$ as it should.
We conclude that at $\varphi=\pi/4$, the fake superpotential reduces to 
\be 
\label{physr}
\framebox{$W_0 = \frac{1}{2}Ê\Scal{ \rho_\zero  + \rho_\un + \rho_\deux + \rho_\trois } \ ,$}
\ee
in agreement with \cite{Andrianopoli:2007gt}. 
It is straightforward to compute deviations to \eqref{physr} away from the locus $\varphi=\frac{\pi}{4}$,
in Taylor series of  $\varepsilon \equiv \cos^2 2 \varphi$.
For the physical solution $W_0$, we find
\be
\label{Worder1}
\begin{split}
W_0 & = \frac12 \sigma_1 
\left[ 1 - \frac{2 \sigma_4}{\sigma_1 \sigma_3 - 4 \sigma_4} \varepsilon
+\frac{2 \sigma_{4}^2 \left(\sigma_{1}^4
   \sigma_{4}-\sigma_{1}^3 \sigma_{2} \sigma_{3}+4 \sigma_{1}^2
   \sigma_{3}^2-24 \sigma_{1} \sigma_{3} \sigma_{4}+48
   \sigma_{4}^2\right)}{(\sigma_{1} \sigma_{3}-4 \sigma_{4})^4}\varepsilon^2
+ \dots
\right]\ ,
\end{split}
\ee
where $\sigma_i\equiv \sigma_i(\rho_\m)$ are the elementary symmetric functions
of $\rho_\n$. 
More generally, the expansion of $W_\n$
to any order involves polynomials in $\sigma_i$ multiplying increasing powers of
of $\varepsilon \sigma_4/(\sigma_1 \sigma_3 - 4 \sigma_4)^3$. 
In contrast, the perturbation of the degenerate 
branch $W_\pm$ involves a power series in $\sqrt{\vareps}$.

\subsection{At $\rho_\deux = \rho_\trois $ - $S^2 T$ model \label{secs2t}}

At the point $\rho_\deux = \rho_\trois $, irrespective of the value of $\vareps$,
the discriminant factorizes into the square of a linear factor in $W^2$, with double
root at
\be
\label{W23}
W_{2,3}^2 =  \frac{\left(\rho_\zero^2-\rho_\un^2\right)^2}{4
   \left(\rho_\zero^2+\rho_\un^2+(2-4 \varepsilon) \rho_\zero
   \rho_\un\right)} \ ,
   \ee
and a quartic polynomial in $W^2$, 
\be
\begin{split}
&{\Sigma'_2} \left(\left({\Sigma'_1}-4 \rho_\deux^2\right)^2-4
   {\Sigma'_2}\right)^2
+ W^2 \left[ 36 \lozenge  {\Sigma'_2} \left(4
   \rho_\deux^2-{\Sigma'_1}\right)-\lozenge  \left(4
   \rho_\deux^2-{\Sigma'_1}\right)^3\right. \\ & \left. 
\qquad   - \left(\left({\Sigma'_1}-4
   \rho_\deux^2\right)^2-4 {\Sigma'_2}\right) 
  \left(4 {\Sigma'_2} \left(8
   \rho_\deux^2-5 {\Sigma'_1}\right)+{\Sigma'_1} \left({\Sigma'_1}-4
   \rho_\deux^2\right)^2\right)\right]\\
+&W^4 \left[-27 \lozenge ^2+6
   \lozenge  \left(16 \rho_\deux^4-32 \rho_\deux^2 {\Sigma'_1}+7 {\Sigma'_1}^2-12
   {\Sigma'_2}\right)+8 {\Sigma'_2} \left(16 \rho_\deux^4-40 \rho_\deux^2
   {\Sigma'_1}+15 {\Sigma'_1}^2\right)\right. \\ & \left. 
\qquad   +\left({\Sigma'_1}-4 \rho_\deux^2\right)^2
   \left(16 \rho_\deux^4+24 \rho_\deux^2 {\Sigma'_1}-15 {\Sigma'_1}^2\right)+16
   {\Sigma'_2}^2\right]\\
+&16 W^6 \left[3 \lozenge  \left(2
   \rho_\deux^2+{\Sigma'_1}\right)-\left(2 \rho_\deux^2-{\Sigma'_1}\right)
   \left(16 \rho_\deux^4-3 {\Sigma'_1}^2-4 {\Sigma'_2}\right)\right]
-64 W^8
   \left[\lozenge -\left({\Sigma'_1}-2
   \rho_\deux^2\right)^2\right]
   \label{fourDegree}
\end{split}
\ee
where we have denoted ${\Sigma'_1} = \rho_\zero^2+\rho_\un^2,{\Sigma'_2} =\rho_\zero^2\rho_\un^2$.
At $\vareps=0$,  all  roots become elementary and match on the roots in \eqref{sixroots}.
It should be noted that the degeneracy between $W_{2,3}$ is not lifted on the
$\rho_\deux=\rho_\trois$ locus, where $W_{2,3}$ is given exactly by \eqref{W23}, and
happens to be independent of $\rho_\deux$. 
As discussed in Section \ref{sectrunc}, the locus $\rho_\deux=\rho_\trois$ 
corresponds to the two-modulus model
of $\N=2$ supergravity, with prepotential $F=-S T^2$. While the
roots of the quartic polynomial \eqref{fourDegree} can be computed explicitely, their
expression is unilluminating.

\subsection{At $\rho_\un= \rho_\deux = \rho_\trois$  - $S^3$ model \label{secs3}}
At the point $\rho_\un= \rho_\deux = \rho_\trois$, the quartic polynomial
\eqref{fourDegree} further factorizes into a linear factor in $W^2$ vanishing at the same
point \eqref{W23}, which now occurs with  multiplicity 3, 
\be
\label{W123}
W_{1,2,3}^2 =  \frac{\left(\rho_\zero^2-\rho_\un^2\right)^2}{4
   \left(\rho_\zero^2+\rho_\un^2+(2-4 \varepsilon) \rho_\zero
   \rho_\un\right)}  \ ,
 \ee
times an irreducible cubic polynomial in $W^2$, 
\be
\label{threeDegree}
 \left(\rho_\zero^3-9 \rho_\zero
   \rho_\un^2\right)^2-12 
    \left(\rho_\zero^4  + 18 (2 \varepsilon-1) \rho_\zero \rho_\un^3-3 \rho_\zero^2
   \rho_\un^2\right)W^2
   +48  \left(\rho_\zero^2+3
   \rho_\un^2\right)W^4 -64 W^6 = 0\ .
\ee
The roots of the cubic polynomial are given by 
\bea
\label{rcu}
W^2_0 &= &  \frac{\rho_\zero^2}{4} 
\Scal{3x^2+1+3   D x + 3  D^{-1} \left(x(x+1)^2 -4 \vareps x^2\right)}\ ,\\
W^2_+ &=&   \frac{\rho_\zero^2}{4}
 \Scal{3x^2+1-3 (-1)^{1/3}D x+ 3 (-1)^{2/3} D^{-1} 
 \left(x(x+1)^2 -4 \vareps x^2\right)}\ ,\nn\\
W^2_- &=&   \frac{\rho_\zero^2}{4} 
\Scal{ 3x^2+1+3 (-1)^{2/3}D x-3 (-1)^{1/3} 
   D^{-1} \left(x(x+1)^2 -4 \vareps x^2\right) }\ ,\nn\
\eea
 where $x\equiv \rho_\un/\rho_\zero$ and $D$ is the real quantity
\be
D\equiv \left( (x+1 )^3 - 2
 \varepsilon (3 x^2+1)-2
 \sqrt{\varepsilon (1-\varepsilon) \left(
 3x^4+8 x^3(1-2\vareps)+6 x^2-1\right)}
 \right)^{1/3} \ . 
\ee
At $\varepsilon=0$, $D=x+1$ so that the three roots become elementary,
\bea
W^2_0 &= &
\frac14(\rho_\zero+3\rho_\un)^2 - \frac{\vareps\rho_\zero \rho_\un (\rho_\zero+3\rho_\un)^2}{3(\rho_\zero+\rho_\un)^2}
+ O(\vareps^2)\ , \nn \\
W^2_{1,2,3} &=& 
    \frac{1}{4}  (\rho_\zero-\rho_\un)^2 +\frac{ \vareps  \rho_\zero\rho_\un(\rho_\zero-\rho_\un)^2}
{(\rho_\zero+\rho_\un)^2}+O(\vareps)\ ,\nn\\
W^2_\pm &=&\frac14\rho_\zero(\rho_\zero-3\rho_\un) \pm \frac{\I \rho_\zero \rho_\un\sqrt{\vareps}}{2} \sqrt{9- \frac{12\rho_\zero}{\rho_\zero+\rho_\un}}  + O(\vareps)\ .
\eea
As the notation suggests, the roots $W_{1,2,3}$ match on to \eqref{sixroots} 
as $\vareps=0$, while the roots $W_{0,1,2,3}$ 
match onto the roots of the quartic polynomial \eqref{fourDegree} at $\rho_\un=\rho_\deux=\rho_3$.
Using
\be
\vareps = \frac12 + \frac{\Pfaff(Z)+\Pfaff(\bZ)}{4\rho_\zero \rho_\un\rho_\deux \rho_\trois}
= \frac12 + \frac{ Z_\zero \bZ_\un^3 + \bZ_\zero Z_\un^3}
{4| Z_\zero Z_\un^3|}\ ,
\ee
one may rewrite the physical root as
\be
\framebox{$
W_0 = \frac12 \sqrt{
| Z_\zero|^2 + 3 \scal{ | Z_\un|^2 + L_+ + L_-   }}\ ,$}
\ee
where $L_+$ and $L_-\equiv |Z_\zero Z_\un| D$ are given by 
\be
\label{eqL}
L^3_\pm = |Z_\un|^4 \Scal{Ê| Z_\un|^2 + 3 | Z|^2 } - \frac{1}{2}Ê\Scal{ÊZ {\bar Z_\un}^3 
+ \bar Z {Z_\un}^3 } \Scal{Ê|Z|^2 + 3 |Z_\un|^2 } \pm \frac{\sqrt{-\lozenge}}{2}Ê \bigl| Z {\bar Z_\un}{}^3 - \bar Z { Z_\un}^3 \bigr| \ .
\ee
%
As discussed in Section \ref{sectrunc}, this expression provides the fake superpotential 
for non-BPS extremal black holes in  the one-modulus model
of $\N=2$ supergravity, with prepotential $F=-S^3$.


\subsection{Near $\varphi=0$}
At  $\varphi=0$, \ie $\varepsilon=1$, 
the degree 6 polynomial \eqref{sixDegree} again becomes solvable,
with roots at 
\bea
\label{sixroots2}
W_{\pm}^2 &=& 
   \frac{\rho_\zero \rho_\un \rho_\deux \rho_{\trois} \lozenge}
   {4 (\rho_\un
   \rho_\deux-\rho_\zero \rho_{\trois}) (\rho_\un \rho_{\trois}-\rho_\zero
   \rho_\deux) (\rho_\zero \rho_\un-\rho_\deux \rho_{\trois})} \nn \\
W_0^2 &=& \frac14 \Scal{\rho_\zero-\rho_\un-\rho_\deux-\rho_\trois}^2\ ,\qquad
W_1^2 = \frac14 \Scal{\rho_\zero-\rho_\un+\rho_\deux+\rho_\trois}^2\ ,\\
W_2^2 &=& \frac14 \Scal{\rho_\zero+\rho_\un-\rho_\deux+\rho_\trois}^2\ ,\qquad
W_3^2 = \frac14 \Scal{\rho_\zero+\rho_\un+\rho_\deux-\rho_\trois}^2\ .\nn
\eea
Using again \eqref{Lozrho} and \eqref{polP}, one checks that the corresponding value of $\alpha$
is $\alpha=0$, the same value encountered at $\varphi=\pi/4$. The quadratic equation 
\eqref{polyW} gives again 
${\rm x}_\n=1-(\rho_\n/W)$
so that the physical condition ${\rm x}_\n$ now selects amongst $W_{0,1,2,3}$ the root 
\be
\label{physr2}
\framebox{$
\varphi=0:\ \qquad
W = \frac12 \Scal{ \rho_\zero  + \rho_\un + \rho_\deux + \rho_\trois }  - {\rm min}(\varrho_\m)\ .$}
\ee
For example, in  the case of the canonical $\N=8$ ordering  $\rho_\zero \ge \rho_\un \ge \rho_\deux \ge \rho_\trois$, the physical root at $\varphi=0$ is $W_3$, while in the case of the $S^3$ 
model the physical root at $\varphi=0$ is still $W_0$. 

Restricting to the locus $\rho_\deux=\rho_\trois$, the expansion near $\vareps=1$
is given by 
\bea
W^2_{0} &=& \frac14 (\rho_\zero-\rho_\un-2\rho_\deux)^2
-\frac{\rho_\zero \rho_\un \rho_\deux (-\rho_\zero+\rho_\un+2
   \rho_\deux)^2}{(\rho_\zero-\rho_\un) (\rho_\zero (2
   \rho_\un+\rho_\deux)-\rho_\deux (\rho_\un+2 \rho_\deux))}(\vareps-1) + \dots\nn \ ,\nn \\
   W^2_{1} &=& \frac14 (\rho_\zero-\rho_\un+2\rho_\deux)^2+
\frac{\rho_\zero \rho_\un \rho_\deux (\rho_\zero-\rho_\un+2
   \rho_\deux)^2}{(\rho_\zero-\rho_\un) (\rho_\zero (2
   \rho_\un-\rho_\deux)+\rho_\deux (\rho_\un-2 \rho_\deux))}(\vareps-1)+\dots\ ,\nn \\
   W^2_{2,3} &=& \frac14 (\rho_\zero+\rho_\un)^2
+\frac{\rho_\zero \rho_\un(\rho_\zero+\rho_\un)^2}{(\rho_\zero-\rho_\un)^2} (\vareps-1)+\dots
\ ,\nn \\
W_{\pm}^2 &=&\frac{\rho_\zero \rho_\un \left((\rho_\zero+\rho_\un)^2-4
   \rho_\deux^2\right)}{4 \rho_\zero \rho_\un-4 \rho_\deux^2}  \\
&&    \mp  \frac{\rho_\zero \rho_\un \rho_\deux }{2
   \left(\rho_\deux^2-\rho_\zero \rho_\un\right)^2}
   \sqrt{(1-\vareps)\Scal{ (\rho_\zero+\rho_\un)^2-4  \rho_\deux^2}
   \Scal{  4  \left(\rho_\deux^2-\rho_\zero \rho_\un\right)^2- \rho_\deux^2 (\rho_\zero-\rho_\un)^2}}
+\dots \nn
\eea
In order to identify the branches consistently with the previous labeling,  we have taken 
advantage of the solutions \eqref{W23}, \eqref{rcu}, which are valid at any $\vareps$.


\section{Explicit solutions \label{secsol}}

\subsection{D0-D4, BPS}

A prototype of a BPS solution is given by the D0-D4 black hole

\bea
\phi &=& -\frac12 \log( H_0 H^1 H^2 H^3/4)\ ,\quad\\
S &=& i \sqrt{\frac{H_0 H^1}{H^2 H^3}}\ ,\quad
T = i \sqrt{\frac{H_0 H^2}{H^3 H^1}}\ ,\quad
U = i \sqrt{\frac{H_0 H^3}{H^1 H^2}}\ ,\quad\\
\zeta^0 &=& \frac{\sqrt2}{H_0}\ ,\quad
\tzeta_i = -\frac{\sqrt2}{H^i}\ ,\quad
H_0 = \sqrt2 + Q_0 \rho\ ,\quad
H^i = \sqrt2 + P^I \rho
\eea
with $\tzeta_0=\zeta^i=\sigma=0$.

The mass and BH entropy are given by
\be
2 G M = \frac{1}{2\sqrt2} | Q_0 + P^1 + P^2+ P^3 |  \ ,\qquad
S_{\rm \scriptscriptstyle BH} = \sqrt{Q_0 P^1 P^2 P^3}
\ee
consistent with the BPS mass formula at $S=T=U=i$.  

One can check that the Noether matrix is nilpotent of degree 3, with Jordan form $[3^2 ,1^2]$.
The $SO(4,4)$ metric in the Jordan basis takes the form
\be
\left(
\begin{array}{cccccccc}
 -\frac{Q_{0}}{2 P^{1}} & 0 & 0 & 0 & 0 & 0 & 0 & 0 \\
 0 & -\frac{P^{2}}{2 P^{3}} & 0 & 0 & 0 & 0 & 0 & 0 \\
 0 & 0 & 0 & 0 & -\frac{P^{2} P^{3}}{2} & 0 & 0 & 0 \\
 0 & 0 & 0 & \frac{P^{2} P^{3}}{2} & 0 & 0 & 0 & 0 \\
 0 & 0 & -\frac{P^{2} P^{3}}{2} & 0 & 0 & 0 & 0 & 0 \\
 0 & 0 & 0 & 0 & 0 & 0 & 0 & -\frac{P^{1} Q_{0}}{2} \\
 0 & 0 & 0 & 0 & 0 & 0 & \frac{P^{1} Q_{0}}{2} & 0 \\
 0 & 0 & 0 & 0 & 0 & -\frac{P^{1} Q_{0}}{2} & 0 & 0
\end{array}
\right)
 \label{jordanBPSmetric}
\ee 
Since $P^i Q_0>0$ for all $i$, the signature is
$[(+-+)^2 (-)^2]$, and so BPS black holes correspond to the nilpotent 
orbit $[(+-+)^2 (-)^2]_{I}$.

\subsection{D0-D4, non-BPS $Z_*=0$}

A prototype of the non-BPS, $Z_*=0$ solution of type (b) is the $D0-D4$ solution

The simplest solution with these properties is 
\be
\begin{split}
\phi &= -\frac12 \log( H_0 H^1 H^2 H^3/4)\ ,\quad\\
S &= i \sqrt{\frac{H_0 H^1}{H^2 H^3}}\ ,\quad
T = i \sqrt{\frac{H_0 H^2}{H^3 H^1}}\ ,\quad
U = i \sqrt{\frac{H_0 H^3}{H^1 H^2}}\ ,\quad\\
\zeta^0 &= \frac{\sqrt2}{H_0}\ ,\quad
\tzeta_1 = -\frac{\sqrt2}{H^1}\ ,\quad
\tzeta_2 = \frac{\sqrt2}{H^2}\ ,\quad
\tzeta_3 = \frac{\sqrt2}{H^3}\ ,\quad\\
H_0 &= \sqrt2 + Q_0 \rho\ ,\quad
H^1 = \sqrt2 + P^1 \rho\ ,\quad
H^2= \sqrt2 - P^2\rho \ ,\quad
H^3 = \sqrt2 - P^3 \rho
\end{split}
\ee
with $\tzeta_0=\zeta^i=\sigma=0$, and charges such that 
\be
\label{signc2}
Q_0>0\ ,\qquad P^1>0\ ,\qquad P^2<0\ ,\qquad P_3<0 \ .
\ee
The mass and entropy are given by
 \be 2 G M =
\frac{1}{\sqrt2} (  P^1 - P^2- P^3 + Q_0 ) \ ,\qquad S_{\rm \scriptscriptstyle BH}
= 2\pi\sqrt{Q_0 P^1 P^2 P^3} 
\ee 

The Noether charge is still  nilpotent of
order 3 and with Jordan form $[3^2 ,1^2]$. Moreover, in the Jordan
basis the metric once again takes the form
\eqref{jordanBPSmetric}. Thus, it corresponds to the
nilpotent orbit $[(+-+)^2 (-)^2]_{II}$.  
The other cases $(c,d)$ are obtained by permutations, and correspond to
the nilpotent orbits $[(-+-)^2 (+)^2]_{I,II}$. 

\subsection{D0-D4, non-BPS $Z\neq 0$}

The simplest non-BPS, $Z\neq 0$ solution with these properties is \cite{Ceresole:2007wx,Hotta:2007wz,Gimon:2007mh}
\be
\begin{split}
\phi &= -\frac12 \log( -H_0 H^1 H^2 H^3/4)\ ,\quad\\
S &= i \sqrt{-\frac{H_0 H^1}{H^2 H^3}}\ ,\quad
T = i \sqrt{-\frac{H_0 H^2}{H^3 H^1}}\ ,\quad
U = i \sqrt{-\frac{H_0 H^3}{H^1 H^2}}\ ,\quad\\
\zeta^0 &= \frac{\sqrt2}{H_0}\ ,\quad
\tzeta_1 = -\frac{\sqrt2}{H^1}\ ,\quad
\tzeta_2 = -\frac{\sqrt2}{H^2}\ ,\quad
\tzeta_3 = -\frac{\sqrt2}{H^3}\ ,\quad\\
H_0 &= -\sqrt2 + Q_0 \rho\ ,\quad
H^1 = \sqrt2 + P^1 \rho\ ,\quad
H^2= \sqrt2 + P^2\rho \ ,\quad
H^3 = \sqrt2 + P^3 \rho
\end{split}
\ee
with $\tzeta_0=\zeta^i=\sigma=0$, and 
for definiteness, we focus on case (e) above, 
\be
\label{signc3}
Q_0<0\ ,\qquad P^1>0\ ,\qquad P^2>0\ ,\qquad P^3>0 \ .
\ee
The mass and entropy are given by
 \be 2 G M =
\frac{1}{\sqrt2} (  P^1 + P^2+ P^3 - Q_0 ) \ ,\qquad S_{\rm \scriptscriptstyle BH}
= 2\pi\sqrt{- Q_0 P^1 P^2 P^3} 
\ee 
The scalar potential has two flat
directions at the horizon, generated by the vectors $P_i \pa_{B^i}
-  P_j \pa_{B^j}$, $i,j=1\dots 3$.

The Noether charge is still  nilpotent of
order 3 and with Jordan form $[3^2 ,1^2]$. 
The $SO(4,4)$ metric in the Jordan basis is still given by
\eqref{jordanBPSmetric}, but now has  signature 
$[(+-+),(-+-),+,-]$


\begin{thebibliography}{99}

\bibitem{Ferrara:1995ih}
  S.~Ferrara, R.~Kallosh and A.~Strominger,
  ``$\N=2$ extremal black holes,''
  Phys.\ Rev.\  D {\bf 52} (1995) 5412
  \eprint{hep-th/9508072}.

\bibitem{Ferrara:1996um}
  S.~Ferrara and R.~Kallosh,
  ``Universality of supersymmetric attractors,''
  Phys.\ Rev.\  D {\bf 54} (1996) 1525
  \eprint{hep-th/9603090}.

\bibitem{Ferrara:1997tw}
  S.~Ferrara, G.~W.~Gibbons and R.~Kallosh,
  ``Black holes and critical points in moduli space,''
  Nucl.\ Phys.\  B {\bf 500} (1997) 75
  \eprint{hep-th/9702103}.

\bibitem{Sen:2005wa}
  A.~Sen,
  ``Black hole entropy function and the attractor mechanism in higher
  derivative gravity,''
  JHEP {\bf 0509} (2005) 038
  \eprint{hep-th/0506177}.
  
  
\bibitem{Goldstein:2005hq}
  K.~Goldstein, N.~Iizuka, R.~P.~Jena and S.~P.~Trivedi,
  ``Non-supersymmetric attractors,''
  Phys.\ Rev.\  D {\bf 72} (2005) 124021
  \eprint{hep-th/0507096}.

\bibitem{Dabholkar:2006tb}
  A.~Dabholkar, A.~Sen and S.~P.~Trivedi,
  ``Black hole microstates and attractor without supersymmetry,''
  JHEP {\bf 0701} (2007) 096
  \eprint{hep-th/0611143}.

\bibitem{Denef:2000nb}
  F.~Denef,
  ``Supergravity flows and D-brane stability,''
  JHEP {\bf 0008} (2000) 050
  [arXiv:hep-th/0005049].


\bibitem{Bates:2003vx}
  B.~Bates and F.~Denef,
  ``Exact solutions for supersymmetric stationary black hole composites,''
  arXiv:hep-th/0304094.


\bibitem{Khuri:1995xq}
  R.~R.~Khuri and T.~Ort\'\i n,
  ``A Non-Supersymmetric Dyonic Extreme Reissner-Nordstrom Black Hole,''
  Phys.\ Lett.\  B {\bf 373} (1996) 56
  [arXiv:hep-th/9512178].


\bibitem{Ortin:1996bz}
  T.~Ort\'{\i}n,
  ``Extremality versus supersymmetry in stringy black holes,''
  Phys.\ Lett.\  B {\bf 422} (1998) 93
  \eprint{hep-th/9612142}.


\bibitem{Tripathy:2005qp}
  P.~K.~Tripathy and S.~P.~Trivedi,
  ``Non-supersymmetric attractors in string theory,''
  JHEP {\bf 0603} (2006) 022
  \eprint{hep-th/0511117}.

\bibitem{Ceresole:2007wx}
  A.~Ceresole and G.~Dall'Agata,
  ``Flow equations for non-BPS extremal black holes,''
  JHEP {\bf 0703}, 110 (2007)
  \eprint{hep-th/0702088}.

\bibitem{Andrianopoli:2007gt}
  L.~Andrianopoli, R.~D'Auria, E.~Orazi and M.~Trigiante,
  ``First order description of black holes in moduli space,''
  JHEP {\bf 0711} (2007) 032
  \eprintN{0706.0712}.

\bibitem{Lopes Cardoso:2007ky}
  G.~Lopes Cardoso, A.~Ceresole, G.~Dall'Agata, J.~M.~Oberreuter and J.~Perz,
  ``First-order flow equations for extremal black holes in very special
  geometry,''
  JHEP {\bf 0710} (2007) 063
  \eprintN{0706.3373}.

\bibitem{Perz:2008kh}
  J.~Perz, P.~Smyth, T.~Van Riet and B.~Vercnocke,
  ``First-order flow equations for extremal and non-extremal black holes,''
  JHEP {\bf 0903} (2009) 150
  \eprintN{0810.1528}.



\bibitem{Hotta:2007wz}
  K.~Hotta and T.~Kubota,
  ``Exact solutions and the attractor mechanism in non-BPS black holes,''
  Prog.\ Theor.\ Phys.\  {\bf 118} (2007) 969
  \eprintN{0707.4554}.


\bibitem{Gimon:2007mh}
  E.~G.~Gimon, F.~Larsen and J.~Simon,
  ``Black holes in supergravity: the non-BPS branch,''
  JHEP {\bf 0801} (2008) 040
  \eprintN{0710.4967}.

\bibitem{Bellucci:2008sv}
  S.~Bellucci, S.~Ferrara, A.~Marrani and A.~Yeranyan,
  ``$stu$ black holes unveiled,''
  \eprintN{0807.3503}.
  
\bibitem{Gimon:2009gk}
  E.~G.~Gimon, F.~Larsen and J.~Simon,
  ``Constituent Model of Extremal non-BPS Black Holes,''
  JHEP {\bf 0907} (2009) 052
  \eprintN{0903.0719}.

\bibitem{Gaiotto:2007ag}
  D.~Gaiotto, W.~W.~Li and M.~Padi,
  ``Non-supersymmetric attractor flow in symmetric spaces,''
  JHEP {\bf 0712} (2007) 093
  \eprintN{0710.1638}.

\bibitem{Gunaydin:2005mx}
  M.~G\"unaydin, A.~Neitzke, B.~Pioline and A.~Waldron,
  ``BPS black holes, quantum attractor flows and automorphic forms,''
  Phys.\ Rev.\  D {\bf 73} (2006) 084019
  \eprint{hep-th/0512296}.

\bibitem{Berkooz:2008rj}
  M.~Berkooz and B.~Pioline,
  ``5D black holes and non-linear sigma models,''
  JHEP {\bf 0805} (2008) 045
  \eprintN{0802.1659}.

\bibitem{Michel:2008bx}
 Y.~Michel, B.~Pioline and C.~Rousset,
  ``$\N=4$ BPS black holes and octonionic twistors,''
   JHEP {\bf 0811} (2008) 068
   \eprintN{0806.4563}.


\bibitem{Bossard:2009at}
  G.~Bossard, H.~Nicolai and K.~S.~Stelle,
  ``Universal BPS structure of stationary supergravity solutions,''
  \eprintN{0902.4438}.

\bibitem{Bossard:2009my}
  G.~Bossard and H.~Nicolai,
  ``Multi-black holes from nilpotent Lie algebra orbits,''
  \eprintN{0906.1987}.

\bibitem{Bossard:2009mz}
  G.~Bossard,
  ``The extremal black holes of $\N=4$ supergravity from $\so(8,2+n)$ nilpotent
  orbits,''
  \eprintN{0906.1988}.
  
\bibitem{Breitenlohner:1987dg}
  P.~Breitenlohner, D.~Maison and G.~W.~Gibbons,
  ``Four-dimensional black holes from Kaluza--Klein theories,''
  Commun.\ Math.\ Phys.\  {\bf 120} (1988) 295.
 
\bibitem{Clement:1996nh}
  G.~Clement and D.~V.~Galtsov,
  ``Stationary BPS solutions to dilaton-axion gravity,''
  Phys.\ Rev.\  D {\bf 54} (1996) 6136
  \eprint{hep-th/9607043}.

\bibitem{Bergshoeff:2008be}
  E.~Bergshoeff, W.~Chemissany, A.~Ploegh, M.~Trigiante and T.~Van Riet,
  ``Generating geodesic flows and supergravity solutions,''
  Nucl.\ Phys.\  B {\bf 812} (2009) 343
  \eprintN{0806.2310}.


\bibitem{Pioline:2006ni} B.~Pioline, ``Lectures on on black holes,
  topological strings and quantum attractors,'' 
  Class.\ Quant.\ Grav.\  {\bf 23}  (2006) S981 
  \eprint{hep-th/0607227}.


\bibitem{Ferrara:1989ik}
  S.~Ferrara and S.~Sabharwal,
  ``Quaternionic manifolds for type II superstring vacua of Calabi--Yau
  spaces,''
  Nucl.\ Phys.\  B {\bf 332} (1990) 317.
  
\bibitem{Moore:1998pn}
  G.~W.~Moore,
  ``Arithmetic and attractors,''
  \eprint{hep-th/9807087}.

 

\bibitem{Bellucci:2007zi}
  S.~Bellucci, A.~Marrani, E.~Orazi and A.~Shcherbakov,
  ``Attractors with vanishing central charge,''
  Phys.\ Lett.\  B {\bf 655} (2007) 185
 \eprintN{0707.2730}.

\bibitem{Andrianopoli:2009je}
  L.~Andrianopoli, R.~D'Auria, E.~Orazi and M.~Trigiante,
  ``First order description of $D=4$ static black holes and the Hamilton--Jacobi
  equation,''
  \eprintN{0905.3938}.
 

\bibitem{Gunaydin:2007bg}
  M.~G\"unaydin, A.~Neitzke, B.~Pioline and A.~Waldron,
  ``Quantum attractor flows,''
  JHEP {\bf 0709} (2007) 056
 \eprintN{0707.0267}.

\bibitem{Neitzke:2007ke}
  A.~Neitzke, B.~Pioline and S.~Vandoren,
  ``Twistors and black holes,''
  JHEP {\bf 0704} (2007) 038
   \eprint{hep-th/0701214}.

\bibitem{Gunaydin:2007qq}
  M.~G\"unaydin, A.~Neitzke, O.~Pavlyk and B.~Pioline,
  ``Quasi-conformal actions, quaternionic discrete series and twistors: $SU(2,1)$
  and $G_{2(2)}$,''
  Commun.\ Math.\ Phys.\  {\bf 283} (2008) 169
  \eprintN{0707.1669}.


  \bibitem{Ferrara:1997uz}
  S.~Ferrara and M.~G\"unaydin,
  ``Orbits of exceptional groups, duality and BPS states in string theory,''
  Int.\ J.\ Mod.\ Phys.\  A {\bf 13} (1998) 2075
  \eprint{hep-th/9708025}.

\bibitem{Bellucci:2006xz}
  S.~Bellucci, S.~Ferrara, M.~G\"unaydin and A.~Marrani,
  ``Charge orbits of symmetric special geometries and attractors,''
  Int.\ J.\ Mod.\ Phys.\  A {\bf 21} (2006) 5043
  \eprint{hep-th/0606209}.



\bibitem{Collingwood}
D.~Collingwood and W.~ McGovern,
``Nilpotent orbits in semisimple Lie algebras"
Van Nostrand Reinhold Mathematics Series,  New York, 1993.

  \bibitem{Levi}
 D.~\v{Z}.~\DJo, 
 ``Classification of nilpotent elements in simple exceptional real Lie algebras of inner type and description of their centralizers,''
 J.\ of Algebra {\bf 112} (1988) 503. 
  


 \bibitem{CremmerJulia}
  E.~Cremmer and B.~Julia,
  ``The $SO(8)$ Supergravity,''
  Nucl.\ Phys.\  B {\bf 159} (1979) 141.

\bibitem{de Wit:1982ig}
  B.~de Wit and H.~Nicolai,
  ``$\N=8$ Supergravity,''
  Nucl.\ Phys.\  B {\bf 208} (1982) 323.


\bibitem{E8strat}
D.~\v{Z}.~\DJo,
``The closure diagram for nilpotent orbits of the split real form of $E_8$,''
CEJM {\bf 4} (2003) 573.



\bibitem{Kallosh:1996uy}
  R.~Kallosh and B.~Kol,
  ``$E(7)$ symmetric area of the black hole horizon,''
  Phys.\ Rev.\  D {\bf 53} (1996) 5344
  \eprint{hep-th/9602014}.





\bibitem{Andrianopoli:1997wi}
  L.~Andrianopoli, R.~D'Auria, S.~Ferrara, P.~Fre and M.~Trigiante,
  ``E(7)(7) duality, BPS black-hole evolution and fixed scalars,''
  Nucl.\ Phys.\  B {\bf 509} (1998) 463
  \eprint{hep-th/9707087}.


\bibitem{Ferrara:2006em}
  S.~Ferrara and R.~Kallosh,
  ``On $\N = 8$ attractors,''
  Phys.\ Rev.\  D {\bf 73} (2006) 125005
  \eprint{hep-th/0603247}.



\bibitem{Ferrara:2007pc}
  S.~Ferrara and A.~Marrani,
  ``$\N=8$ non-BPS attractors, fixed scalars and magic supergravities,''
  Nucl.\ Phys.\  B {\bf 788} (2008) 63
  \eprintN{0705.3866}.

\bibitem{Ferrara:2007tu}
  S.~Ferrara and A.~Marrani,
  ``On the moduli space of non-BPS attractors for $\N=2$ symmetric manifolds,''
  Phys.\ Lett.\  B {\bf 652} (2007) 111
  \eprintN{0706.1667}.
  
  \bibitem{Gunaydin:1983rk}
  M.~G\"unaydin, G.~Sierra and P.~K.~Townsend,
  ``Exceptional supergravity theories and the magic square,''
  Phys.\ Lett.\  B {\bf 133} (1983) 72.
  
  \bibitem{magicstratVIII}
D.~\v{Z}.~\DJo,
``The closure diagrams for nilpotent orbits of the real form E IX of $E_8$,''
Asian J.\  Math.\  {\bf 5} (2001) 561.

\bibitem{magicstratVII}
D.~\v{Z}.~\DJo,
``The closure diagrams for nilpotent orbits of the real forms E VI and E VII of $E_7$,''
Representation Theory {\bf 5} (2001) 17.

\bibitem{magicstratVI}
D.~\v{Z}.~\DJo,
``The closure diagrams for nilpotent orbits of real forms of $E_6$,''
J.\ Lie Theory {\bf 11} (2001) 381.

\bibitem{magicstratIV}
D.~\v{Z}.~\DJo,
``The closure diagrams for nilpotent orbits of real forms of $F_4$ and $G_2$,''
J.\ Lie Theory {\bf 10} (2000) 491.


\bibitem{Ceresole:1995ca}
  A.~Ceresole, R.~D'Auria and S.~Ferrara,
  ``The symplectic structure of $\N=2$ supergravity and its central extension,''
  Nucl.\ Phys.\ Proc.\ Suppl.\  {\bf 46} (1996) 67
  \eprint{hep-th/9509160}.

\bibitem{DokovicSO}
 D.~\v{Z}.~\DJo, N. Lemire, J. Sekiguchi, 
``The closure ordering of adjoint nilpotent orbits in $\so(p,q)$,
Tohoku Mat.\ J.\ {\bf 53}  (2001) 395.





\bibitem{Ceresole:2009iy}
  A.~Ceresole, G.~Dall'Agata, S.~Ferrara and A.~Yeranyan,
  ``First order flows for N=2 extremal black holes and duality invariants,''
  \eprintN{0908.1110}.



 
  

  
\end{thebibliography}
\end{document}